\documentclass[longauth]{aa}
\usepackage{colortbl}
\usepackage{color}
\usepackage{booktabs}
\usepackage[normalem]{ulem}
\useunder{\uline}{\ul}{}
\usepackage{longtable}
\usepackage[utf8]{inputenc}
\usepackage{graphicx}
\usepackage[varg]{txfonts}    
\usepackage{textcomp}
\usepackage{multirow}
\usepackage{caption}
\usepackage{footnotehyper}
\usepackage{subcaption}
\usepackage[shortlabels]{enumitem}
\usepackage{hyperref}
\hypersetup{colorlinks=true,
            citecolor=blue,
            linkcolor=blue,
            urlcolor=blue}

\begin{document} 

\title{A large topographic feature on the surface of the trans-Neptunian object (307261) 2002 MS$_4$ measured from stellar occultations\thanks{Tables B.1, B.2, B.3, B.4, and B.5 are only available in electronic form at the CDS via anonymous ftp to \url{cdsarc.cds.unistra.fr} (130.79.128.5) or via xxxxxx}}
\titlerunning{stellar occultation events by (307261) 2002 MS$_4$}
\authorrunning{Rommel, F. L., et al.}

\author{F. L. Rommel\inst{1,2,3} 
\and
F. Braga-Ribas\inst{3,1,2} 
\and
J. L. Ortiz\inst{4} 
\and
B. Sicardy\inst{5} 
\and
P. Santos-Sanz\inst{4} 
\and
J. Desmars\inst{6,7} 
\and
J. I. B. Camargo\inst{1,2} 
\and
R.~Vieira-Martins\inst{1,2} 
\and
M. Assafin\inst{8,2} 
\and
B. E. Morgado\inst{8,2,1} 
\and
R. C. Boufleur\inst{1,2} 
\and
G. Benedetti-Rossi\inst{9,5,2} 
\and
A.~R.~Gomes-Júnior\inst{10,9,2} 
\and
E. Fernández-Valenzuela\inst{11} 
\and
B. J. Holler\inst{12} 
\and
D. Souami\inst{5,13,14}\thanks{Fulbright Visiting Scholar (2022 - 2023) at University of California, Berkeley} 
\and
R. Duffard\inst{4} 
\and
G. Margoti\inst{3,2} 
\and
M. Vara-Lubiano\inst{4} 
\and
J. Lecacheux\inst{5} 
\and
J. L. Plouvier\inst{15} 
\and
N. Morales\inst{4} 
\and
A. Maury\inst{16} 
\and
J. Fabrega\inst{17} 
\and
P. Ceravolo\inst{18} 
\and
E. Jehin\inst{19} 
% ...
}

\institute{National Observatory/MCTI, R. General José Cristino 77, CEP20921-400, Rio de Janeiro, Brazil\\
\email{flaviarommel@on.br}
\and
Interinstitutional e-Astronomy Laboratory - LIneA $\&$ INCT do e-Universo, Rio de Janeiro, Brazil
\and
Federal University of Technology - Paraná (UTFPR - Curitiba), Av. Sete de Setembro 3165, Curitiba, Paraná, Brazil
\and
Institute of Astrophysics of Andalucía, IAA-CSIC, Glorieta de la Astronomía s/n, 18008 Granada, Spain
\and
LESIA, Paris Observatory, PSL University, CNRS, Sorbonne University, Univ. Paris Diderot, Sorbonne Paris Cité, 5 place Jules Janssen, 92195 Meudon, France
\and
Polytechnic Institute of Advanced Sciences-IPSA, 63 boulevard de Brandebourg, 94200 Ivry-sur-Seine, France
\and
Institute of Celestial Mechanics and Ephemeris Calculation (IMCCE), Paris Observatory, PSL Research University, CNRS, Sorbonne University, UPMC Univ Paris 06, Univ. Lille, 77, Av. Denfert-Rochereau, 75014 Paris, France
\and
Valongo Observatory, Federal University of Rio de Janeiro (UFRJ), Ladeira Pedro Ant\^onio 43 - Saúde, 20080-090 Rio de Janeiro, Brazil
\and
Orbital Dynamics and Planetology Group, São Paulo State University (UNESP), Av. Ariberto Pereira da Cunha 333, Guaratinguetá, 12516-410 São Paulo, Brazil
\and
Institute of Physics, Federal University of Uberlândia, Uberlândia, Minas Gerais, Brazil
\and
Florida Space Institute, University of Central Florida, 12354 Research Parkway, Partnership 1, Orlando, Florida, United States of America
\and
Space Telescope Science Institute, Baltimore, Maryland, United States of America
\and
Department of Astronomy, and of Earth and Planetary Science, 501, Campbell Hall, University of California, Berkeley, California 94720, United States of America
\and
naXys, Department of Mathematics, University of Namur, Rue de Bruxelles 61, 5000 Namur, Belgium
\and
Domaine de la Blaque Observatory,  83670 Varages, France
\and
San Pedro de Atacama Celestial Explorations (SPACE), Chicache 81, Ayllu de Solor, 1410000 San Pedro de Atacama, Chile
\and
Panamanian Observatory in San Pedro de Atacama (OPSPA), Chile
\and
Anarchist Mt. Observatory, British Columbia, Canada
\and
STAR Institute, University of Liège, Allée du 6 aoûtl 19C, 4000 Liège, Belgium
\pagebreak
% ...
}

  \date{Last version: \today}

 
  \abstract
  {The physical characterization of trans-Neptunian objects is essential for improving our understanding of the formation and evolution of our Solar System. Stellar occultation is a ground-based technique that can be successfully used to determine some of the TNOs' fundamental physical properties with high precision, such as size and shape.
  }
  {This work is aimed at constraining the size, shape, and geometric albedo of the dwarf planet candidate (307261) 2002 MS$_4$ through the analysis of nine stellar occultation events. Using multichord detection, we also study the object's topography by analyzing the obtained limb and residuals between the observed chords and the best-fit ellipse.
}
   {We predicted and organized the observational campaigns of nine stellar occultations by 2002 MS$_4$ between 2019 and 2022, resulting in two single-chord events, four double-chord detections, and three events with between 3 and 61 positive chords. We derived the occultation light curves using differential aperture photometry, from which the star ingress and egress instants were calculated. Using 13 selected chords from the 8 August 2020 event, we determined the global elliptical limb of 2002 MS$_4$. The best-fit ellipse, combined with the object's rotational information from the literature, sets constraints on the object's size, shape, and albedo. Additionally, we developed a new method to characterize the topography features on the object's limb.
   }
   {The global limb has a semi-major axis of $\mathnormal{a'}$ = 412 $\pm$ 10 km, a semi-minor axis of $\mathnormal{b'}$= 385 $\pm$ 17 km, and the position angle of the minor axis is 121$^\circ$ $\pm$ 16$^\circ$. From this instantaneous limb, we obtained 2002 MS$_4$'s geometric albedo of $p_V$ = 0.1 $\pm$ 0.025,  using $H_V$~=~3.63~$\pm$~0.05~mag and a projected area-equivalent diameter of 796 $\pm$ 24 km. Significant deviations from the fitted ellipse in the northernmost limb were detected  from multiple sites, highlighting three distinct topographic features: one 11 km depth depression, followed by a 25$^{+4}_{-5}$ km height elevation next to a crater-like  depression, with an extension of 322 $\pm$ 39 km and 45.1 $\pm$ 1.5 km deep.  
   }
 {Our results indicate the presence of an object that is $\approx$138 km smaller in diameter than that derived from thermal data, possibly indicating the presence of a thus-far unknown satellite. However, within the error bars, the geometric albedo in the V-band is in agreement with the results published in the literature, even with the radiometric-derived albedo. This stellar occultation has allowed for the first multichord measurement of a large topography in a TNO.
 }

   \maketitle

\section{Introduction}

Trans-Neptunian objects (TNOs) are small Solar System bodies that orbit the Sun with a semi-major axis larger than that of Neptune \citep{Jewitt2008}. Due to the low spatial density of material in this orbital region and the significant distance from the Sun, their global physical-chemical composition has been largely unaffected since their formation. Therefore, they are considered to be remnants of the primordial disk and a valuable source of information about the primitive solar nebula and the evolution of our planetary system \citep{Gladman2008, Morbidelli2008, Nesvorny2012}. In addition, knowledge of the size-frequency distribution of TNOs allows for constraints to be placed on  Solar System formation models \citep{Petit2008}. Mainly due to the faintness and small angular sizes seen from Earth, our knowledge of the fundamental physical properties of the TNO population is still scarce and fragmented \citep{Stansberry2008, Lellouch2013, Lacerda2014}. Since the discovery of (15760) Albion in 1992 \citep{Jewitt1993}, thousands of objects have been observed in this orbital region. However, the size and albedo of only 178 Centaurs (objects with unstable orbits between Jupiter and Neptune) and TNOs have been determined using thermal observations \citep{Mueller2020}. On the other hand, spacecraft visits can fully characterize these objects,  as in the case of the visit of the New Horizons mission \citep{Weaver2008} to the Pluto system \citep{Stern2015, Stern2020, Spencer20b} and (486958) Arrokoth \citep{Stern2019, Benecchi19, Buie2020a, Spencer20a}, which allowed for detailed studies. However, the aforementioned approaches require significant investment and cannot be used to study a larger number of objects.

Stellar occultation is an efficient ground-based  method to study dozens of these distant bodies. It consists of observing a background star while a small body passes in front of it and blocks the stellar flux for several seconds. An updated list of Lucky Star stellar occultation detections (that we are aware of) can be found in the SOSB Database\footnote{Stellar Occultation by Small Bodies database is available on \url{http://occultations.ct.utfpr.edu.br/results}} \citep{Braga-Ribas2019}. These observations provide an instantaneous limb of the object that can be combined with information derived from other observational methods to better characterize the small body \citep{Ortiz2020b}. 

In this work, we predicted, observed, and analyzed nine stellar occultations by the large TNO (307261) 2002 MS$_4$ (hereafter denoted as MS4 for brevity). It was discovered by the Near-Earth Asteroid Tracking (NEAT)\footnote{More information available at \url{https://sbn.psi.edu/pds/resource/neat.html}} program on June 18, 2002, and is classified as a hot classical TNO due to its high orbital inclination \citep{Gladman2008, Van_Laerhoven2019}. Furthermore, MS4 is a candidate to be a dwarf planet due to its thermally derived equivalent diameter \citep{Vilenius2012}. Physical and orbital parameters taken from previously published works are listed in Table \ref{table:1}.  

\begin{table}[!h]
\caption{Orbital and physical properties of MS4 from the literature.}             
\label{table:1}      
\centering                          
\begin{tabular}{l c|l c}  \hline
        \multicolumn{2}{c|}{\textbf{Orbital properties}\tablefootmark{a}} & \multicolumn{2}{|c}{\textbf{Physical properties}} \\ \hline
        a & 41.8 au        & D\tablefootmark{b} & 934 $\pm$ 47 km\\
        q & 35.75 au        & H\textsubscript{V}   & 4.0 $\pm$ 0.6\tablefootmark{b}$/$ 3.63\tablefootmark{c}\\
        i & 17.7$^\circ$   & p\textsubscript{V}\tablefootmark{b}    &0.051$^{+0.036}_{-0.022}$\\
        e & 0.14           & Ap\textsubscript{mag}\tablefootmark{d}   & 20.39 mag \\ \hline
    \end{tabular}
    \tablefoot{\tablefoottext{a}{Orbital elements from JPL Small-Boby Database Browser web page (\url{https://ssd.jpl.nasa.gov/tools/sbdb\_lookup.html\#/?sstr=2002MS4})}. \tablefoottext{b}{Physical properties obtained by \cite{Vilenius2012}: area-equivalent diameter (\textbf{D}) and geometric albedo at V-band (\textbf{p\textsubscript{V}}); \textbf{H\textsubscript{V}:} average absolute magnitude at V-band;  \cite{Stansberry2008} obtained D = 726.05 $\pm$ 123.05 km and p\textsubscript{V} = 0.084$^{+0.038}_{-0.023}$ for a value of H\textsubscript{V} = 4.0 using Spitzer data only.} \tablefoottext{c}{Information from \cite{Tegler2016}.}\tablefoottext{d}{\textbf{Ap\textsubscript{mag}:} object's average apparent visual magnitude on August 8, 2020, from JPL website \url{https://ssd.jpl.nasa.gov/horizons/app.html\#/}.}}
\end{table}

\section{Predictions and observations}
\label{sec:2}

We performed classical astrometric runs to refine MS4's ephemeris at the Pico dos Dias (Brazil), La Silla (Chile), Calar Alto (Spain), and Pic du Midi (France) observatories between 2009 and 2019. The updated ephemeris and the \textit{Gaia} Data Release 1 catalog \citep{Gaia16a, Gaia16b, Gaia18} significantly improved the prediction of the July 9, 2019 occultation, resulting in our first occultation by MS4. Furthermore, the astrometry derived from this occultation data improved the subsequent predictions. 

In 2019, we observed four stellar occultations by MS4 from Argentina, Brazil, Canada, and Chile (see Tables \ref{table:other_pos_sites} and \ref{table:other_neg_sites}). On July 26, we obtained a multichord detection from three well-separated sites and about eight hours later, a single-chord occultation of a different star from Canada. On July 9 and August 19, we detected two positive chords and a set of negatives. The astrometric results from 2019 data (Table \ref{table:astrometric_occ}) were used to calculate new ephemeris and predict the subsequent events using the Numerical Integration of the Motion of an Asteroid (NIMA) tool described on \cite{Desmars2015b}\footnote{MS4's ephemeris (NIMA v9) is publicly available on \url{https://lesia.obspm.fr/lucky-star/obj.php?p=692}.}

The first observation in 2020 was a double chord from South Africa on July 26, which confirmed the accuracy of MS4's ephemeris at an eight-milliarcsec (mas) level. Therefore, we organized an extensive campaign and successfully observed, from 61 sites, an occultation of a bright star (V = 14.62 mag) on August 8, 2020. As described in this work, we derived valuable physical information from this multichord event observed in North Africa, Europe, and Western Asia. After the 8 August 2020 campaign, three other events were observed on 24 February 2021, 14 October 2021, and 10 June 2022. Single, double, and triple detections, respectively (Tables \ref{table:other_pos_sites} and \ref{table:other_neg_sites}). Those data and equipment information collection processes were carried out through the Occultation Portal\footnote{More information on \url{https://occultation.tug.tubitak.gov.tr/about/}} \citep{Kilic2022}. Table \ref{table:star_properties} shows the relevant information about all the occulted stars from the \textit{Gaia} Data Release 3 catalog (GDR3,  \citeauthor{Gaia2021} \citeyear{Gaia2021}).

The default procedure for all events was to update the prediction and send alerts to potential observers within or close to the predicted shadow path. However, an exception was made for the 8 August 2020 occultation due to favorable circumstances. A campaign web page with helpful information for the observers was built\footnote{The campaign web page is available in \url{https://lesia.obspm.fr/lucky-star/campaigns/2020-08-08_2002MS4.html}}. Also, alerts were sent to all individuals with access to portable or professional telescopes and near or inside the predicted shadow path, resulting in such a large number of positive detections.

\begin{table*}[!ht]
\caption{Target stars designation and geocentric coordinates at closest approach instant (UT) sorted by occultation date (day-month-year).}
\label{table:star_properties}
\resizebox{\textwidth}{!}{%
\begin{tabular}{cccccccccc}
\hline
\textbf{Date} &
  \textbf{\begin{tabular}[c]{@{}c@{}}Gaia DR3\\ Designation \\ \end{tabular}} &
  \textbf{\begin{tabular}[c]{@{}c@{}}Propagated\\ right ascension\\ (hh mm ss.sssss)\end{tabular}} &
  \textbf{\begin{tabular}[c]{@{}c@{}}Error \\ (mas)\end{tabular}} &
  \textbf{\begin{tabular}[c]{@{}c@{}}Propagated\\ declination \\ (º \textquotesingle \hspace{1pt} \textquotesingle\textquotesingle)\end{tabular}} &
  \textbf{\begin{tabular}[c]{@{}c@{}}Error \\ (mas)\end{tabular}} &
  \textbf{\begin{tabular}[c]{@{}c@{}}V\tablefootmark{a} \\ (mag)\end{tabular}} &
  \textbf{\begin{tabular}[c]{@{}c@{}}K\tablefootmark{a} \\ (mag)\end{tabular}} &
  \textbf{\begin{tabular}[c]{@{}c@{}}S$_{\rm Diam}$\tablefootmark{b}\\ (km) \end{tabular}} &
  \textbf{\begin{tabular}[c]{@{}c@{}}$\Delta_{MS4}$\tablefootmark{c}\\ (au) \end{tabular}} \\ \hline
09-07-2019                  & 4253196402592965504 & 18 45 19.24565 & 0.15 & -06 24 13.0031 & 0.12 & 15.00 & 14.15 & 0.19 & 45.62 \\ \hline
\multirow{2}{*}{26-07-2019} & 4253186506987951104 & 18 44 07.57274 & 0.54 & -06 26 40.1240 & 0.46 & 17.78 & 16.27 & 0.08 & 45.67 \\ \cline{2-10} 
                            & 4253186477047835648 & 18 44 06.31756 & 0.13 & -06 26 43.8948 & 0.11 & 15.45 & 11.66 & 0.98 & 45.68 \\ \hline
19-08-2019                  & 4253181804071259648 & 18 42 43.51905 & 0.24 & -06 32 34.0868 & 0.19 & 16.51 & 16.59 & 0.05 & 45.88 \\ \hline
26-07-2020                  & 4253244201379441792 & 18 48 18.07372 & 0.12 & -06 13 31.6134 & 0.12 & 14.76 & 12.61 & 0.47 & 45.60 \\ \hline
08-08-2020                  & 4253248324549054464 & 18 47 29.96384 & 0.12 & -06 16 31.4727 & 0.10 & 14.62 & 11.13 & 1.19 & 45.70 \\ \hline
24-02-2021                  & 4253709191700784896 & 18 56 35.98731 & 0.25 & -06 30 23.1569 & 0.23 & 16.51 & 12.96 & 0.53 & 47.05 \\ \hline
14-10-2021                  & 4252495635735083264 & 18 50 30.76176 & 0.31 & -06 24 13.3375 & 0.27 & 15.83 & 13.44 & 0.34 & 46.52 \\ \hline
10-06-2022                  & 4253907305577009664 & 19 00 15.44628 & 0.23 & -05 42 42.9960 & 0.21 & 15.1  & 13.00 & 0.39 & 45.48 \\ \hline
\end{tabular}
}
\tablefoot{It is essential to mention that none of the stars have a duplicity flag in the \textit{Gaia} DR3 catalog. \tablefoottext{a}{The magnitudes retrieved from NOMAD catalog and used in SORA to calculate the} \tablefoottext{b}{stellar diameter ($S_{\rm Diam}$)} at the \tablefoottext{c}{MS4’s geocentric distance ($\Delta_{MS4}$).}}
\end{table*}

The data came from a wide range of telescopes, from small portable ones (apertures between 13 cm and 30 cm) to large facilities such as the 4.1 m telescope at the Southern Astrophysical Research (SOAR, Chile), the 2 m Liverpool telescope at Roque de Los Muchachos (Spain), the 1.6 m telescope at Pico dos Dias (Brazil), and the 1.5 m telescope at Sierra Nevada (Spain) observatories. Most observers did not use filters to maximize photon fluxes on the CCD and so, they were able to  a better signal-to-noise ratio (S/N). Even though some observers used the Global Positioning System (GPS) to acquire the time, the most common time source was the computer clock synchronization with a Network Time Protocol (NTP). A compilation of all the participating observers and instruments is presented in Appendix \ref{appendix1}. All the predictions and observational efforts were developed in the framework of the European Research Council (ERC) \textit{Lucky Star} project\footnote{\url{https://lesia.obspm.fr/lucky-star/}}. 

\section{Data reduction, analysis, and results}
\label{sec:3}

The great diversity of telescopes and detectors was reflected in five data formats\footnote{\textit{avi} = Audio Video Interleave. Information for the \textit{adv} (Astronomical Digital Video) can be found in \url{https://www.iota-es.de/JOA/JOA2020_3.pdf}. The documentation for the simple image sequence format known as \textit{ser} can be found in \url{http://www.grischa-hahn.homepage.t-online.de/astro/ser/}. \textit{cpa} is a compressed image file associated with \textsc{prism} (\url{http://www.prism-astro.com/fr/index.html}). The most recent Flexible Image Transport System-FITS documentation can be found in \url{https://fits.gsfc.nasa.gov/fits_standard.html}}: \textit{avi}, \textit{adv} \citep{Pavlov2020}, \textit{ser}, \textit{cpa}, and FITS. Most \textit{avi}, \textit{adv}, and \textit{ser} video files were converted to FITS images using \textsc{tangra}\footnote{\url{http://www.hristopavlov.net/Tangra3/}} software. However, from some videos, the images were extracted using a proprietary \textsc{python 3} script based on \textsc{astropy} v4.0.1 \citep{Astropy_2013}. When calibration images were available, the raw images were corrected from any bias, dark, and flat-field using standard procedures with the Image Reduction and Analysis Facility (\textsc{iraf}, \citeauthor{Butcher}\citeyear{Butcher}).

We applied aperture photometry on the target and some comparison stars on all the FITS files using the Package for the Reduction of Astronomical Images Automatically (\textsc{praia}, \citeauthor{Assafin11} \citeyear{Assafin11}, \citeyear{Assafin2023}). The chosen photometric apertures considered the maximization of the S/N. The light curve obtained for the target star, which includes less than 1\% of flux contributions from MS4, is then divided by the averaged light curves of the comparison stars to account for sky transparency variations in the data. Also, the flux outside the occultation is normalized to unity by fitting a polynomial function. Finally, the ingress and egress times were derived using the standard chi-square method ($\chi^2$ test) between the observed and a synthetic light curve implemented in the Stellar Occultation Reduction and Analysis package, v0.2.1 (\textsc{sora}, \citeauthor{Gomes-Junior2022}\citeyear{Gomes-Junior2022}). The synthetic light curve considers a sharp-edge model convolved with Fresnel diffraction, finite exposure time,  CCD bandwidth, and stellar diameter at MS4's distance (details about this procedure are available in \citeauthor{Gomes-Junior2022} \citeyear{Gomes-Junior2022} and references therein). The stellar diameters projected at MS4 distance were calculated according to \cite{Kervella2004}'s formalism and are listed in Table \ref{table:star_properties}.  Organized by the occultation date, Table \ref{tab:occ_times} contains the ingress and egress times (UTC) with 1$\sigma$ uncertainties for each station with a positive detection. Appendix \ref{appendix2} presents the normalized and the synthetic light curves used to obtain the occultation timings.

\begin{table}[!h]
\caption{Ingress and egress instants with $1\sigma$ error bars.}
\centering
\label{tab:occ_times}
\resizebox{0.76\linewidth}{!}{
\begin{tabular}{lll}
\hline
\multicolumn{1}{c}{\textbf{Sites}} &
  \multicolumn{1}{c}{\textbf{\begin{tabular}[c]{@{}c@{}}Ingress\\ (hh:mm:ss.ss $\pm$ ss.ss)\end{tabular}}} &
  \multicolumn{1}{c}{\textbf{\begin{tabular}[c]{@{}c@{}}Egress\\ (hh:mm:ss.ss $\pm$ ss.ss)\end{tabular}}} \\ \hline
\multicolumn{3}{c}{\textbf{09 July 2019}}                                    \\ \hline
OPSPA                  & 04:23:28.65 $\pm$ 0.19   & 04:23:58.26 $\pm$ 0.38   \\ \hline
ASH2                   & 04:23:27.2 $\pm$ 1.1     & 04:23:56.73 $\pm$ 2.1    \\ \hline
Pico dos Dias          & 04:22:02.182 $\pm$ 0.006 & 04:22:29.44 $\pm$ 0.67   \\ \hline
\multicolumn{3}{c}{\textbf{26 July 2019 - South America}}                    \\ \hline
OPSPA                  & 02:47:15.6 $\pm$ 7.8     & 02:48:04.0 $\pm$ 3.7     \\ \hline
ASH2                   & 02:47:21.0 $\pm$ 3.8     & 02:47:54.8 $\pm$ 5.3     \\ \hline
Paranal                & 02:47:34.33 $\pm$ 0.12   & 02:48:07.9 $\pm$ 1.2     \\ \hline
Pico dos Dias          & 02:45:55.310 $\pm$ 0.099 & 02:46:17.581 $\pm$ 0.093 \\ \hline
\multicolumn{3}{c}{\textbf{26 July 2019 - North America}}                    \\ \hline
Osoyoos                & 10:15:16.65 $\pm$ 0.40   & 10:15:53.12 $\pm$ 0.33   \\ \hline
\multicolumn{3}{c}{\textbf{19 August 2019}}                                  \\ \hline
Penticton              & 07:37:19.2 $\pm$ 2.6     & 07:37:55.2 $\pm$ 1.3     \\ \hline
Osoyoos                & 07:37:17.55 $\pm$ 0.55   & 07:37:53.7 $\pm$ 2.0     \\ \hline
\multicolumn{3}{c}{\textbf{26 July 2020}}                                    \\ \hline
Pretoria               & 23:15:28.46 $\pm$ 0.12   & 23:15:54.27 $\pm$ 0.15   \\ \hline
Johannesburg           & 23:15:28.940 $\pm$ 0.092 & 23:15:55.89 $\pm$ 0.18   \\ \hline
\multicolumn{3}{c}{\textbf{08 August 2020}}                                  \\ \hline
Varages                & 20:43:44.858 $\pm$ 0.058 & 20:43:53.359 $\pm$ 0.028 \\ \hline
TAROT North            & 20:43:39.5 $\pm$ 2.0     & 20:43:51.5 $\pm$ 2.0     \\ \hline
Méo station            & 20:43:39.488 $\pm$ 0.021 & 20:43:53.960 $\pm$ 0.027 \\ \hline
Caussols               & 20:43:39.27 $\pm$ 0.13   & 20:43:54.18 $\pm$ 0.56   \\ \hline
Lleida                 & 20:44:01.4 $\pm$  6.6    & 20:44:25.13 $\pm$  0.12  \\ \hline
Cannes                 & 20:43:37.23 $\pm$ 0.21   & 20:43:55.64 $\pm$ 0.39   \\ \hline
Nice                   & 20:43:36.444 $\pm$ 0.029 & 20:43:52.834 $\pm$ 0.071 \\ \hline
Valbonne               & 20:43:37.376 $\pm$ 0.055 & 20:43:54.07 $\pm$ 0.13   \\ \hline
\v Crni Vrh            & 20:43:09.71 $\pm$ 0.94   & 20:43:26.58 $\pm$ 0.35   \\ \hline
Mátraszentistván       & 20:42:46.866 $\pm$ 0.015 & 20:43:04.739 $\pm$ 0.045 \\ \hline
Ljubljana              & 20:43:06.70  $\pm$ 0.20  & 20:43:25.20 $\pm$ 0.33   \\ \hline
Budapest               & 20:42:49.59 $\pm$ 0.13   & 20:43:08.61 $\pm$ 0.20   \\ \hline
Trieste                & 20:43:07.7    $\pm$ 3.3  & 20:43:27.41 $\pm$  0.85  \\ \hline
Catalonia              & 20:43:57.15 $\pm$  0.21  & 20:44:17.84 $\pm$  0.14  \\ \hline
Bologna                & 20:43:17.40 $\pm$ 0.26   & 20:43:38.77 $\pm$  0.67  \\ \hline
Massa 02               & 20:43:20.52 $\pm$ 0.39   & 20:43:43.83 $\pm$ 0.32   \\ \hline
Castelvecchio Pascoli  & 20:43:18.742 $\pm$ 0.088 & 20:43:44.1 $\pm$ 1.4     \\ \hline
Borgo a Mozzano        & 20:43:19.224 $\pm$ 0.078 & 20:43:42.80 $\pm$ 0.12   \\ \hline
San Marcello Pistoiese & 20:43:16.200 $\pm$ 0.072 & 20:43:42.6 $\pm$ 1.2     \\ \hline
Sta. Maria a Monte     & 20:43:17.68 $\pm$ 0.12   & 20:43:42.80 $\pm$ 0.11   \\ \hline
Signa                  & 20:43:15.61 $\pm$ 0.57   & 20:43:40.38 $\pm$ 0.67   \\ \hline
Lastra a Signa         & 20:43:13.815 $\pm$ 0.084 & 20:43:41.9 $\pm$ 1.2     \\ \hline
Khmelnytskyi           & 20:42:01.81 $\pm$ 0.82   & 20:42:49.7 $\pm$ 15.7    \\ \hline
Sevilla                & 20:44:29.03 $\pm$ 0.13   & 20:44:59.40 $\pm$ 0.19   \\ \hline
Huelva                 & 20:44:32.74 $\pm$ 0.37   & 20:45:02.72 $\pm$ 0.28   \\ \hline
Cluj-Napoca            & 20:42:25.0 $\pm$ 1.2     & 20:42:56.61 $\pm$ 0.30   \\ \hline
Fuensanta de Martos    & 20:44:22.0 $\pm$ 3.0     & 20:44:55.0 $\pm$ 3.1     \\ \hline
Fiastra                & 20:43:05.1 $\pm$ 1.8     & 20:43:37.89 $\pm$ 0.52   \\ \hline
Dragsina               & 20:42:32.938 $\pm$ 0.087 & 20:43:05.82 $\pm$ 0.32   \\ \hline
Ibiza                  & 20:43:56.6 $\pm$ 1.8     & 20:44:28.55 $\pm$ 0.24   \\ \hline
Alhendín               & 20:44:17.89 $\pm$ 0.57   & 20:44:54.83 $\pm$ 0.18   \\ \hline
Granada (150 cm)       & 20:44:17.571 $\pm$ 0.008 & 20:44:52.08 $\pm$ 0.51   \\ \hline
Granada (90 cm)        & 20:44:16.53 $\pm$ 0.51   & 20:44:51.755 $\pm$ 0.026 \\ \hline
Estepona               & 20:44:24.37 $\pm$ 0.45   & 20:45:00.4  $\pm$ 1.9    \\ \hline
Marbella               & 20:44:26.1  $\pm$ 4.3    & 20:44:59.64 $\pm$ 0.22   \\ \hline
Rome                   & 20:43:08.37 $\pm$ 0.37   & 20:43:43.79 $\pm$ 0.37   \\ \hline
Hvar                   & 20:42:51.69 $\pm$ 0.21   & 20:43:27.96 $\pm$ 0.35   \\ \hline
Bacau                  & 20:42:12.60 $\pm$ 0.30   & 20:42:50.49 $\pm$ 0.72   \\ \hline
Sassari                & 20:43:19.35 $\pm$ 0.35   & 20:43:56.60 $\pm$ 0.50   \\ \hline
Guarcino               & 20:43:03.86 $\pm$ 0.15   & 20:43:41.13 $\pm$ 0.15   \\ \hline
Kharkiv (70 cm)        & 20:41:40.00 $\pm$ 0.13   & 20:42:17.16 $\pm$ 0.12   \\ \hline
Kharkiv (36 cm)        & 20:41:39.6 $\pm$ 1.9     & 20:42:18.3 $\pm$ 1.2     \\ \hline
Ceccano                & 20:43:02.39 $\pm$ 0.41   & 20:43:40.76 $\pm$ 0.37   \\ \hline
Brasov                 & 20:42:14.2 $\pm$ 2.7     & 20:42:51.5 $\pm$ 1.6     \\ \hline
Bârlad                 & 20:42:09.23 $\pm$ 0.55   & 20:42:47.00 $\pm$ 0.49   \\ \hline
Valenii de Munte       & 20:42:15.14 $\pm$ 0.24   & 20:42:52.61 $\pm$ 0.50   \\ \hline
Ploiesti               & 20:42:14.25 $\pm$ 2.1    & 20:42:53.2 $\pm$ 1.5     \\ \hline
Odesa                  & 20:42:00.00  $\pm$ 0.14  & 20:42:38.49 $\pm$ 0.11   \\ \hline
Sorrento               & 20:43:00.66  $\pm$ 0.30  & 20:43:37.7 $\pm$ 1.1     \\ \hline
Agerola                & 20:42:59.603 $\pm$ 0.050 & 20:43:37.842 $\pm$ 0.093 \\ \hline
Algiers                & 20:43:50.844 $\pm$ 0.021 & 20:44:29.115 $\pm$ 0.075 \\ \hline
La Palma               & 20:45:35.032 $\pm$ 0.014 & 20:46:08.313 $\pm$ 0.018 \\ \hline
Tijarafe               & 20:45:35.68  $\pm$ 0.19  & 20:46:08.08 $\pm$ 0.11   \\ \hline
Ariana                 & 20:43:23.87  $\pm$ 0.29  & 20:43:56.80 $\pm$ 0.25   \\ \hline
Artemis                & 20:45:31.938 $\pm$ 0.022 & 20:46:01.297 $\pm$ 0.024 \\ \hline
TAR 1                  & 20:45:32.360 $\pm$ 0.057 & 20:46:01.595 $\pm$ 0.063 \\ \hline
Catania                & 20:43:04.59 $\pm$ 0.92   & 20:43:34.564 $\pm$ 0.085 \\ \hline
Kuban                  & 20:41:36.2 $\pm$ 1.2     & 20:42:00.63 $\pm$ 0.31   \\ \hline
Çanakkale              & 20:42:23.39 $\pm$ 0.56   & 20:42:38.830 $\pm$ 0.072 \\ \hline
\multicolumn{3}{c}{\textbf{24 February 2021}}                                \\ \hline
OPSPA                  & 08:41:37.00 $\pm$ 0.61   & 08:42:09.182 $\pm$ 0.082 \\ \hline
ASH2                   & 08:41:36.82 $\pm$ 0.98   & 08:42:08.43 $\pm$ 0.30   \\ \hline
\multicolumn{3}{c}{\textbf{14 October 2021}}                                 \\ \hline
Osoyoos                & 03:23:30.14 $\pm$ 0.40   & 03:24:35.33 $\pm$ 0.67   \\ \hline
Flagstaff              & 03:25:54.61 $\pm$ 0.44   & 03:26:39.33 $\pm$ 0.73   \\ \hline
\multicolumn{3}{c}{\textbf{10 June 2022}}                                    \\ \hline
La Palma               & 05:30:08.475 $\pm$ 0.091 & 05:30:43.30 $\pm$ 0.13   \\ \hline
Artemis                & 05:30:02.427 $\pm$ 0.066 & 05:30:37.51 $\pm$ 0.49   \\ \hline
Tree Gate Farm         & 05:34:47.18  $\pm$ 0.33  & 05:35:22.12 $\pm$ 0.28   \\ \hline
\end{tabular}
}
\tablefoot{The times are sorted from northernmost to southernmost observatories regarding the object's probed latitude and divided by date.}
\end{table}

If not differentiated, large TNOs such as MS4 may reach one of the hydrostatic equilibrium shapes: the Jacobi three-axial ellipsoid or the Maclaurin oblate spheroid \citep{Chandrasekhar1987, Tancredi2008}. The apparent global limb of the body is then an ellipse projected in the sky plane, defined by M = 5 free parameters: center offset relative to the ephemeris (\textit{f} and \textit{g}), the semi-major axis ($a'$), the semi-minor axis ($b'$), or equivalently, the oblateness [$\epsilon'= (a'-b')/a'$], and the position angle (PA) of $b'$. The PA counts positively, starting from the celestial north and increasing to the east. We converted the ingress and egress times into a stellar position for each stellar occultation event, with \textit{f} and \textit{g} increasing toward celestial east and north, respectively. At this point, we can fit a limb model to these points, which provides, among others, the position of MS4's center in the sky plane and, thus, an ephemeris offset. 

Among the nine stellar occultation events, only three allow for an elliptical fit to the chords, namely: 9 July 2019, 8 August 2020, and 10 June 2022. We started our fitting procedure with the 61 chords acquired on 8 August 2020 (Sect. \ref{sec:8august}), and we then used the residuals of the elliptic limb fit to search for topographic features on MS4 (Sect. \ref{sec:topography}). Finally, we compared the resulting global ellipse with the chords observed in the other events (Sect. \ref{sec:other_events_limb}).

\subsection{8 August 2020}
\label{sec:8august}

Three circumstances triggered an extensive observational campaign for this occultation: i) the bright target star (G~=~14.6~mag from GDR3 catalog), ii) a milliarcsecond-level accuracy of MS4's ephemeris stemming from previously detected  occultations, and iii) a shadow path crossing densely populated regions. Accordingly, the observational campaign motivated the participation of 116 telescopes from Europe, North Africa, and Western Asia. As a result, we received  61 positive and 40 negative data sets. The other 15 locations had bad weather conditions, and observers could not acquire data. The number of effective chords is smaller than the 61 positives due to overlapping observations from nearby observatories along the object's limb.

\begin{figure}[!htb]
    \centering
    \includegraphics[width=0.95\linewidth]{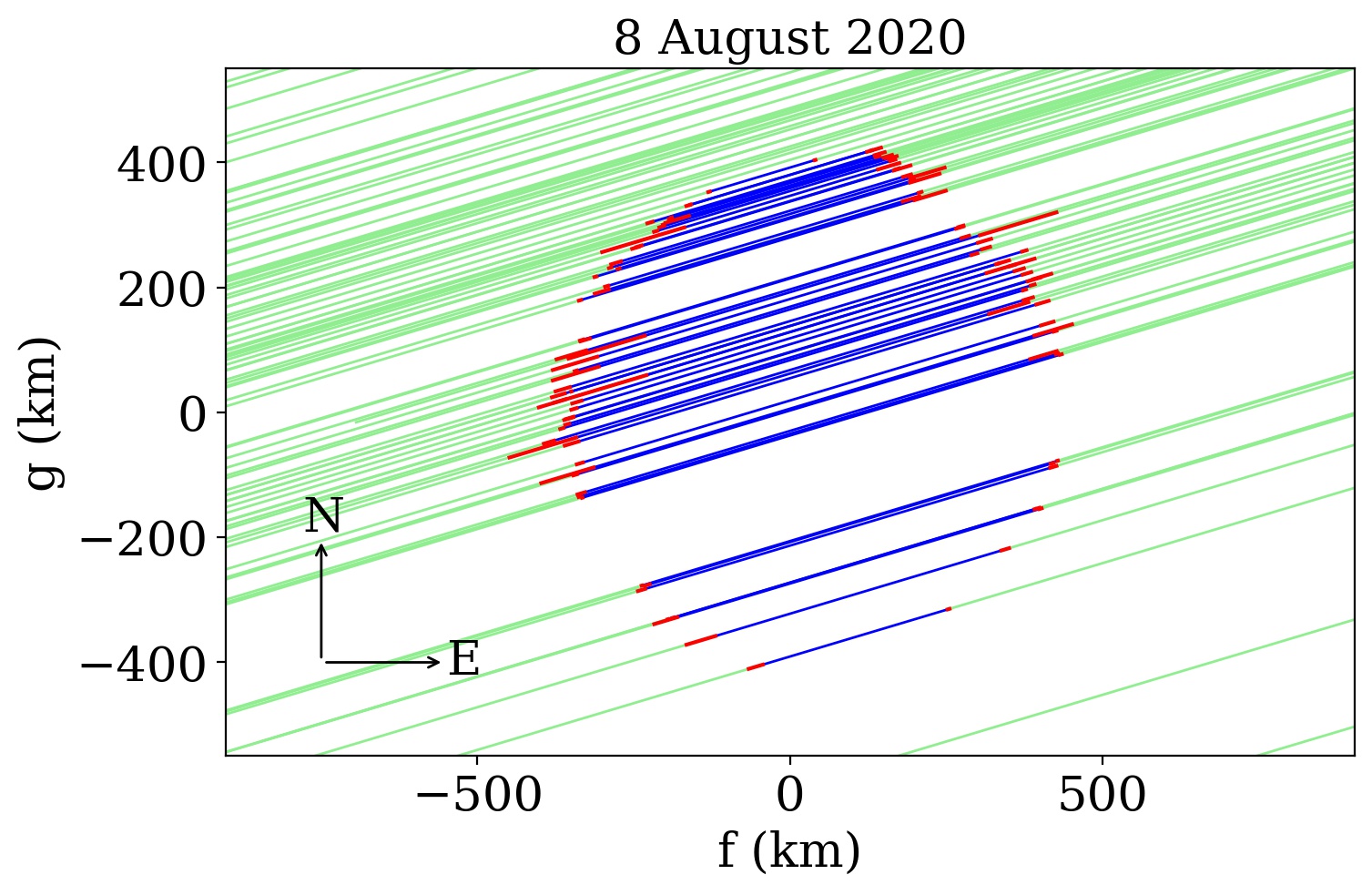}
    \caption{Chords measured during the 8 August 2020 event show the detection of MS4’s limb in blue with 1$\sigma$ error bars (red segments). Six positive chords with large error bars were suppressed from this plot for better visualization: TAROT, Lleida, Khmelnytskyi, Fuensanta de Martos, Kharkiv T36, and Marbella. The green lines represent positions compatible with the total target's flux within the noise (i.e., no secondary occultation). The order of positive chords, from north to south, is the same as in Table \ref{tab:occ_times}.
    }
    \label{fig:mall_aug_chords}
\end{figure}

We then submitted all the images to the procedure described at the beginning of this section. As absolute time acquisition is essential to achieve good results, we checked each data set and applied offsets when: i) the observer reported time issues during the acquisition, ii) the camera acquisition software is known to have a systematic offset, and iii) overlapped chords that do not match the ingress and egress instants.  In this last case, the time shifts applied to the original positive data were based on comparison with close-by chords. All the time shifts are presented in Table \ref{tab:aug_pos_sites}, and the corrected instants are in Table \ref{tab:occ_times}. Figure \ref{fig:mall_aug_chords} shows all the positives (blue lines) and their uncertainties in ingress and egress times (red segments).  

A simple elliptical model cannot effectively reproduce the observed profile projected at the sky plane. However, a global profile can be obtained by selecting 13 positive chords among the 61 positives. The first selection criterion was the time source, namely, data acquired with GPS were preferred. Among the 17 GPS data sets, we discarded the following: 1) Varages because it presented a gradual emersion in the light curve (see Sect. \ref{sec:topography}); 2) Nice due to reported time issues during the acquisition; 3)  Khmelnytskyi and Sevilla because it probed the latitude where a large topography is suspected (see Sect. \ref{sec:topography}); 4) Guarcino because the GPS was connected to the computer and it presented a large offset with respect to all close-by chords; 5) Artemis because its length does not match the length of Catania's chord (at 1$\sigma$ level), which probed exactly the same object's profile but is a bit larger. 6) Finally, Caussols, Ariana, and Kuban have larger uncertainties than other chords that probed the same region. In addition to the eight GPS data sets, we selected another five positive chords acquired from Méo station, Mátraszentistván, Hvar, Agerola, and La Palma. The main criteria for selecting the mentioned NTP chords were: data acquired with professional telescopes, low dispersion in the light curves, and the smallest uncertainties of the probed limb.

\begin{figure}[!htb]
    \centering
    \includegraphics[width=0.95\linewidth]{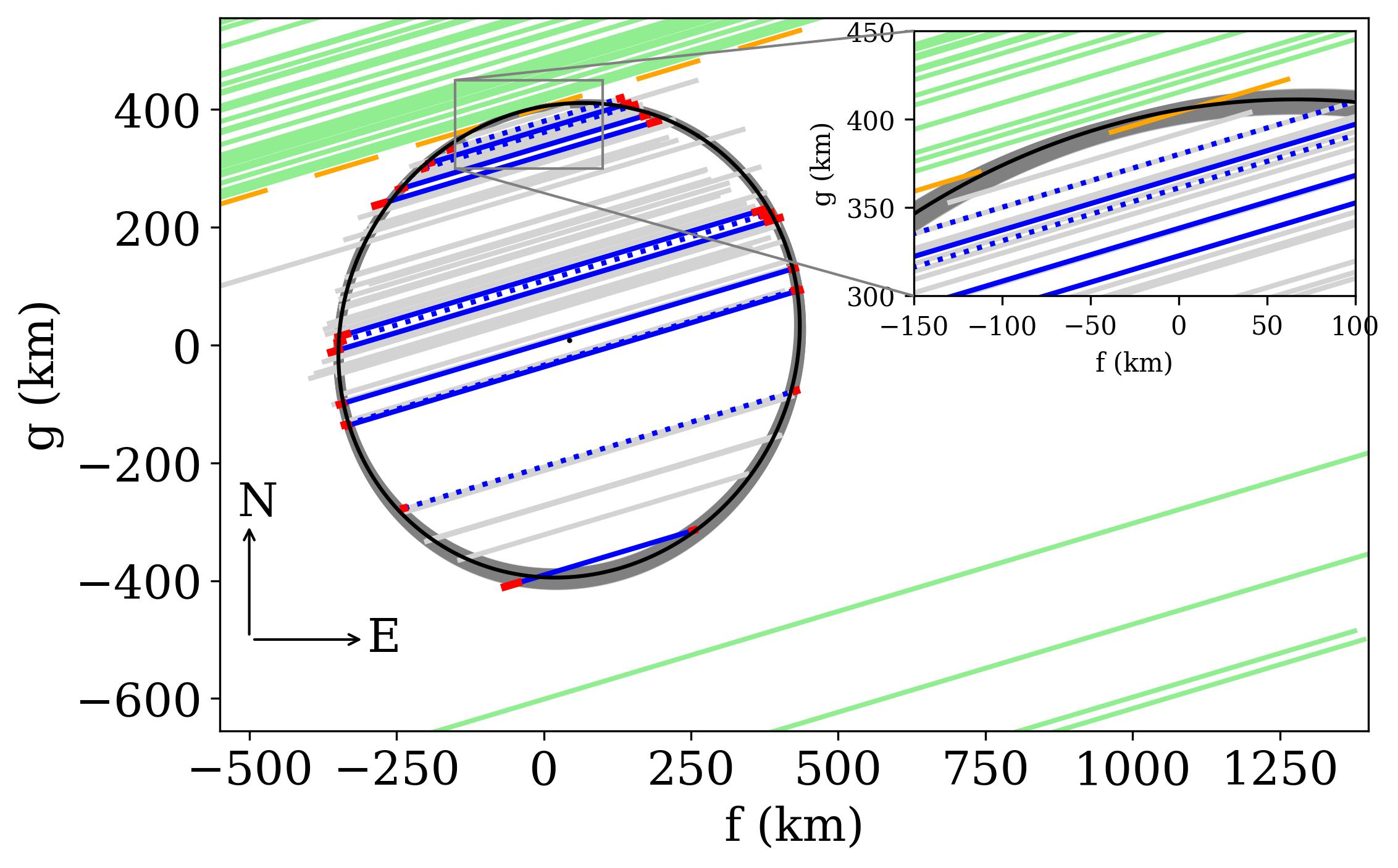}
    \caption{Thirteen selected chords (blue), where GPS data are presented in solid lines and NTP by dashed lines. Gray segments show the other positive chords not used in limb-fitting. The black ellipse shows the best elliptical limb, and the gray region the solutions within $3\sigma$.  The orange segments represent each image acquired from the Montsec station and the light green segments show other negative chords.}
    \label{fig:fixed_ellipse_chords}
\end{figure}

Figure \ref{fig:fixed_ellipse_chords} presents the 13 selected positive chords (blue) over-plotted to the other positives (gray segments).  Solid lines represent GPS chords, while dashed lines show NTP data. The selected data are ordered from north to south, as follows: Méo station (FRA), Valbonne (FRA), Mátraszentistván (HUN), Catalonia (ESP), Massa (ITA), Rome (ITA), Hvar (HRV), Sassari (ITA), Odesa (UKR), Agerola (ITA), Algiers (DZA), La Palma (ESP), and Çanakkale (TUR). They provide N = 26 independent points at the sky plane to fit the five ellipse parameters. The global elliptical limb is determined by minimizing the classical $\chi^2$ function. The quality of the result is given by the $\chi^2$ per degree of freedom $\chi^2_{\rm pdf}=\chi^2/(N-M) \approx 1$ for satisfactory fits, where $N$ is the number of points and $M$ is the number of fitted parameters \citep{Gomes-Junior2022}. A set of empirical tests\footnote{See Eq. 11 in \cite{Gomes-Junior2022} for details about the function that considers topography in the limb fitting.} assuming topography values between 0 and 10 km were performed and revealed a good fit ($\chi^2_{\rm pdf}=0.92$) when features of 7 km were considered. This result agrees with the theoretical approach proposed by \cite{Johnson1973}, which gives a lower limit of 6 to 7 km for topography on MS4 (see Sect. \ref{sec:4}).

Among the elliptical solutions inside the $3\sigma$ region, we excluded those that crossed or approached the negative grazing chords within the tolerance level of 7 km (radial direction). Therefore, although the solutions cross the negative chord as seen from Montsec (Fig. \ref{fig:fixed_ellipse_chords}), they are inside the 7 km assumed range. The area equivalent radius was calculated using the relation $R_{\rm equiv}=a'\sqrt{1-\epsilon'}$. Finally, the limb solution in Table \ref{tab:ellipse_parameters} represents the best-fitted elliptical limb at the sky plane with uncertainties at the 3$\sigma$ level. 

\begin{table}[!h]
    \centering
    \caption{Parameters of the best-fitted ellipse (at  3$\sigma$ level) derived from the 13 selected positive chords from the August 8 event. }
    \label{tab:ellipse_parameters}
    \begin{tabular}{c l||c l||c l}
    \hline
     \multicolumn{6}{c}{MS4's global elliptical limb\tablefootmark{a}} \\
    \hline
        \textit{f}  & 43 $\pm$ 6 km & $a'$ & 412 $\pm$ 10 km          & PA & 121 $\pm$ 16$^\circ$\\
        \textit{g} & 7 $\pm$ 9 km   & $\epsilon'$ & 0.066 $\pm$ 0.034 &$R_{\rm equiv}$ & 398 $\pm$  12 km\\ \hline
    \end{tabular}
    \tablefoot{\tablefoottext{a}{These solutions admit topographic features up to 7 km and are limited to the north by the negative chord from the Montsec station (orange segments in Fig. \ref{fig:fixed_ellipse_chords}).}}
\end{table}

\begin{figure}[!htb]
    \centering\includegraphics[width=\linewidth]{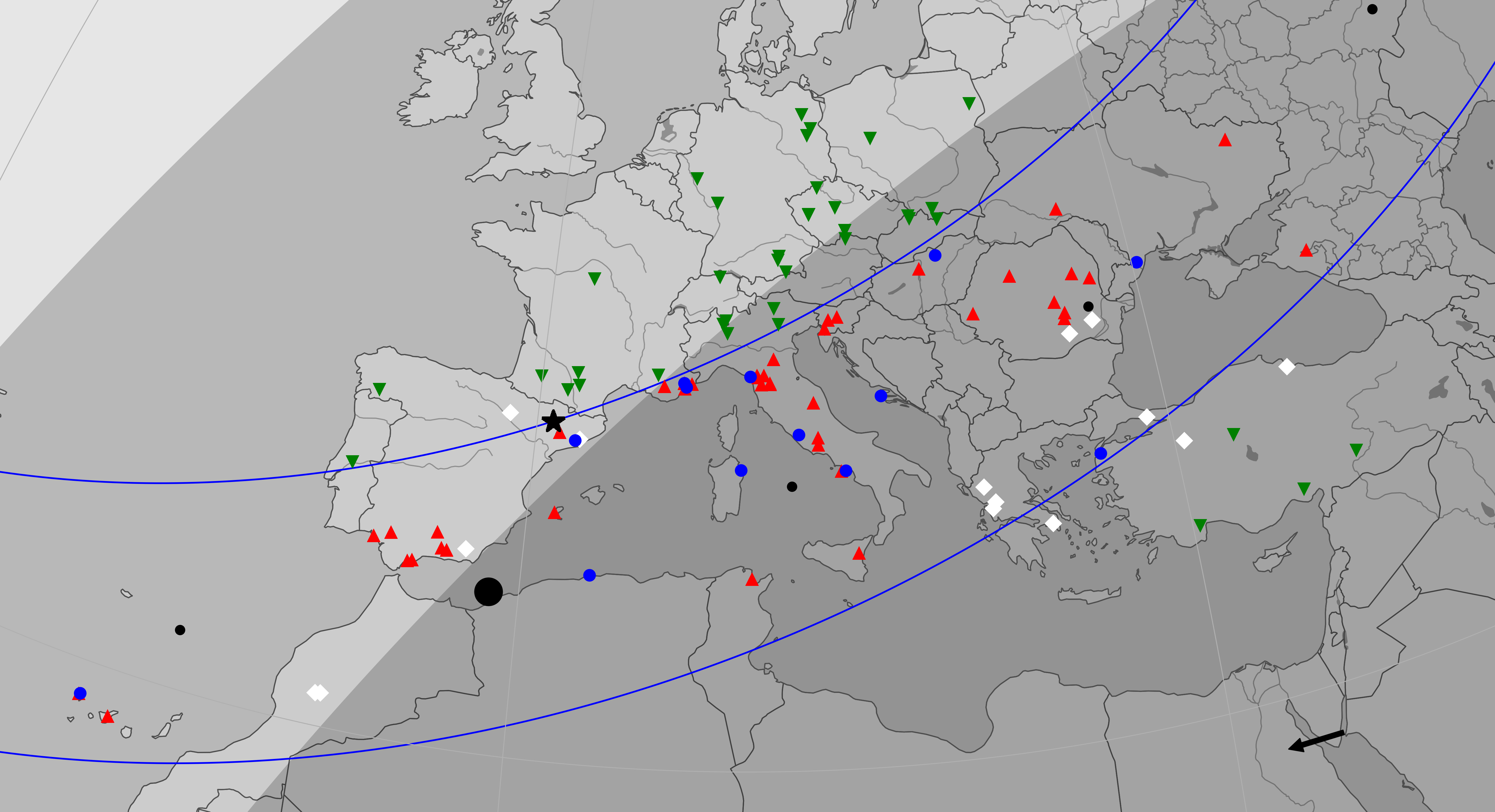}
    \caption{Post-occultation map showing the location of each station that participated in the 8 August 2020 occultation campaign. Solid lines delimit the observed shadow path, and the black dots mark the shadow position every minute, with the largest at the moment of the geocenter's closest approach. The shadow follows the direction given by the black arrow. The blue dots indicate the 13 selected chords, while the other positives are in red triangles. A black star marks the close-by negative chord acquired from Montsec station. Green triangles and white diamonds represent stations with negative data and bad weather, respectively.}
    \label{fig:postmap_aug2020}
\end{figure}

A general view of the August 2020 event is shown in Fig. \ref{fig:postmap_aug2020}. The blue lines represent the observed shadow path. It is worth mentioning that the 13 selected chords are in blue circles, and the other positives are in red triangles. We note that a black star marks the negative detection by Montsec station, while the green triangles mark the other negatives observations.

\subsection{Topographic features}
\label{sec:topography}

Determining the topographic limits for small bodies in the outer Solar System is challenging. The first attempts were theoretical; for instance, \cite{Johnson1973} proposed a method to determine topography limits for planetary satellites using their global density and composition. On the other hand, from an observational point of view, just a few small bodies orbiting the Sun beyond Neptune had their topography limits set using stellar occultations \citep{Dias-Oliveira2017, Leiva2017}. Also, with the advance in interplanetary spacecraft technology, the surface of a few objects was studied using in situ images. For instance, using Voyager's images of Uranus's largest satellites \cite{Schenk2020} found superficial features up to 11 km. Likewise, New Horizons' flyby over the Pluto system \citep{Moore2016, Nimmo2017} and (486958) Arrokoth \citep{Spencer20a} revealed superficial structures on the same scale. 

Applying the theoretical approach mentioned above for MS4 and assuming an icy body with a density between $\rho=1.0-2.0$~g.cm$^{-3}$, the lower limit for superficial features is 6 to 7 km. If material strength may increase toward the nucleus, the surface might support more prominent features. Also, considering the sizes of the structures observed by spacecraft in other objects, assuming features up to 7 km on MS4's surface is reasonable.

\begin{figure}[!ht]
    \centering
    \includegraphics[width=\linewidth]{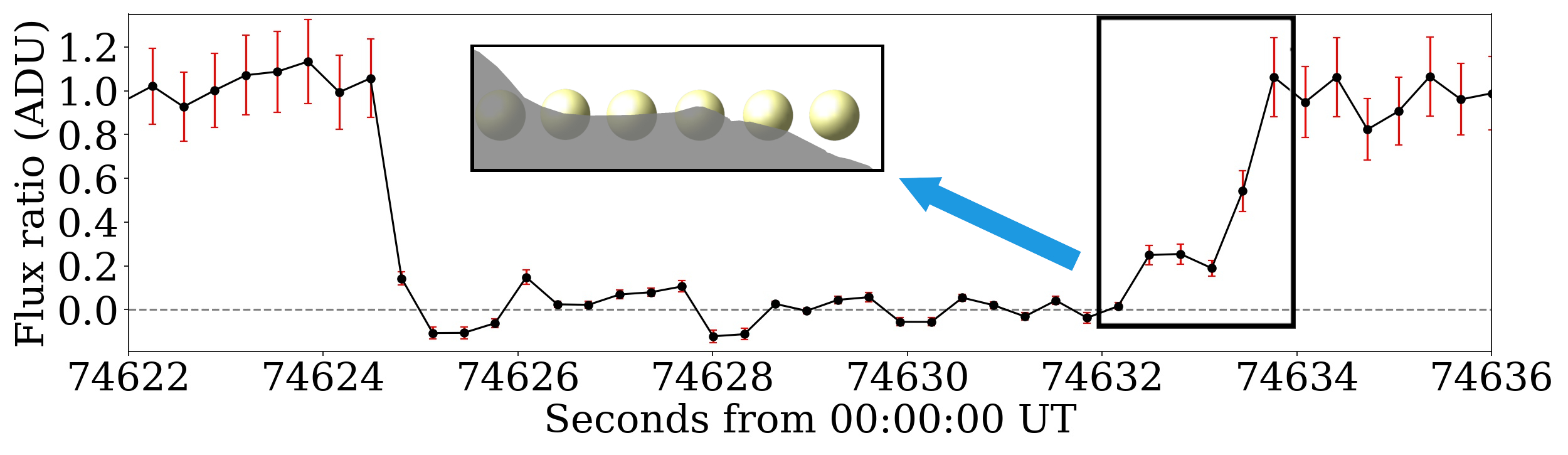}
    \caption{ Normalized stellar flux in each frame acquired in Varages station as a function of time in black dots, with photometric uncertainties in red. The insert selects the egress region and illustrates a possible explanation for such a signal (see text).}
    \label{fig:varages_ff}
\end{figure}

The first evidence of topography was observed in the Varages light curve.  This data set does not have dead time between the images, and each exposure translates into a resolution of 1.97~km into the sky plane. The Fresnel diffraction and stellar diameter at MS4 geocentric distance are at the same level, 1.54~km and 1.19~km, respectively. The mentioned light curve presents a sharp ingress and a gradual egress above the noise level, as shown in Fig. \ref{fig:varages_ff}. The feature did not appear in the other high S/N light curves, thus weakening the possibility of a secondary star. Therefore, the most plausible explanation is a topography where a portion of the star appeared a few frames before egress, corresponding to $\approx$10~km-long feature in the chord's direction. The insert in Fig. \ref{fig:varages_ff} displays the stellar position in each frame, represented by yellow circles, relative to a proposed limb in gray.

The second evidence of topography on MS4's observed limb comes when investigating the residuals of the average ellipse presented before. Some groups of points between PA = -5º and PA = 120º presented a large offset regarding the global limb. Therefore, to better describe this local limb, we built a model based on a combination of symmetric parabolas, the simplest function that can reproduce the observed features. Equations \ref{equation:depre} and \ref{equation:elev} provide the models used to fit the group of points with negative and positive dispersion ($R_D$), respectively. The $y$ term defines the parabola's depth and height, when positive indicates depression. The $x$ term is related to the parabola's curvature. PA is the position angle, and $z$ accounts for the parabola's distance from the plot's origin. The model's $R_D$ values outside the topography region are defined as being zero.
\begin{equation}
\label{equation:depre}
Depression = \begin{cases}
      0 & R_D\geq 0,\\
      x (PA-z)^2 - y & R_D<0.\\
    \end{cases}
\end{equation}
\begin{equation}
\label{equation:elev}
Elevation = \begin{cases}
      0 & R_D\leq 0,\\
      x (PA-z)^2 + y & R_D>0.\\
    \end{cases}
\end{equation}
Then, the model is built by summing the equations,
\begin{equation*}
Model = Depression + Elevation + Depression,
\end{equation*}
where one depression corresponds to the Varages egress region (shown here as negative position angles for better viewing).

The fitting was made using a high-level \textsc{python} interface named \textsc{lmfit}\footnote{More about this library can be found in the \url{https://lmfit.github.io/lmfit-py/}}, designed for non-linear optimization and curve-fitting problems. First, we used the differential evolution (DE) minimization method \citep{Storn1997} to derive the first estimation of the model's parameters, which can explore large areas of candidate space without getting stuck in a local minimum. Then, to get a representative estimation of the model's uncertainties, we explored the parameter space using the maximum likelihood via Monte Carlo Markov chain sampler: \textit{emcee}\footnote{Documentation available on \url{https://emcee.readthedocs.io/en/stable/}}\citep{Foreman2013}. The center of each feature was limited as follows: main depression between $40^\circ \pm 20^\circ$, elevation between $7.5^\circ \pm 7.5^\circ$, while the Varages depression was fixed on PA = -3.9$^\circ$. The model section between -5$^\circ$ and 0$^\circ$ does not have significant errors due to the precision of the grazing detection presented in Fig. \ref{fig:varages_ff}. For this angle interval, we do not allow the sampler to estimate for unknown uncertainties. However, for the elevation and the main depression, \textit{emcee} found that unknown uncertainties must be about 4.5 km. This value is reasonable, considering that the 7~km of tolerance assumed during the global fit was not included in the points error bars.

\begin{figure}[!htb]
    \centering
    \includegraphics[width=\linewidth]{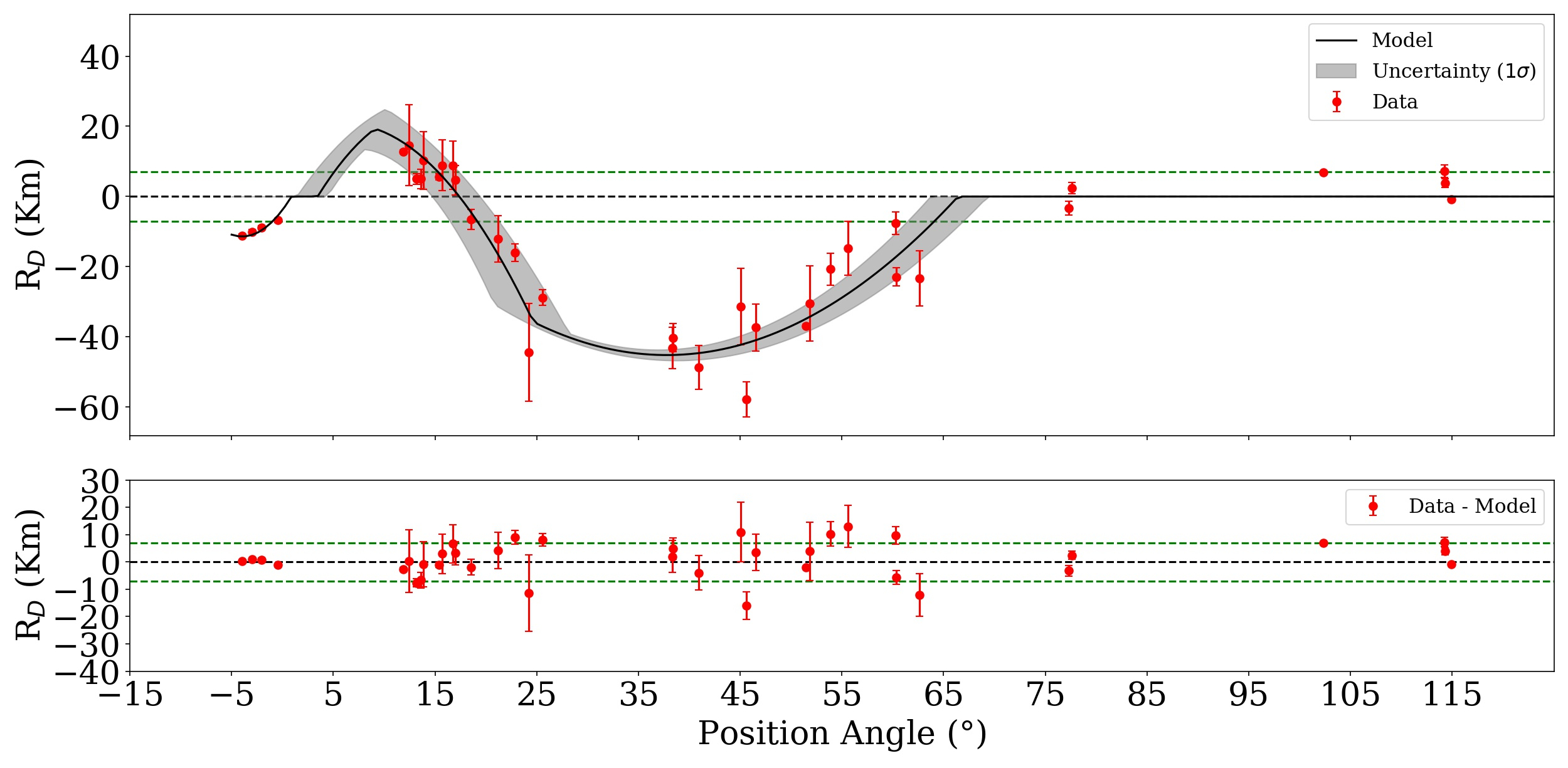}
    \caption{ Limits of the assumed ± 7 km tolerance (see text) shown as horizontal green lines and the black dashed line is the best-fit ellipse. \textbf{Upper panel:} Red points ($R_D$) are the distance of each point from the best-fit ellipse (in the normal-to-the-ellipse direction) as a function of the position angle, and the solid black line is the model with $1\sigma$ uncertainty represented in gray. \textbf{Lower panel:} Residuals after subtracting the model from the data points.}
    \label{fig:features_model}
\end{figure}

\begin{figure}[!ht]
    \centering
    \includegraphics[width=\linewidth]{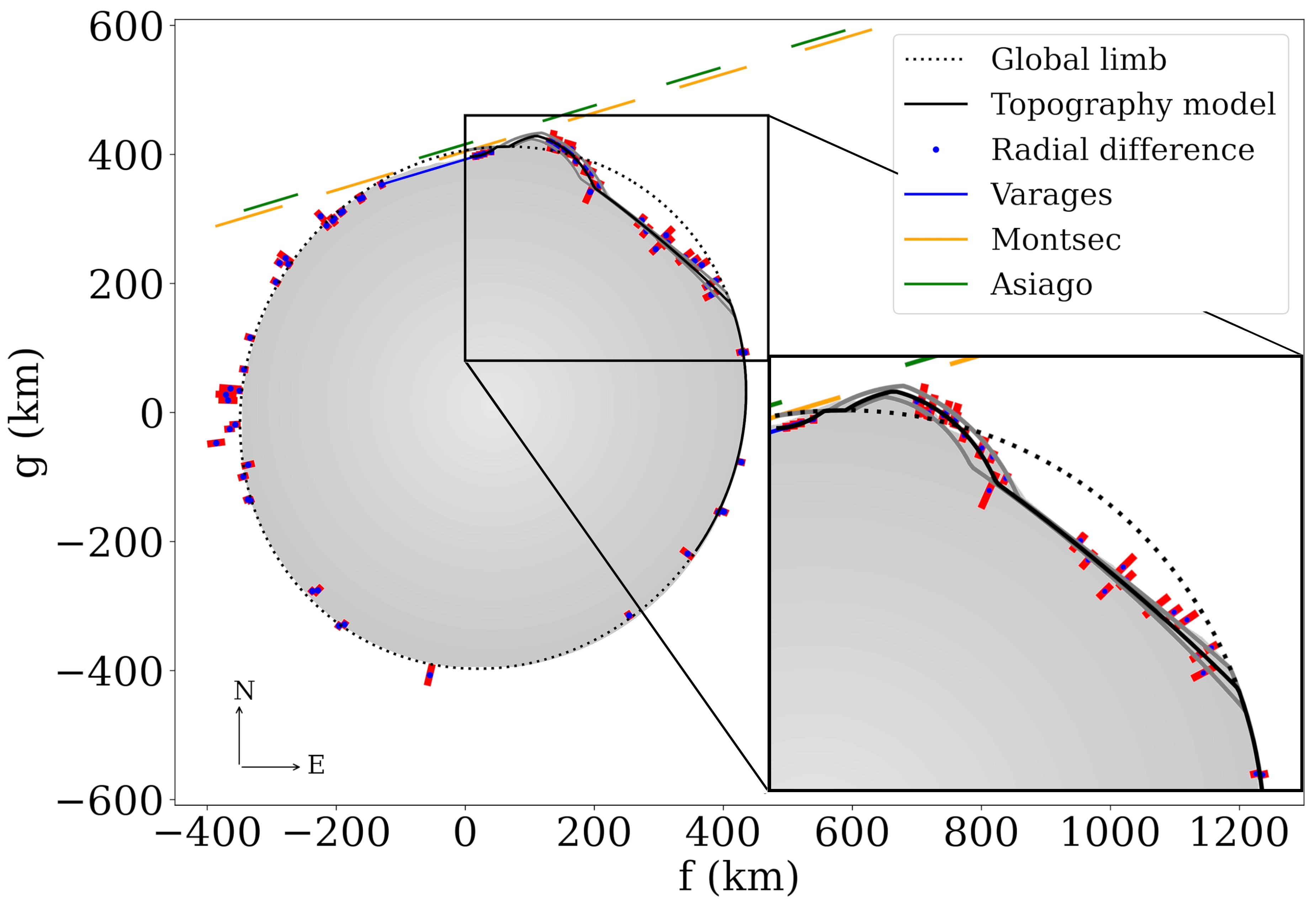}
    \caption{ R$_D$ points projected at the sky plane in blue points, with 1$\sigma$ uncertainties in red. The blue segment is the positive detection from Varages station. Orange and green segments correspond to negative frames acquired in Montsec and Asiago stations, respectively. In black, the dotted line is the best-fitted global limb model described in Sect. \ref{sec:8august}, and the solid line is the model for local topography. The solid gray lines limit the topographic model's 1$\sigma$-error bars. Finally, the filled gray color shows the proposed global limb with topography.}
    \label{fig:features}
\end{figure}

After subtracting the model from the data set, residuals are inside the expected range (bottom panel in Fig. \ref{fig:features_model}). Therefore, according to the model at $1\sigma$ level, MS4's surface has an $\approx$11~km depth depression in the region detected by Varages station, followed by an elevation of 25$^{+4}_{-5}$~km. However, the most impressive feature is the 45.1~$\pm$~1.5~km depth depression with a linear extension of 322 $\pm$ 39 km. Figure \ref{fig:features} presents a general view of the detected limb and summarizes the topography solutions. However, because the depression was likely not in its middle position at the limb, it is likely more extensive and profound than what it seems from this snapshot at a particular rotation phase.  Such prominent superficial features may be caused by collisions with other small objects.

\subsection{Other occultation events}
\label{sec:other_events_limb}

An object's 3D shape is strongly correlated with the body's rotational modulation \citep{Chandrasekhar1987, Tancredi2008}. For example, Maclaurin objects usually have single-peaked rotational light curves with small peak-to-peak amplitudes caused by albedo features. In contrast, the rotational light curve of a Jacobi shape presents double-peaked curves with more pronounced amplitudes (unless they are seen nearly pole-on). Therefore, a reasonable determination of MS4's rotational parameters is crucial to derive an accurate 3D size, shape, albedo, and density. 

However, MS4 has crossed a highly dense field of stars since its discovery.  Therefore, it is complicated to obtain precise photometric measurements because it is usually blended with faint background stars (as seen from Earth). However, in 2011 it passed in front of a dark cloud when it was observed well isolated from other stars in about 100 images. Using those images along with a data set acquired in the Sierra Nevada observatory, \cite{Thirouin2013} derived a single-peaked light curve with an amplitude of 0.05 $\pm$ 0.01 mag and two possibilities for the rotational period: 7.33 h or 10.44 h. Such a small amplitude may indicate that MS4 is a Maclaurin object. It is also possible that it is a triaxial body, such as a Jacobi ellipsoid observed close to a pole-on orientation. Considering the observed small amplitude and small change in the aspect angle, the projected area on the sky plane between 2019 and 2022 should not  change considerably. Nevertheless, if it is a triaxial object, the position angle of the projected ellipse on each observed occultation will present significant changes according to its rotational phase.

\begin{figure}[!ht]
     \centering
     \includegraphics[width=\linewidth]{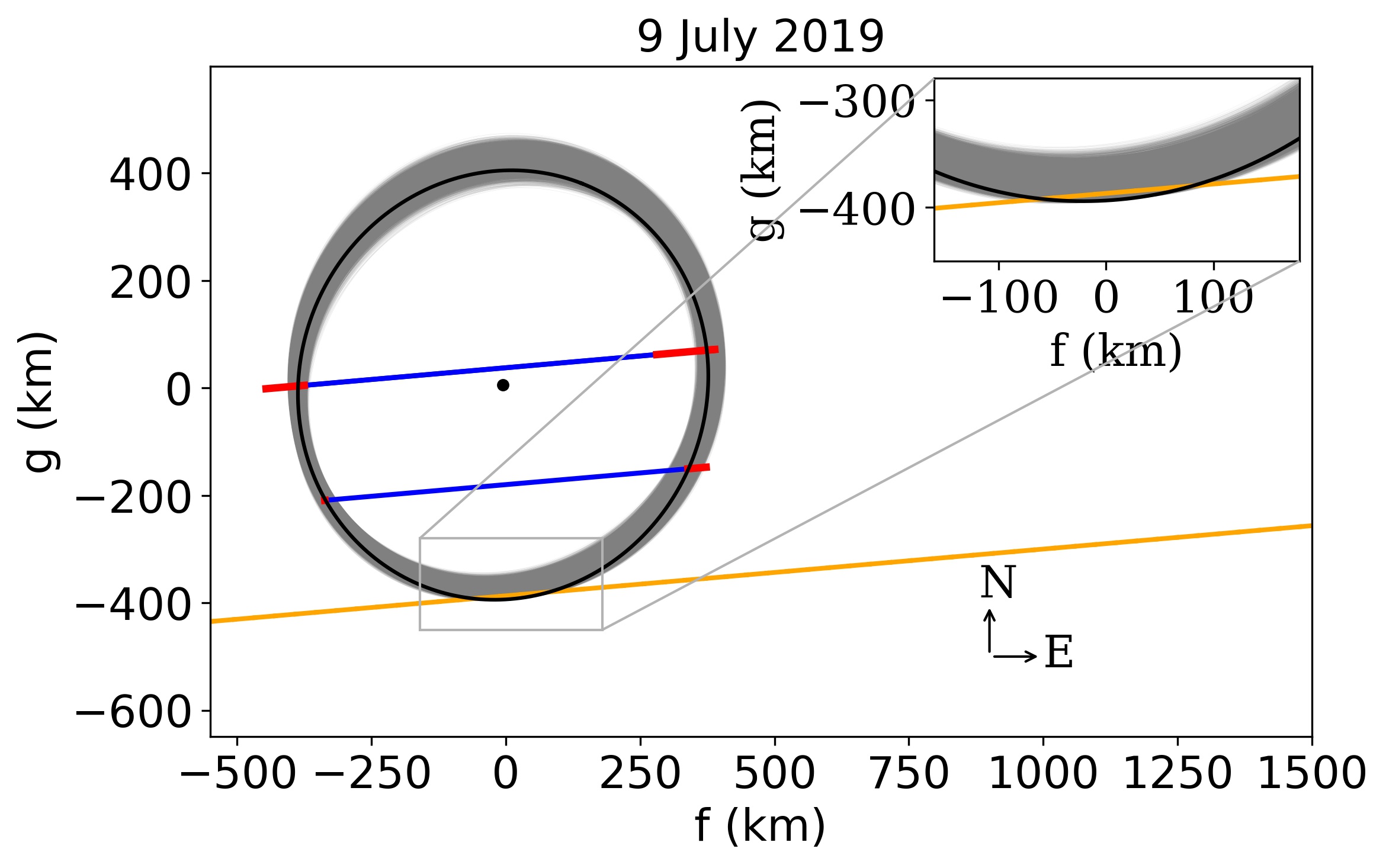}
    \caption{Positive data from the stellar occultation on 9 July 2019 shown as the blue segments. The $1\sigma$ uncertainties are in red, and the negative chord acquired from the Ponta Grossa station is in orange. The best-fitted ellipse is in black, with the center presented by the black dot, and the solutions in the 3$\sigma$ range are in gray. The fit considers topographic features up to 7 km in size. Thus, the ellipses crossing the negative chord are in this range.}
    \label{fig:190709_ellipse_cutted}
\end{figure}

\begin{figure}[!ht]
     \begin{subfigure}[b]{0.24\textwidth}
         \centering
         \caption{09 July 2019 (0.5)}
         \includegraphics[width=\textwidth]{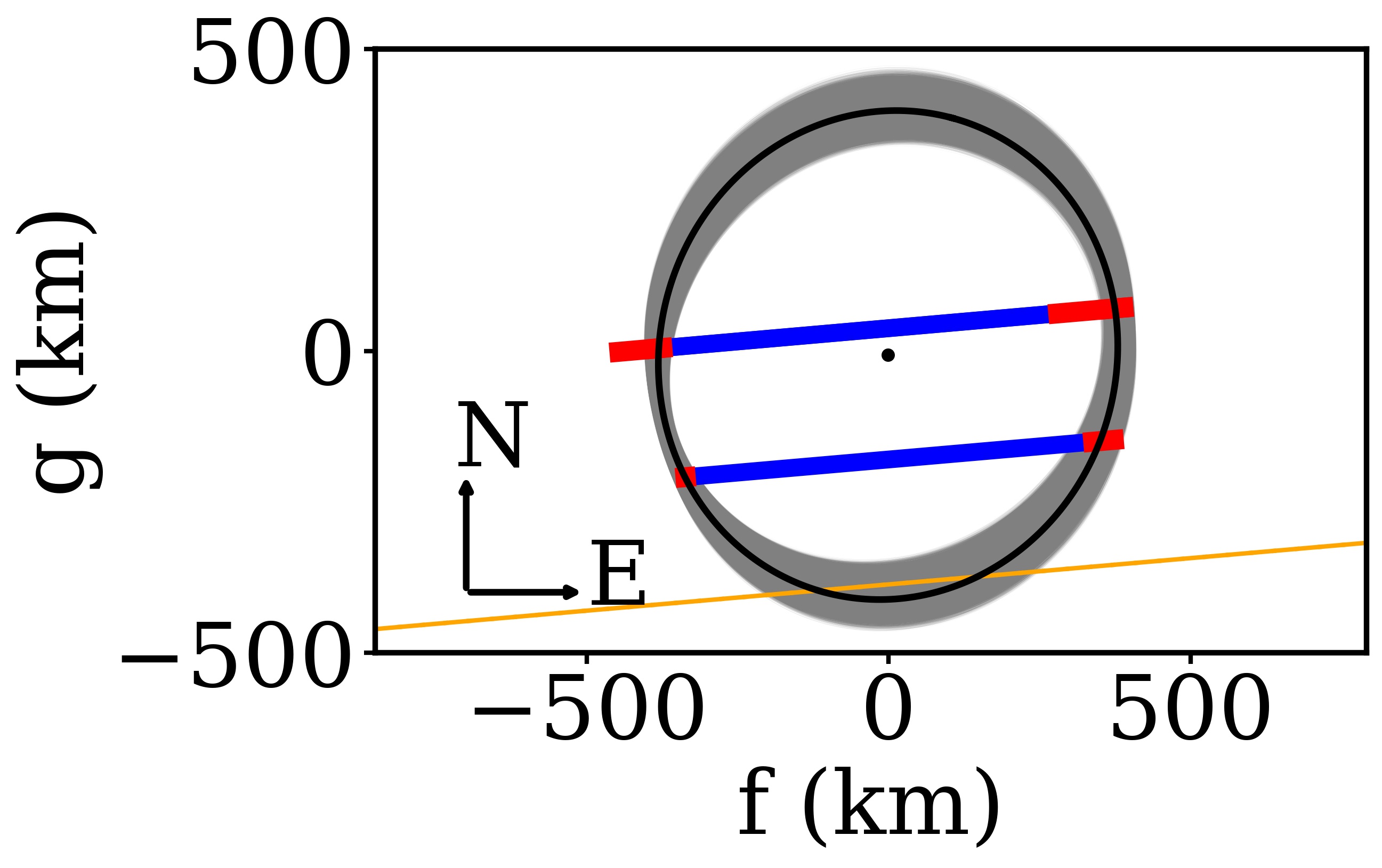}
         \label{fig:190709}
     \end{subfigure}
     \hfill
\begin{subfigure}[b]{0.24\textwidth}
         \centering
         \caption{26 July 2019 - 1 (1.0)}
         \includegraphics[width=\textwidth]{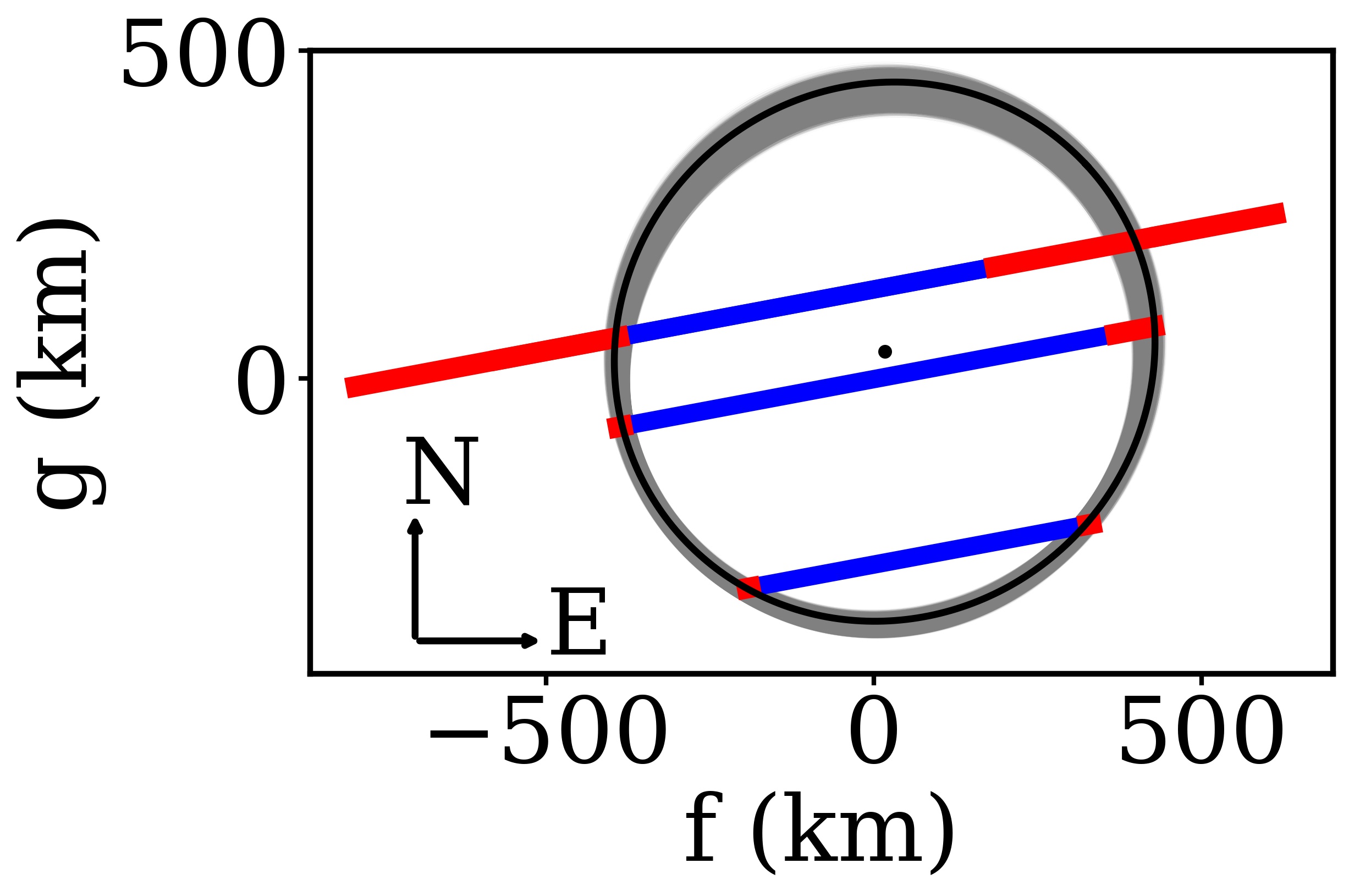}
         \label{fig:190726_1}
     \end{subfigure}
     \hfill
     \begin{subfigure}[b]{0.24\textwidth}
         \centering
         \caption{26 July 2019 - 2 (1.1)}
         \includegraphics[width=\textwidth]{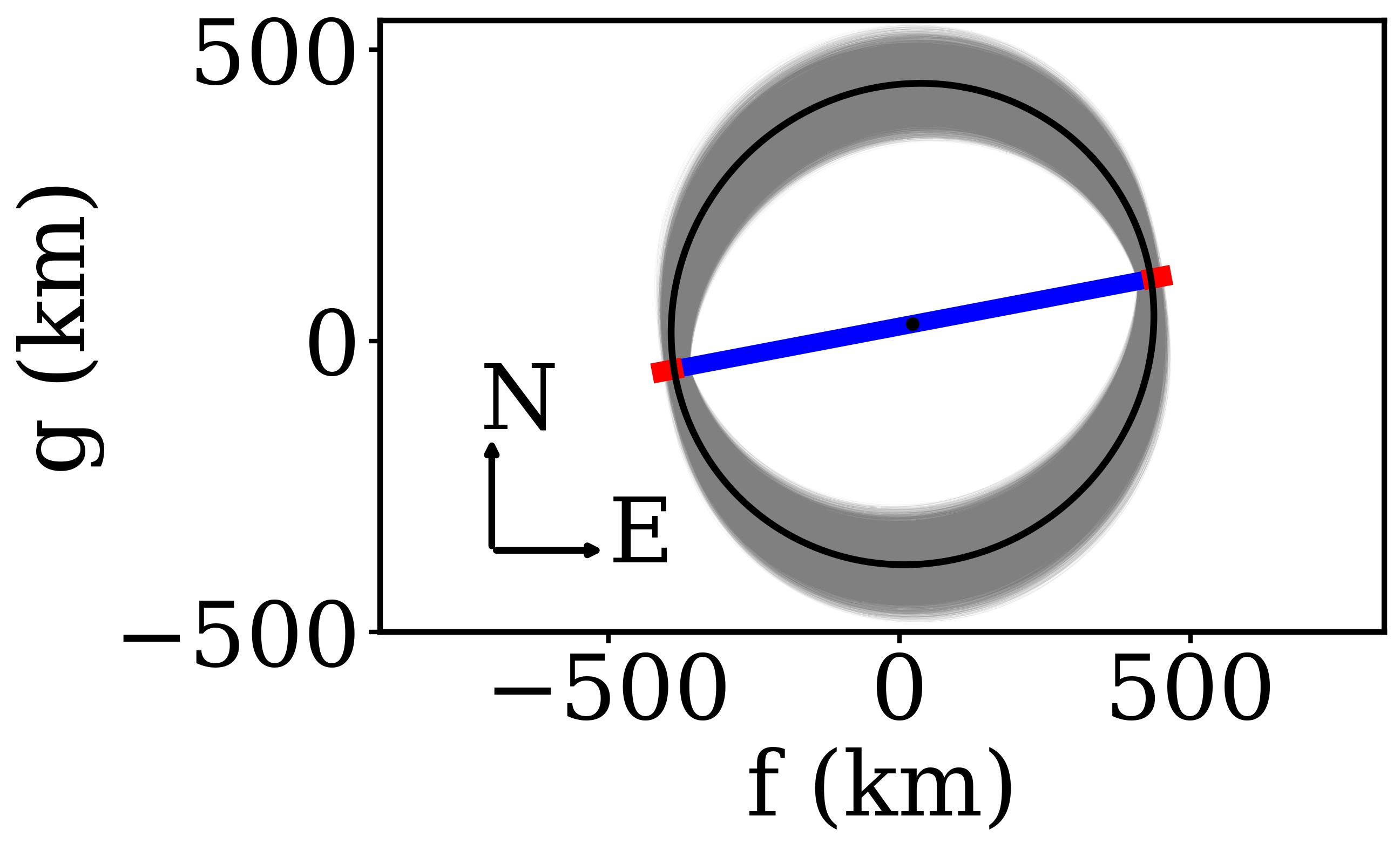}
         \label{fig:190726_2}
     \end{subfigure}
     \hfill
     \begin{subfigure}[b]{0.24\textwidth}
         \centering
         \caption{19 August 2019 (0.32)}
         \includegraphics[width=\textwidth]{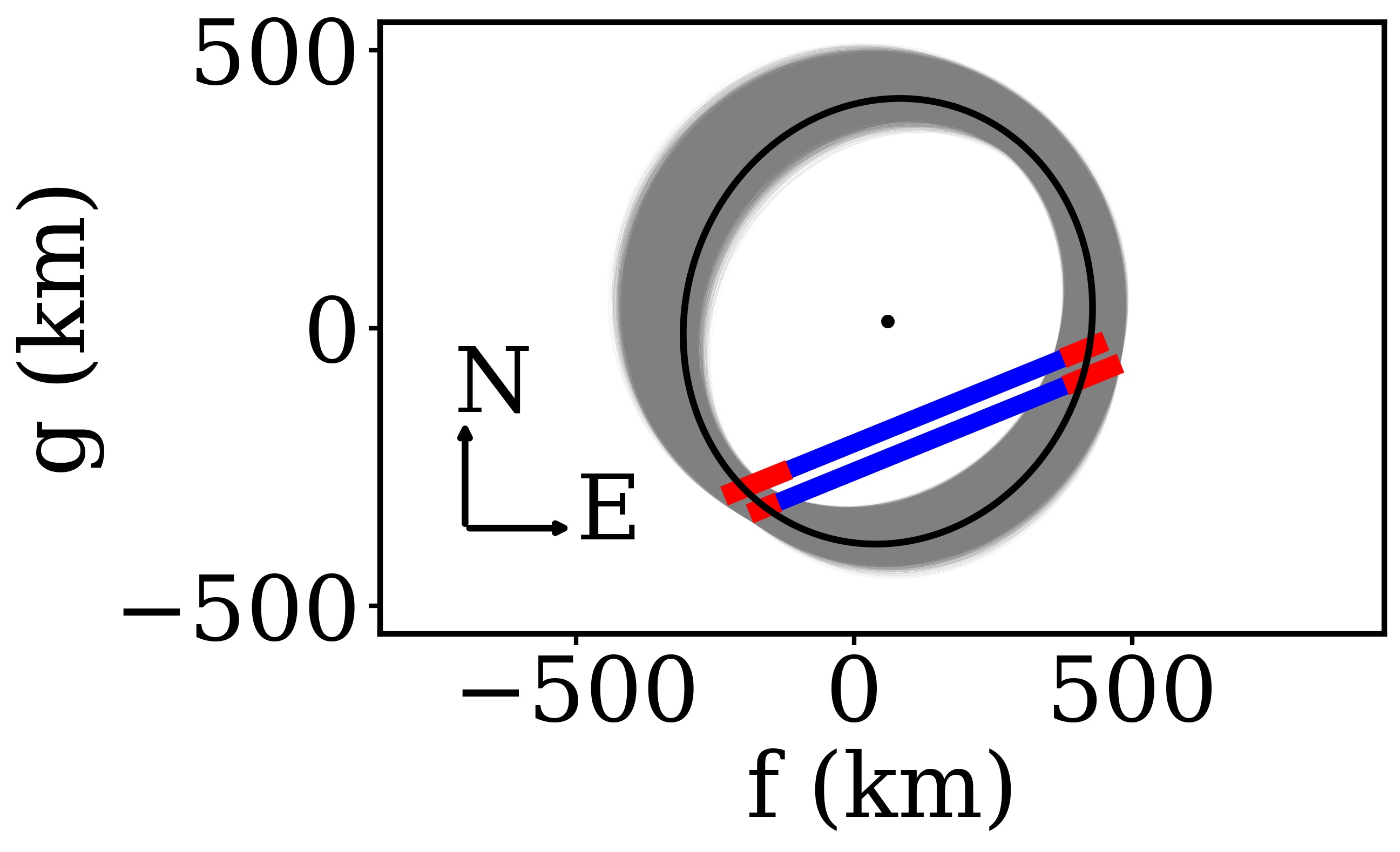}
         \label{fig:190819}
     \end{subfigure}
     \hfill
     \begin{subfigure}[b]{0.24\textwidth}
         \centering
         \caption{26 July 2020 (0.005)}
         \includegraphics[width=\textwidth]{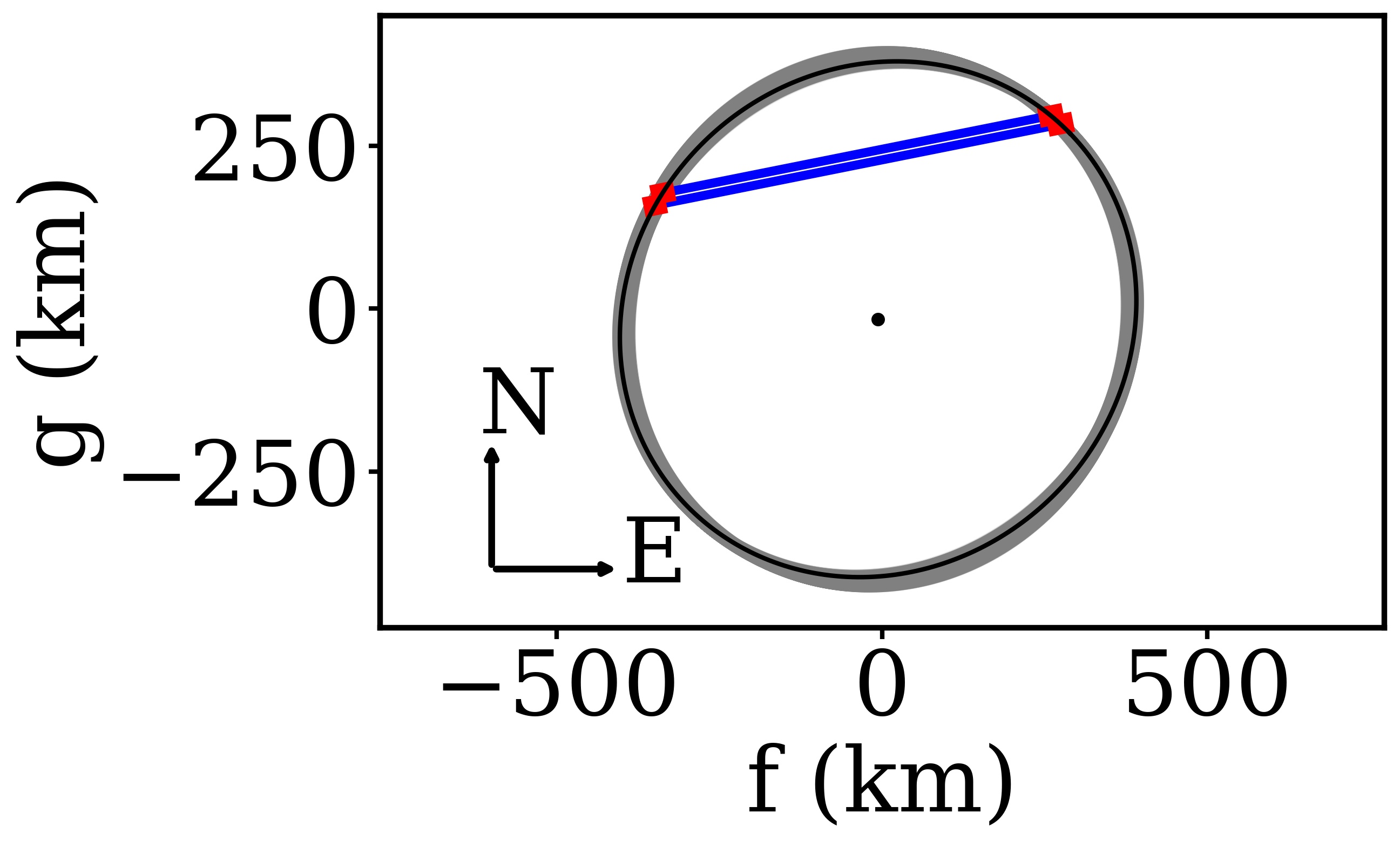}
         \label{fig:200726}
     \end{subfigure}
     \hfill
     \begin{subfigure}[b]{0.24\textwidth}
         \centering
         \caption{24 February 2021 (0.88)}
         \includegraphics[width=\textwidth]{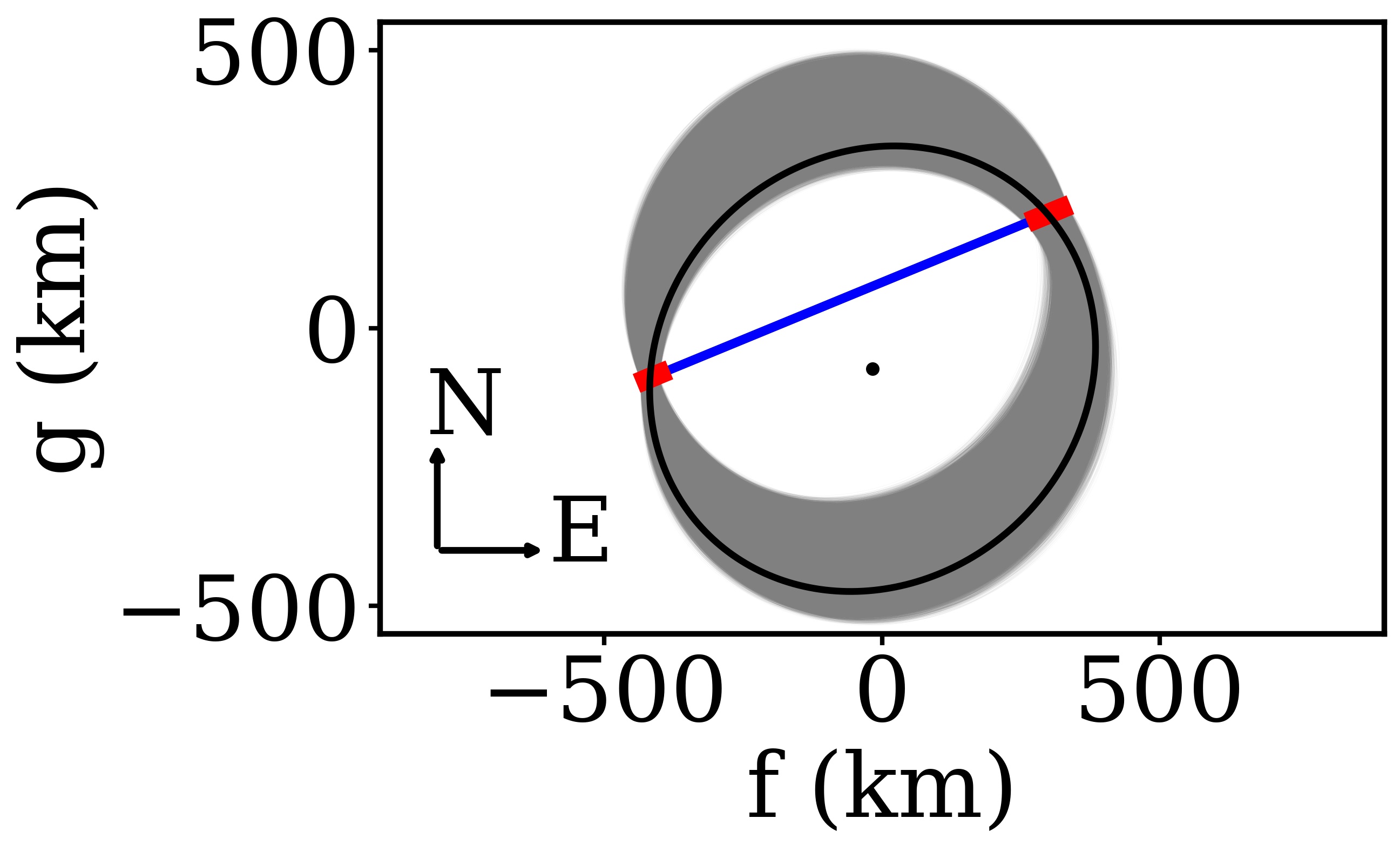}
         \label{fig:210224}
     \end{subfigure}
     \hfill
     \begin{subfigure}[b]{0.24\textwidth}
         \centering
         \caption{14 October 2021 (0.11)}
         \includegraphics[width=\textwidth]{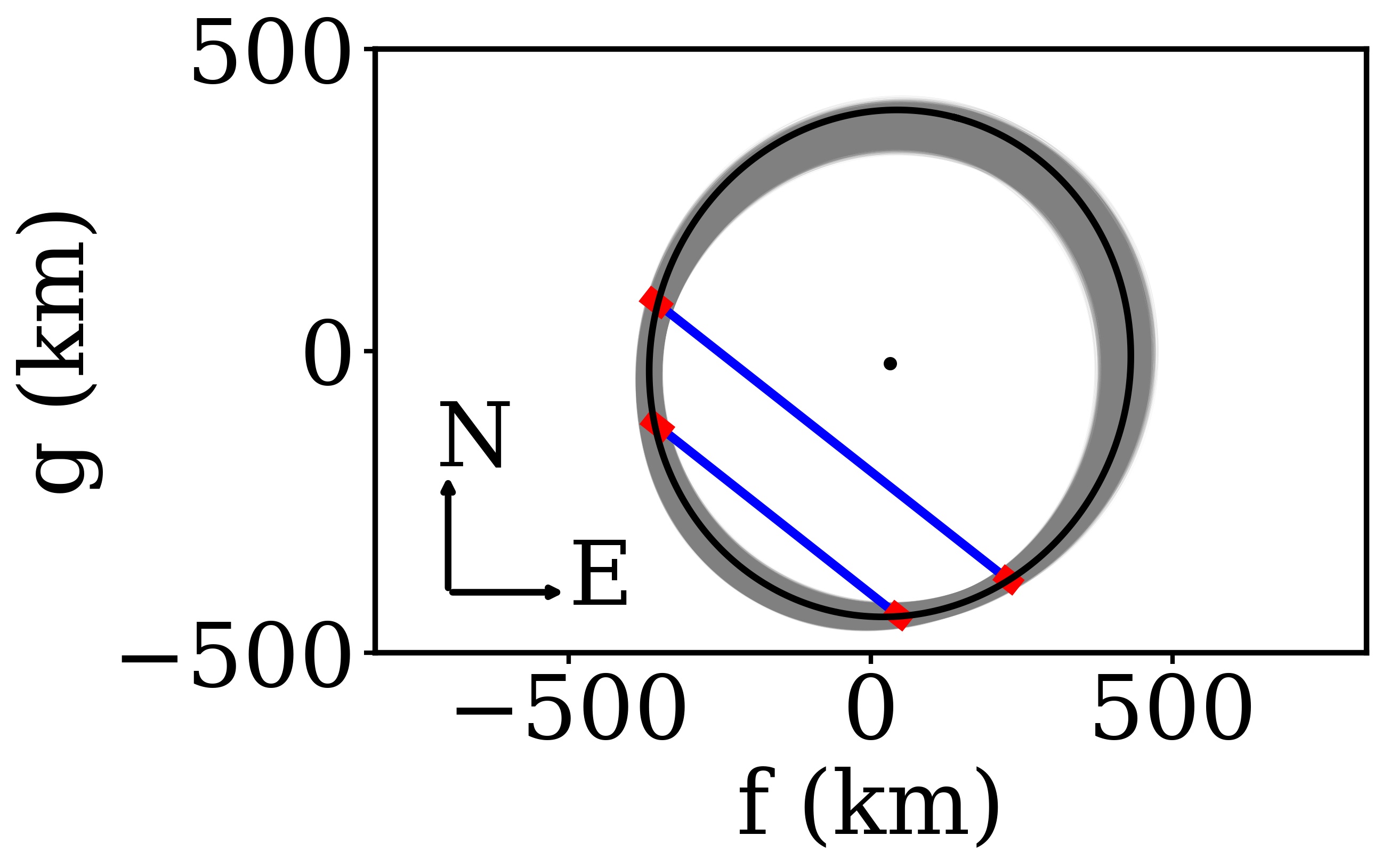}
         \label{fig:211014}
     \end{subfigure}
     \hfill
     \begin{subfigure}[b]{0.24\textwidth}
         \centering
         \caption{10 June 2022 (0.24)}
         \includegraphics[width=\textwidth]{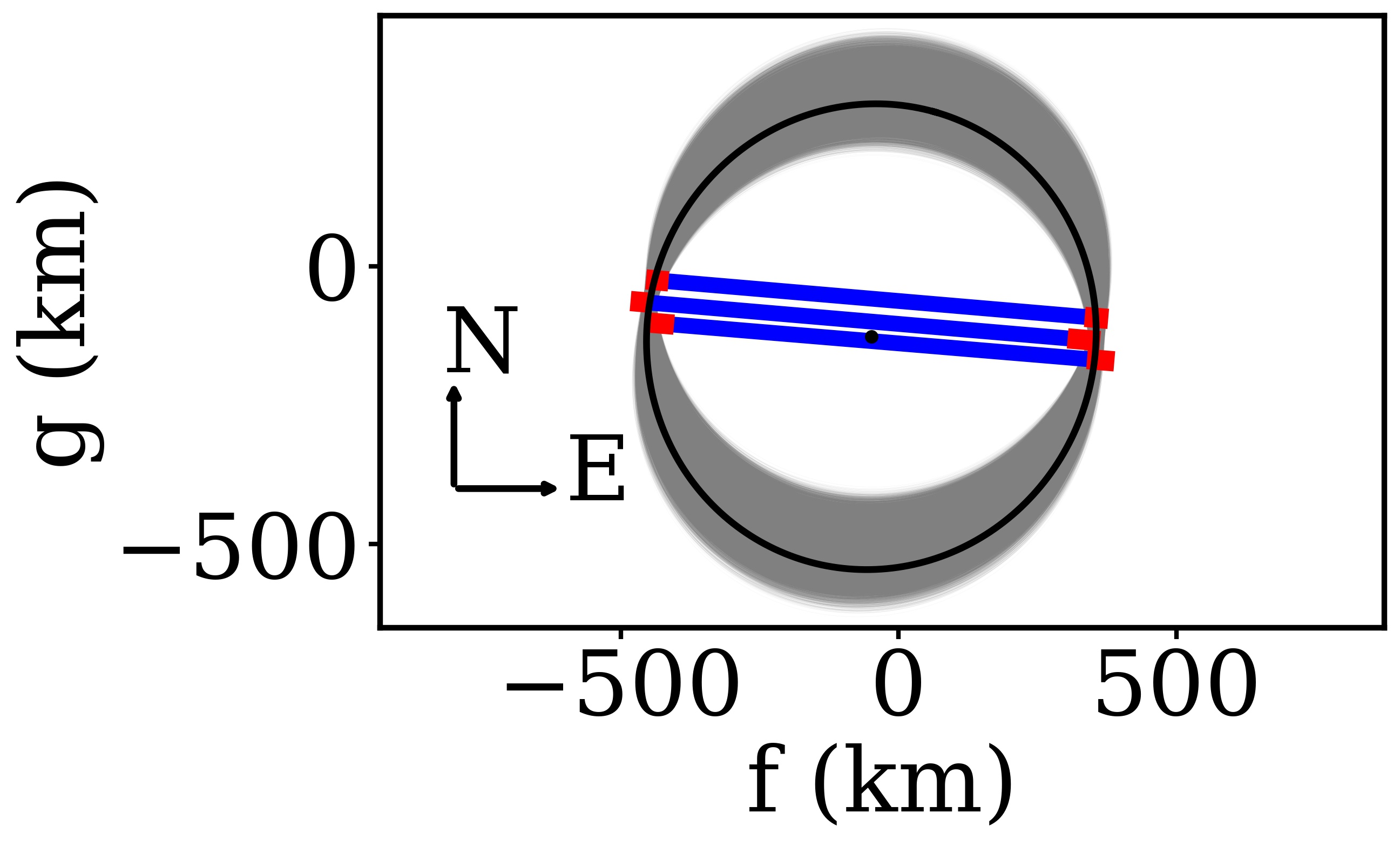}
         \label{fig:220610}
     \end{subfigure}
    \caption{ Results for the additional eight stellar occultation events. Blue segments are the positive detections with $1\sigma$ uncertainties in red. The best elliptical limb is in black, with the center presented by the black dot. The gray region presents all the limb solutions within $3\sigma$. The $\chi^2_{\rm pdf}$ of each fit is presented between parenthesis in the individual labels. For the occultations presented in d, e, and f, the chosen center solution was the closest one to the predicted by NIMA v9 ephemeris.}
    \label{fig:other_ellipses}
\end{figure}

Due to its large diameter and small rotational light curve amplitude, the Maclaurin spheroid is our preferred 3D shape for MS4. Thus, we tried fitting the same 3$\sigma$ solution derived from the 8 August 2020 event (Table \ref{tab:ellipse_parameters}) on the chords obtained on the other occultation events. Using $\chi^2$ minimization, the ellipse was fitted with the center (\textit{f} and \textit{g}) as a free parameter. When two center positions are equally possible (single chord cases),  we present the center solution closer to the position predicted by the NIMA v9 ephemeris. Figures \ref{fig:190709_ellipse_cutted} and \ref{fig:other_ellipses} show the results of the limb fitting for the other eight events. Table \ref{table:astrometric_occ} presents the derived astrometric information, and Appendix \ref{appendix0} shows the post-occultation maps with the observed shadow path and station locations.


\begin{table}[!htb]
\caption{Astrometric information (ICRS) for the geocentric closest approach instant (UT) obtained from the nine stellar occultation events observed between 2019 and 2022, sorted by date (day-month-year).
}
\label{table:astrometric_occ}
\resizebox{\linewidth}{!}{%
\begin{tabular}{cccccc}
\hline
\textbf{Date} &
  \textbf{\begin{tabular}[c]{@{}c@{}}Instant (UT)\\ (hh:mm:ss.ss)\end{tabular}} &
  \textbf{\begin{tabular}[c]{@{}c@{}}Right ascension\\ (hh mm ss.ss)\end{tabular}} &
  \textbf{\begin{tabular}[c]{@{}c@{}}Error\\ (mas)\end{tabular}} &
  \textbf{\begin{tabular}[c]{@{}c@{}}Declination\\ (º \textquotesingle \hspace{1pt} \textquotesingle\textquotesingle)\end{tabular}} &
  \textbf{\begin{tabular}[c]{@{}c@{}}Error\\ (mas)\end{tabular}} \\ \hline
09-07-2019     & 04:23:49.08 & 18 45 19.245981 & 0.23 & -06 24 13.05928 & 0.60 \\ \hline
26-07-2019     & 02:47:08.52 & 18 44 07.573463 & 0.57 & -06 26 40.17686 & 0.51 \\ \hline
26-07-2019     & 10:18:43.02 & 18 44 06.315990 & 0.37 & -06 26 43.7686  & 1.3 \\ \hline
19-08-2019   & 07:41:52.28 & 18 42 43.51613  & 1.0  & -06 32 33.9776  & 1.1 \\ \hline
26-07-2020     & 23:17:56.04 & 18 48 18.075014 & 0.12 & -06 13 31.70897 & 0.12 \\ \hline
08-08-2020   & 20:44:27.26 & 18 47 29.961308 & 0.12 & -06 16 31.34442 & 0.10  \\ \hline
24-02-2021 & 08:45:52.82 & 18 56 35.9873   & 1.1  & -06 30 23.1583  & 2.8 \\ \hline
14-10-2021  & 03:26:05.50 & 18 50 30.768595 & 0.48 & -06 24 13.20676 & 0.52 \\ \hline
10-06-2022     & 05:32:47.30 & 19 00 15.446841 & 0.32 & -05 42 42.8843  & 1.3 \\ \hline
\end{tabular}
}
\end{table}

\subsection{MS4's size, shape, and albedo}
\label{sec:4}
 
As mentioned previously in this work, we consider MS4 to have a Maclaurin shape ($a=b>c$) with an equatorial radius $\mathnormal{a}$, polar radius $\mathnormal{c}$, and true oblateness $\epsilon = (a-c)/a$. Given it is a Maclaurin body, the apparent semi-major axis will be equal to the true semi-major axis ($\mathnormal{a}'=\mathnormal{a}$). In addition, we assume that it was observed with the same aspect angle $\theta$ during all the stellar occultations, which $\theta= 0^\circ$ (resp. $90^\circ$) corresponds to a pole-on (resp. equator-on) viewing. The maximum true oblateness that a Maclaurin object can have is $\epsilon\leq0.417$ \citep{Tancredi2008}, which gives $c = 234$ km. On the other hand, $c > 387$ km if we consider the $\epsilon = \epsilon' = 0.034$, derived from the occultation, as the lower limit for the object's true oblateness. If the above-mentioned conditions are true, those values for the pole radius can be understood as the minimum and maximum values.

\begin{figure}[!ht]
    \centering
    \includegraphics[width=\linewidth]{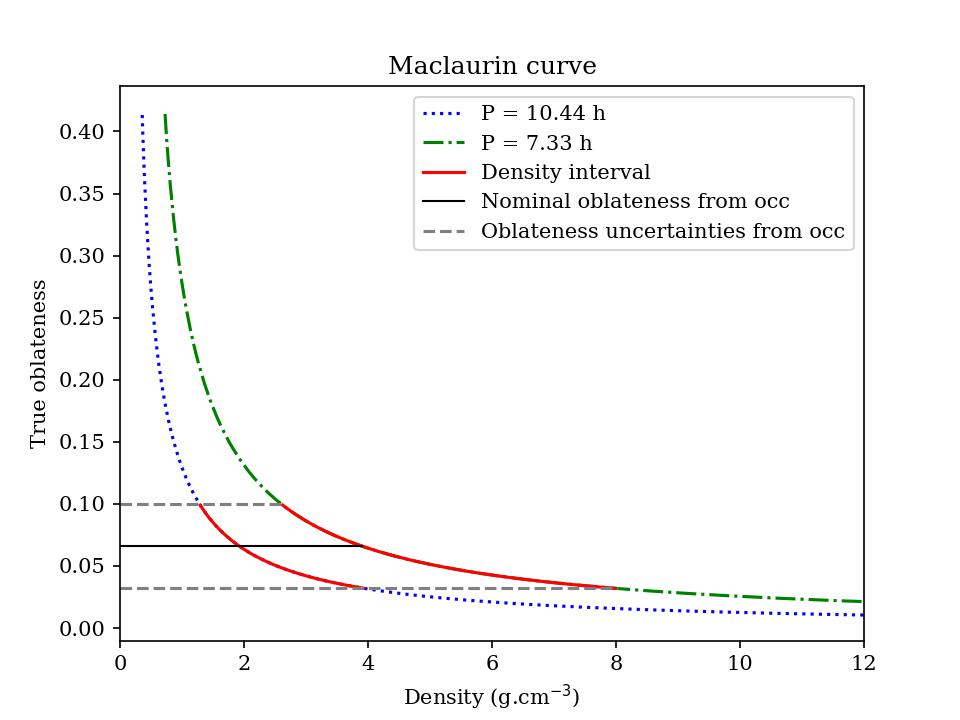}
    \caption{Relation between the true oblateness and the density of a Maclaurin spheroid for the rotational periods of 10.44 h (blue dotted line) and 7.33 h (green dashed line). The solid black line is the nominal oblateness value with uncertainties (gray dashed lines) derived from the multichord stellar occultation event. Red segments present the global density interval for each rotational period assuming that the apparent oblateness corresponds to the real one.}
    \label{fig:maclaurin}
\end{figure}

The Maclaurin spheroid formalism \citep{Braga-Ribas2013} allows us to estimate an object's density from the true oblateness and rotational period. Using the lower limit of true oblateness from the occultation and the two rotational periods in the literature, we obtained an upper limit for global density of 8.0~g.cm$^{-3}$ and 3.9~g.cm$^{-3}$ for periods of 7.33~h and 10.44~h, respectively. These values are too high for objects in the trans-Neptunian region, so it is reasonable to infer that the true oblateness is higher than the observed in the stellar occultation events. As shown in Fig. \ref{fig:maclaurin}, the density decreases as the true oblateness increases. Therefore, we can use the upper limit for the oblateness of a Maclaurin object ($\epsilon$ = 0.417) to obtain the lower limits of MS4's global density, which gives 0.72~g.cm$^{-3}$ and 0.36~g.cm$^{-3}$ for 7.44~h and 10.44~h, respectively.

Finally, a geometric albedo of $p_V$~=~0.1~$\pm$~0.025 was calculated using the equivalent radius from the occultation and the absolute magnitude on V-band of $H_V$~=~3.63~$\pm$~0.05~mag, which was obtained from the published V-magnitudes by \cite{Verbiscer2022}. The obtained absolute magnitude is in agreement with \citep{Tegler2016}. However, we consider the different phase angles of the measurements and the error bar was calculated considering the expected amplitude of the rotational light curve of $\Delta_m = 0.05$~mag \citep{Thirouin2013, Thirouin2013b}.

\section{Discussion and conclusions}
\label{sec:5}

This work presents physical and astrometric information derived from nine stellar occultations by the hot classical TNO (307261) 2002 MS$_4$ observed between 2019 and 2022 from South and North America, Africa, Europe, and Western Asia sites. The most successful campaign took place on 8 August 2020, with 116 telescopes involved and 61 positive chords, which represents a record number of detections of a stellar occultation by a TNO up to date.

The projected elliptical limb of MS4 derived from the 8 August 2020 provides a semi-major axis of 412$~\pm~$10~km, a semi-minor axis of 385$~\pm~$17 km, and an area-equivalent radius of 398$~\pm~$12~km. The obtained diameter is $\approx$138~km smaller than that derived with observations in thermal bands \citep{Vilenius2012}. It may indicate the presence of an unknown satellite as suggested for 2002 TC$_{302}$ in similar circumstances \citep{Ortiz2020a}. Still, the error bars from the thermal diameter are large and can accommodate the difference within 3$\sigma$.

Despite their being inconclusive, a shallow rotational light curve, the derived equivalent diameter, and the agreement between the limb obtained from the nine stellar occultations favor an oblate spheroid (Maclaurin) for the 3D shape of MS4. Furthermore, considering the expected values for TNOs, the density intervals mentioned above are quite large. This indicates that the object's true oblateness is higher than observed in the occultations or that the observed topography is the cause of the brightness variation in the rotational light curves. In the last case, the real rotational period may be double the published values, which provides smaller densities in the Maclaurin curve (Fig. \ref{fig:maclaurin}). In any case, more data are needed to confirm MS4's 3D shape and density. 

In addition, this work presents the first detailed multichord detection of an extensive feature on the surface of a monolithic TNO. A method was developed to identify and measure such detection.  In the northeast region of the observed limb, a $\approx$11~km depth depression was found, followed by an elevation of 25$^{+4}_{-5}$~km, subsequently followed by the most impressive feature, namely, a 45.1~$\pm$ 1.5~km depth depression with a linear extension of 322~$\pm$~39 km. Assuming a straight line that connects the model's initial and endpoints, the largest feature has $\approx40\%$ of the object's equivalent diameter. Such large topography is out of the range of expected global topography (Sect. \ref{sec:topography}) and may indicate a big impact during MS4 history. If so, one hypothesis that can be raised is whether the impact could have created a collisional family, such as that of Haumea \citep{Brown2007, Vilenius2018}, the only known collisional family in the trans-Neptunian region.

A comparison can be made with known craters in the outer Solar System. Among the largest Saturnian satellites, the Voyager and Cassini missions were able to acquire images of Tethys and Iapetus. The largest imaged craters are similar in size ratio to the feature observed in MS4. The Odysseus crater has a rim-to-rim diameter corresponding to $\approx43\%$ of the Tethys's mean diameter. The Turgis\footnote{Turgis diameter was obtained from \url{https://planetarynames.wr.usgs.gov/Feature/14488}.} crater has a diameter of $\approx40\%$ Iapetus' mean diameter \citep{Moore2004, Thomas1991}. However, the most recent and detailed studies about the surface of similar-to-MS4 objects in the trans-Neptunian region were performed for the Pluto-Charon system. Contrary to Saturn's satellites, the largest craters imaged by New Horizons have only 10.5$\%$ and 18.9$\%$ of the Pluto- and Charon-equivalent diameters, respectively \citep{Moore2016}. Therefore, the putative crater on the MS4 limb is the largest observed in the trans-Neptunian region, despite having a similar size ratio to craters found on planetary satellites. 

Finally, even with the unprecedented coverage of a stellar occultation by a TNO, no clear secondary drops in the star flux caused by rings, jets, or satellites were identified. Establishing upper limits for detecting such structures in the occultation light curves is  beyond the scope of this work. In addition, although it is still very unlikely, we cannot rule out the hypothesis that the elevation observed between PA = 0º and PA = 25º was caused by an unknown satellite with a diameter of $\approx$213 km passing in front of or behind the main body. If so, the main body would then have an effective diameter of $\approx$788 km (see Appendix \ref{appendix3}).

\begin{acknowledgements} 
This study was financed in part by the Coordenação de Aperfeiçoamento de Pessoal de Nível Superior – Brazil (CAPES) – Finance Code 001, the National Institute of Science and Technology of the e-Universe project (INCT do e-Universo, CNPq grant 465376/2014-2), the Spanish MICIN/AEI/10.13039/501100011033, the Institute of Cosmos Sciences University of Barcelona (ICCUB, Unidad de Excelencia 'María de Maeztu') through grant CEX2019-000918-M, the "ERDF A way of making Europe" by the “European Union” through grant PID2021-122842OB-C21, and within the “Lucky Star” umbrella that agglomerates the efforts of the Paris, Granada, and Rio teams, which the European Research Council funds under the European Community’s H2020 (ERC Grant Agreement No. 669416). 
The following authors acknowledge the respective CNPq grants: F. L. R. 103096/2023-0; F.B.-R. 314772/2020-0; R.V.-M. 307368/2021-1; J.I.B.C. 308150/2016-3 and 305917/2019-6; M.A. 427700/2018-3, 310683/2017-3, 473002/2013-2; G. M. 128580/2020-8; B.E.M. 150612/2020-6; and O.C.W. 305210/2018-1. 
The following authors acknowledge the respective grants: B.E.M. thanks the CAPES/Cofecub-394/2016-05; G. M. thanks the CAPES grant 88887.705245/2022-00; G.B-R. acknowledges CAPES - FAPERJ/PAPDRJ grant E26/203.173/2016 and the scholarship granted in the scope of the Program CAPES-PrInt, process number 88887.310463/2018-00, Mobility number 88887.571156/2020-00; M.A. acknowledges FAPERJ grant E-26/111.488/2013; A.R.G.Jr acknowledges FAPESP grant 2018/11239-8; O.C.W. and R.S. acknowledges FAPESP grant 2016/24561-0;
K. B. acknowledges the scholarship funded by F.R.S.-FNRS grant T.0109.20 and by the Francqui Foundation; D.N. acknowledges the support from the French Centre National d’Etudes Spatiales (CNES); D. S. thanks to Fulbright Visiting Scholar (2022 - 2023) at the University of California, Berkeley; A.P. and R.S. thanks to the National Research, Development and Innovation Office (NKFIH, Hungary) grants K-138962 and K-125015. Partial funding for the computational infrastructure and database servers is received from the grant KEP-7/2018 of the Hungarian Academy of Sciences; R. D., J.L.O., P. S.-S., N. M., R. H., A. S. L., and J. M. T.-R. acknowledge the MCIN/AEI/10.13039/501100011033 under the grant respective grants: CEX2021-001131-S, PID2019-109467GB-I00, and PID2021-128062NB-I00; T. S. R. acknowledges funding from the NEO-MAPP project (H2020-EU-2-1-6/870377); K.H. was supported by the project R.V.O.: 67985815; A. K. thanks to the IRAP, Midi-Pyrenees Observatory, CNRS, University of Toulouse, France; J. M. O. acknowledges the Portuguese Foundation for Science and Technology (FCT) and the European Social Fund (ESF) through the Ph.D. grant SFRH/BD/131700/2017; J.d.W. and MIT acknowledge the Heising-Simons Foundation, Dr. and Mrs. Colin Masson, and Dr. Peter A. Gilman for Artemis, the first telescope of the SPECULOOS network situated in Tenerife, Spain. The ULiege’s contribution to SPECULOOS has received funding from the ERC under the European Union’s Seventh Framework Programme (FP/2007-2013) (grant number 336480/SPECULOOS); J. L. acknowledges the ACIISII, Consejería de Economía, Conocimiento y Empleo del Gobierno de Canarias, and the European Regional Development Fund (ERDF) under the grant ProID2021010134, also the Agencia Estatal de Investigación del Ministerio de Ciencia e Innovacion (AEI-MCINN) under the grant PID2020-120464GB-100; D. T., R. K., M. H., and T. P. were supported by the Slovak Grant Agency for Science grants number VEGA 2/0059/22, and VEGA 2/0031/22"; M. P. was supported by a grant from the Romanian National Authority for Scientific Research – UEFISCDI, project number PN-III-P1-1.1-TE-2019-1504; P. B., M. M., and M. D. G. thanks the support of the Italian Amateur Astronomers Union (UAI).  
This work has made use of data from the European Space Agency (ESA) mission {\it Gaia} (\url{https://www.cosmos.esa.int/gaia}), processed by the {\it Gaia} Data Processing and Analysis Consortium (DPAC, \url{https://www.cosmos.esa.int/web/gaia/dpac/consortium}). Funding for the DPAC has been provided by national institutions, in particular, the institutions participating in the {\it Gaia} Multilateral Agreement.
The Joan Oró Telescope (TJO) of the Montsec Observatory (OdM) is owned by the Catalan Government and operated by the Institute for Space Studies of Catalonia (IEEC). TCH telescope is financed by the Island Council of Ibiza. İST60 and İST40 are the observational facilities of the Istanbul University Observatory, funded by the Scientific Research Projects Coordination Unit of Istanbul University with project numbers BAP-3685 and FBG-2017-23943 and the Presidency of Strategy and Budget of the Republic of Turkey with the project 2016K12137. TRAPPIST is a project funded by the Belgian FNRS grant PDR T.0120.21 and the ARC grant for Concerted Research Actions, financed by the Wallonia-Brussels Federation. E.J. is an FNRS Senior Research Associate. TRAPPIST-North is funded by the University of Liège and performed in collaboration with the Cadi Ayyad University of Marrakesh. 
This work made use of observations obtained at the 1.6 m telescope on the Pico dos Dias Observatory of the National Laboratory of Astrophysics (LNA/Brazil), at the Copernicus and Schmidt telescopes (Asiago, Italy) of the INAF-Astronomical Observatory of Padova, and at the Southern Astrophysical Research (SOAR) telescope, which is a joint project of the Ministério da Ciência, Tecnologia, e Inovação (MCTI) da República Federativa do Brasil, the U.S. National Optical Astronomy Observatory (NOAO), the University of North Carolina at Chapel Hill (UNC), and Michigan State University (MSU).
We thank the following observers who participated and provided data for the events listed in Appendix A: 
A. Ciarnella,
A. L. Ivanov,
A. Olsen,
A. Ossola
B. Dintinjana, 
B. Hanna,
C. Costa,
G. H. Rudnick,
J. Paul,
J. J. Castellani,
J. Polak, 
K. Guhl,
L. A. Molnar,
L. Perez,
M. Bertini,
M. Bigi, 
M. Rottenborn,
M. Sabil,  
N. V. Ivanova,
P. Langill,
P. Thierry, 
S. Bouquillon, 
S. Lamina, 
T. O. Dementiev, 
V. A. Ivanov, and 
X. Delmotte.
\end{acknowledgements}



\begin{appendix}
\onecolumn
\section{Post occultation maps with sites location}
\label{appendix0}
Here we present the post-occultation maps of each stellar occultation by 2002 MS4 described in this work. 

\begin{figure}[!htb]
    \centering
    \includegraphics[width=\linewidth]{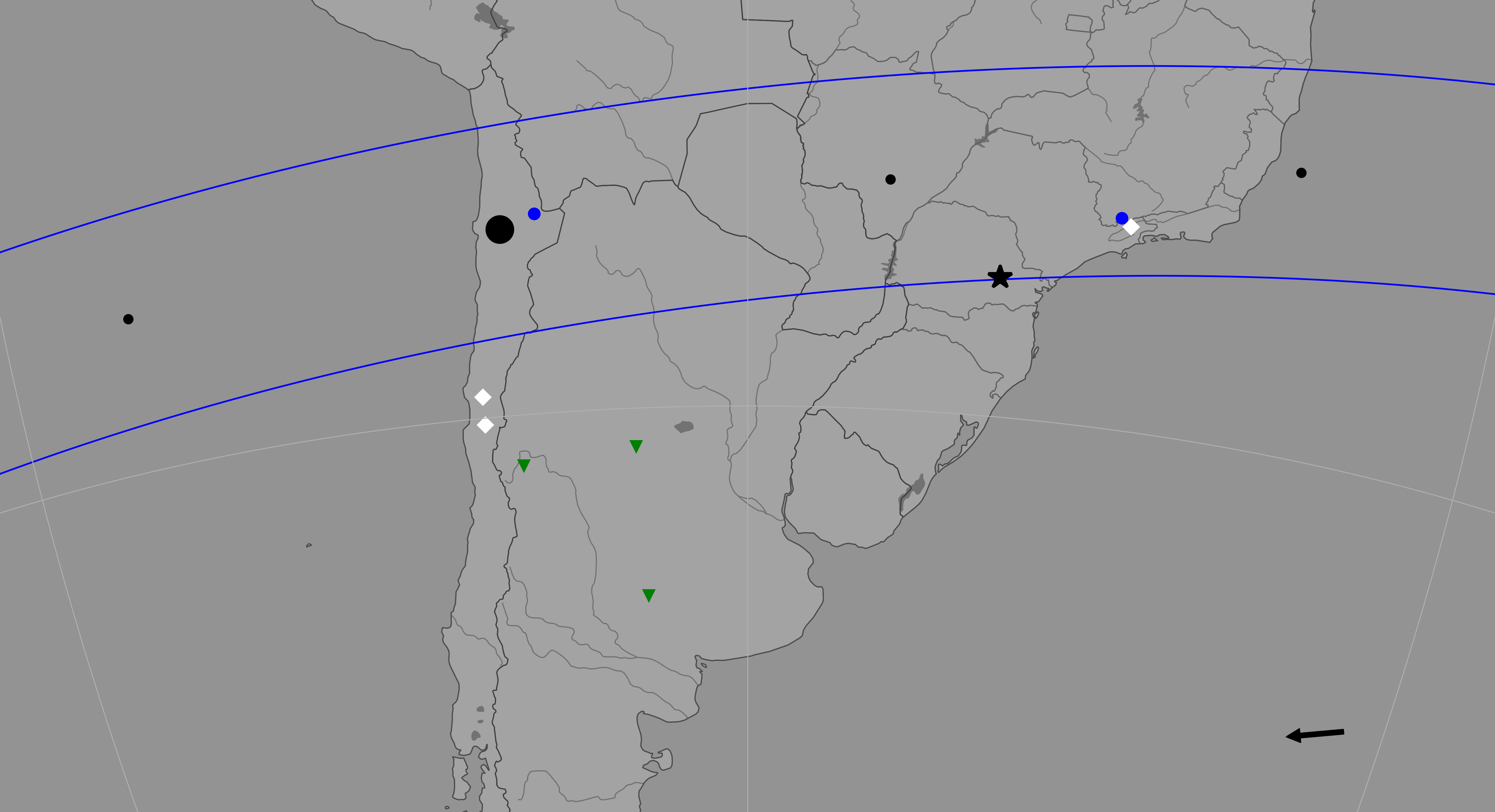}\quad
    \includegraphics[width=\linewidth]{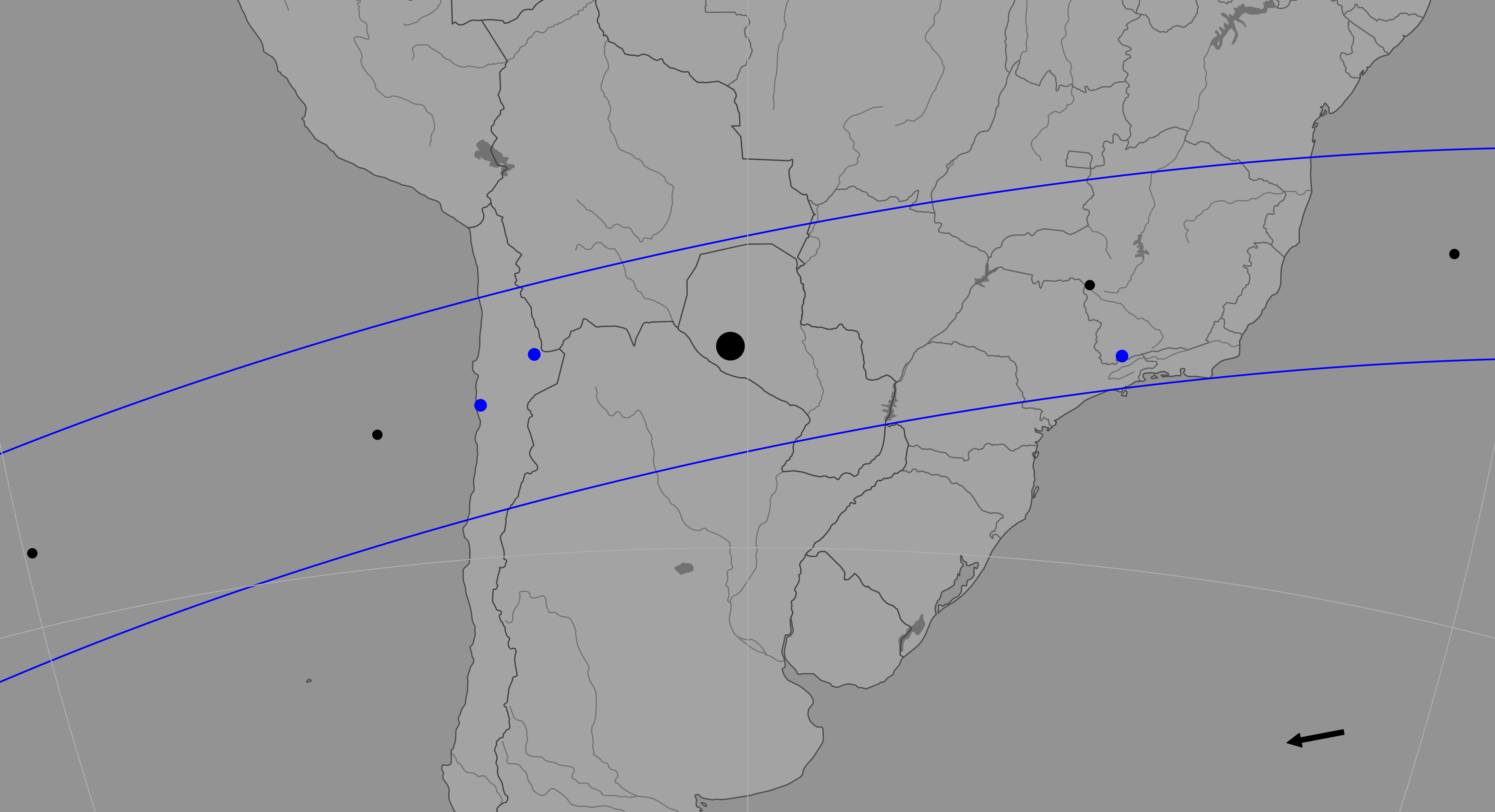}
    \caption{Post-occultation map of the stellar occultations observed on July 09 (upper panel) and July 26, 2019 (lower panel) from South America. Solid blue lines and black dots indicate the observed shadow path every minute, and the largest black bullet represents the closest approach instant. The arrow shows the shadow direction over Earth's surface. Blue dots show the positive detections, green triangles show the negatives, and white diamonds show the stations with bad weather conditions or technical problems.  The black star marks the Ponta Grossa station, a close negative chord that limits the ellipse solutions at the south for the July 09, 2019, event.}
    \label{fig:postmap_2019_P1}
\end{figure}

\begin{figure}[!htb]
    \centering
    \includegraphics[width=\linewidth]{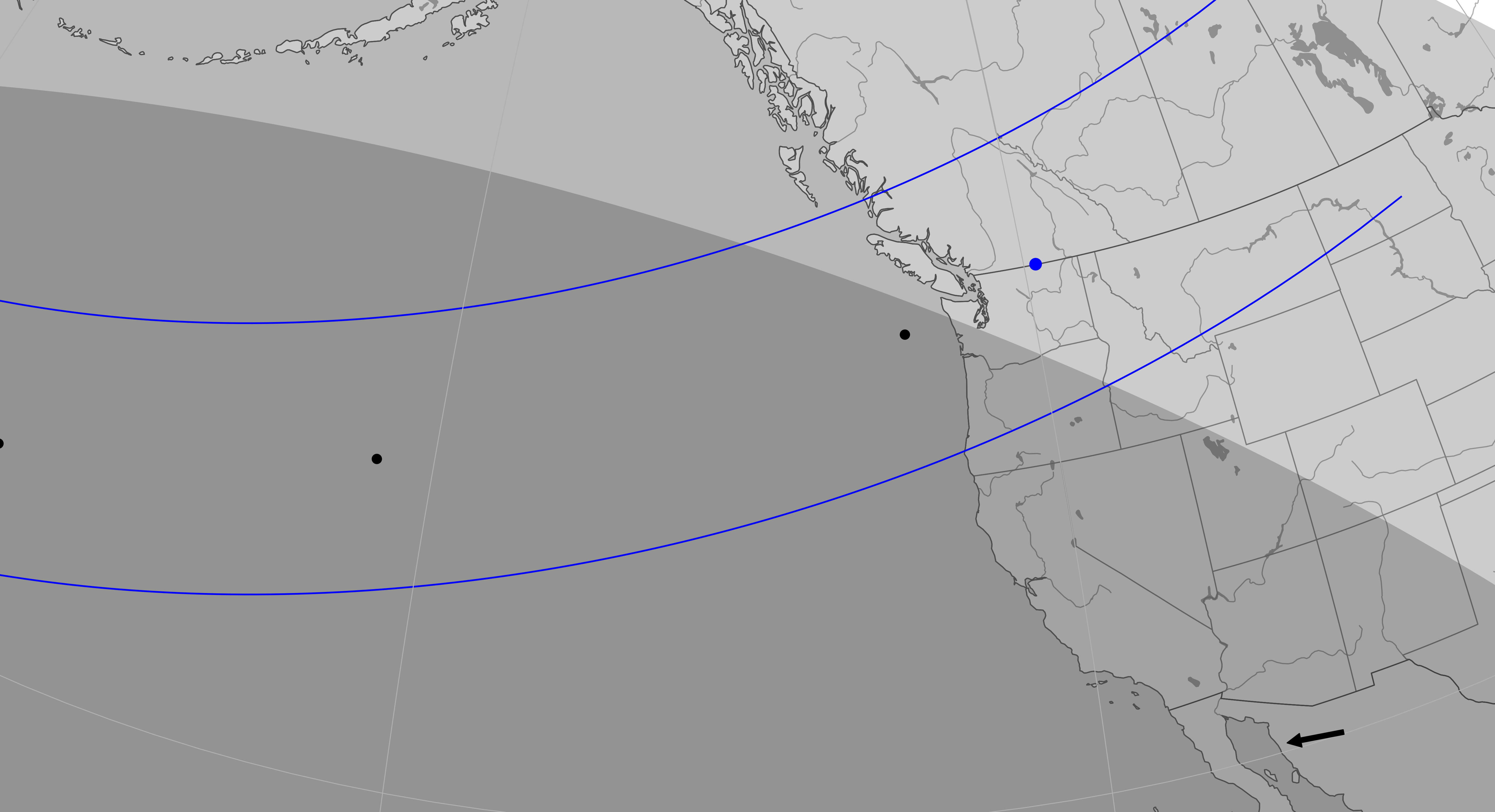}\quad
    \includegraphics[width=\linewidth]{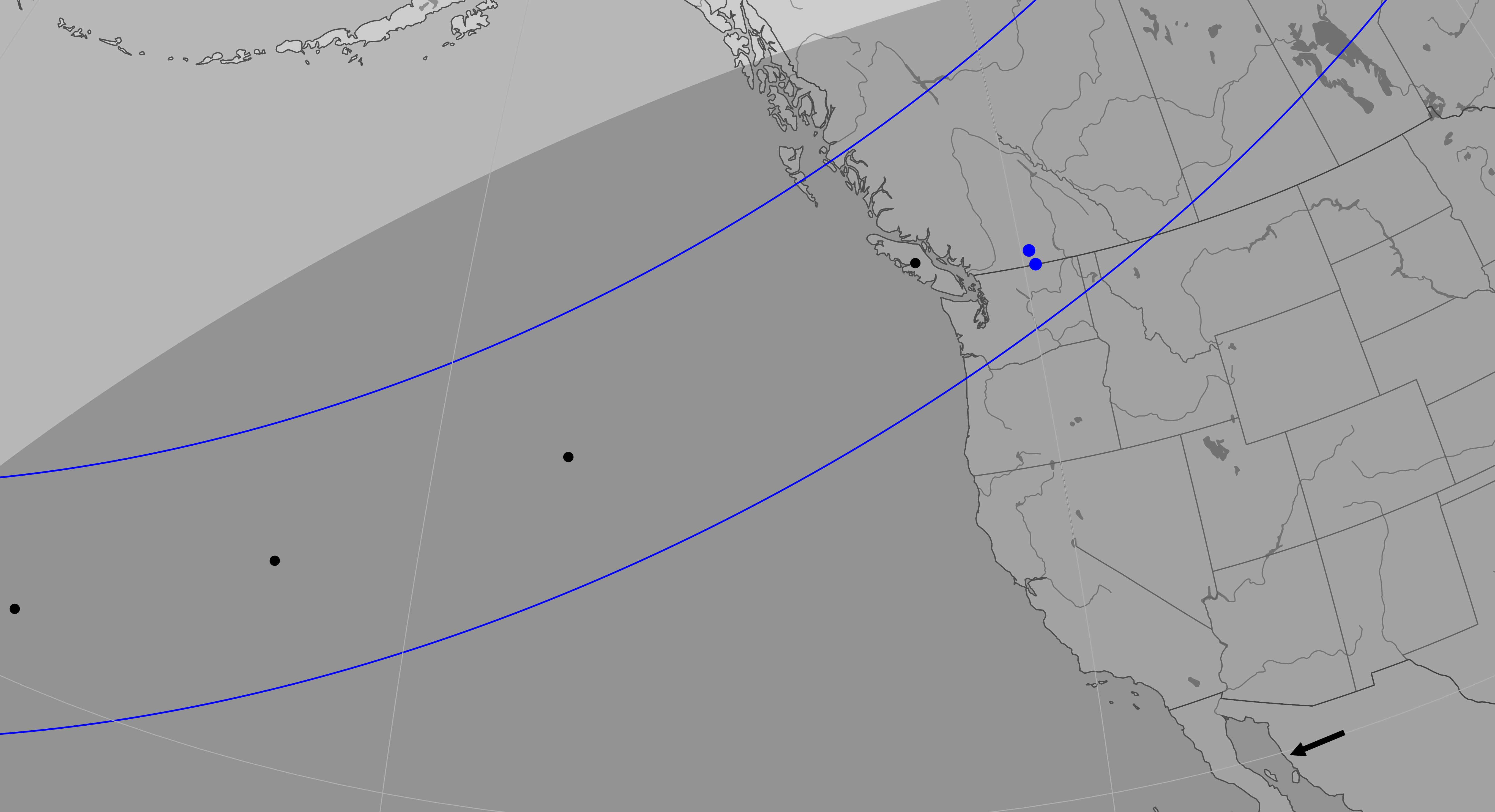}
    \caption{Post-occultation map of the stellar occultations observed on July 26 (upper panel) and August 19, 2019 (lower panel) from North America. Solid blue lines and black dots indicate the observed shadow path every minute. The arrow shows the shadow direction over Earth's surface, and the blue dots show the positive detections.}
    \label{fig:postmap_2019_P2}
\end{figure}

\begin{figure}[!htb]
    \centering
    \includegraphics[width=\linewidth]{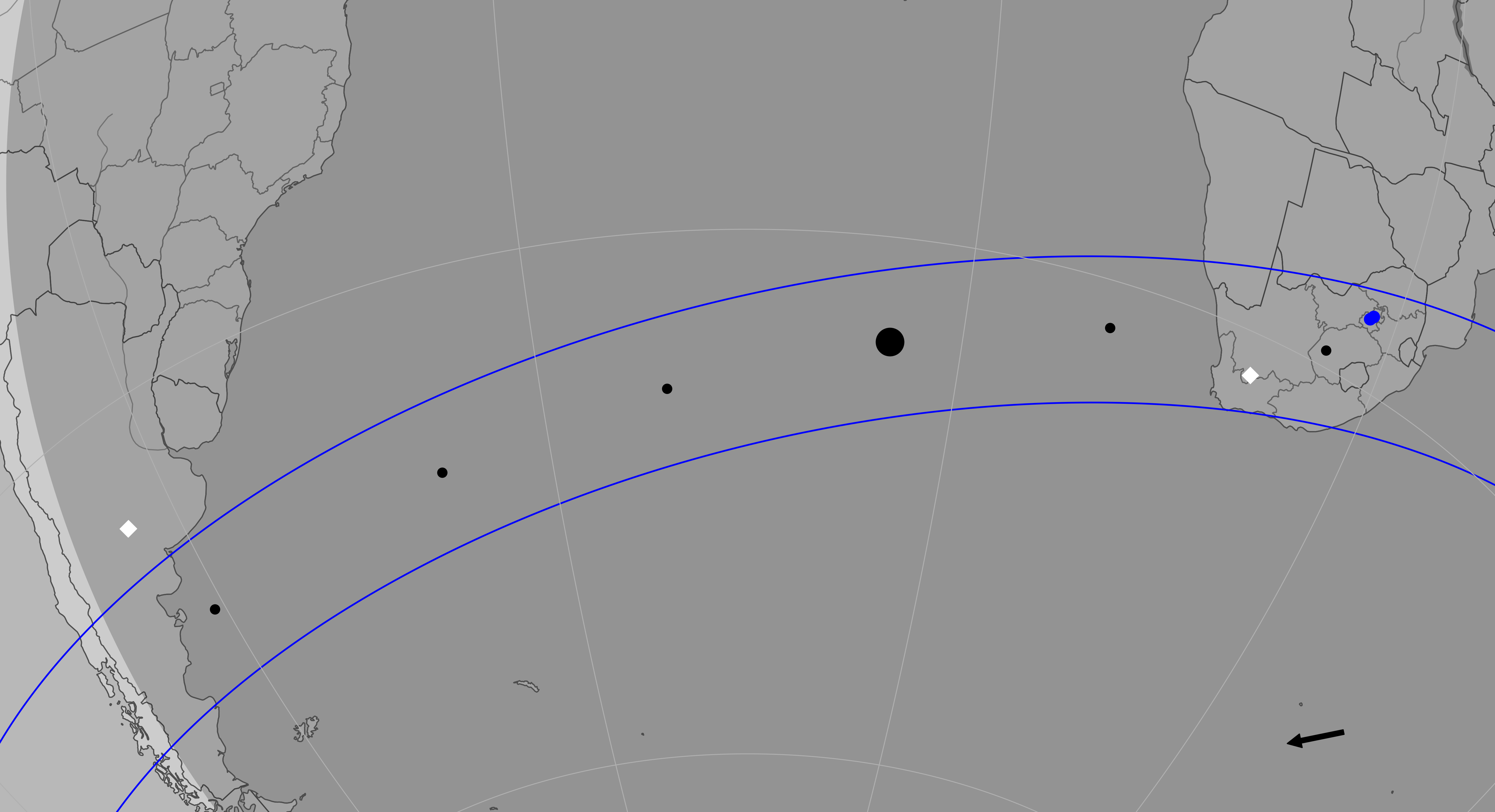}\quad
    \includegraphics[width=\linewidth]{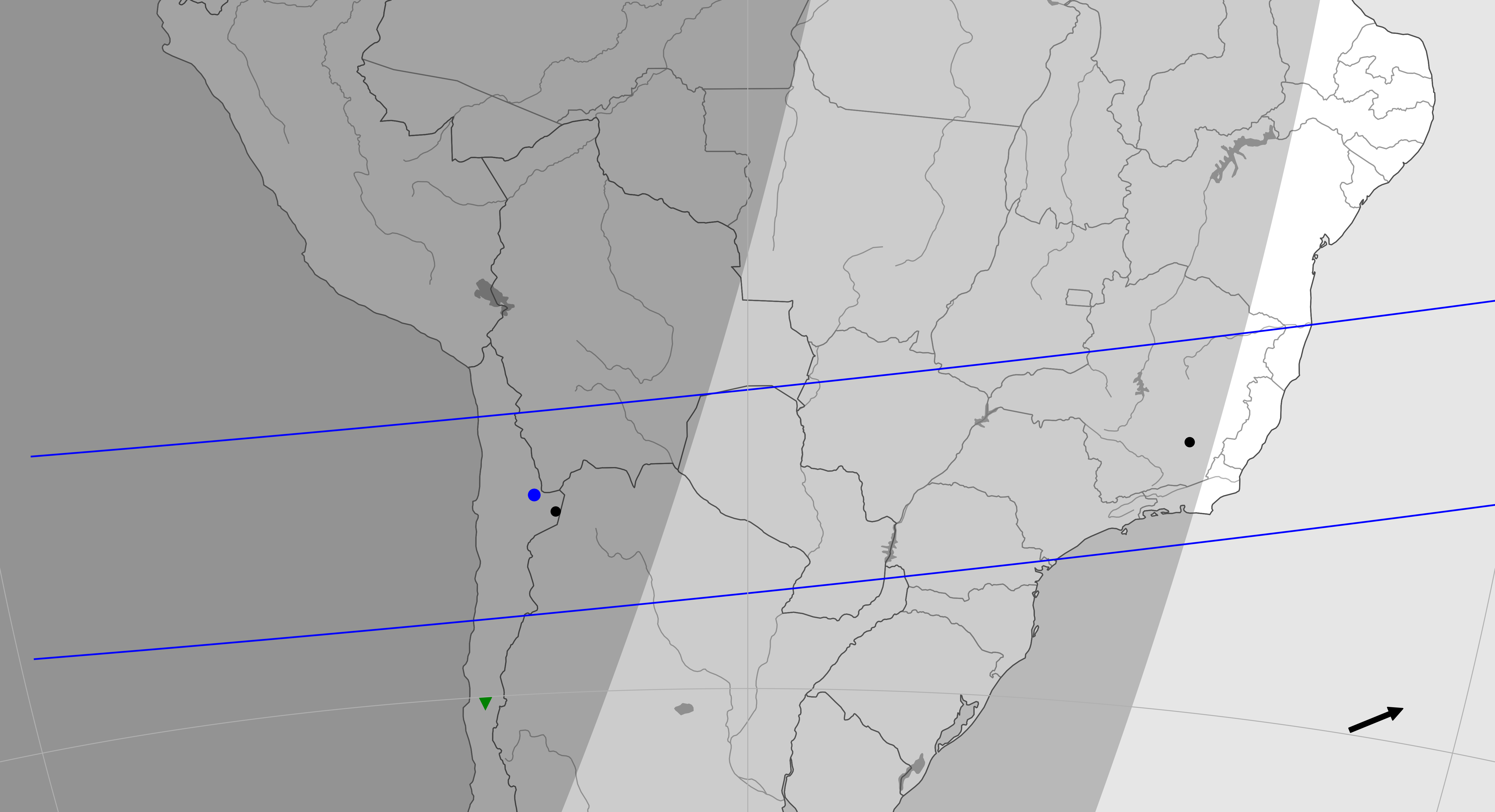}
    \caption{Post-occultation map of the stellar occultations observed on July 26, 2020, from South Africa and South America (upper panel) and on February 24, 2021, also from South America (lower panel). Solid blue lines and black dots indicate the observed shadow path every minute, and the largest black bullet represents the closest approach instant. The arrow shows the shadow direction over Earth's surface. Blue dots show the positive detections, green triangles show the negatives, and white diamonds show the stations with bad weather conditions or technical problems.}
    \label{fig:postmap_2020_2021}
\end{figure}

\begin{figure}[!htb]
    \centering
    \includegraphics[width=\linewidth]{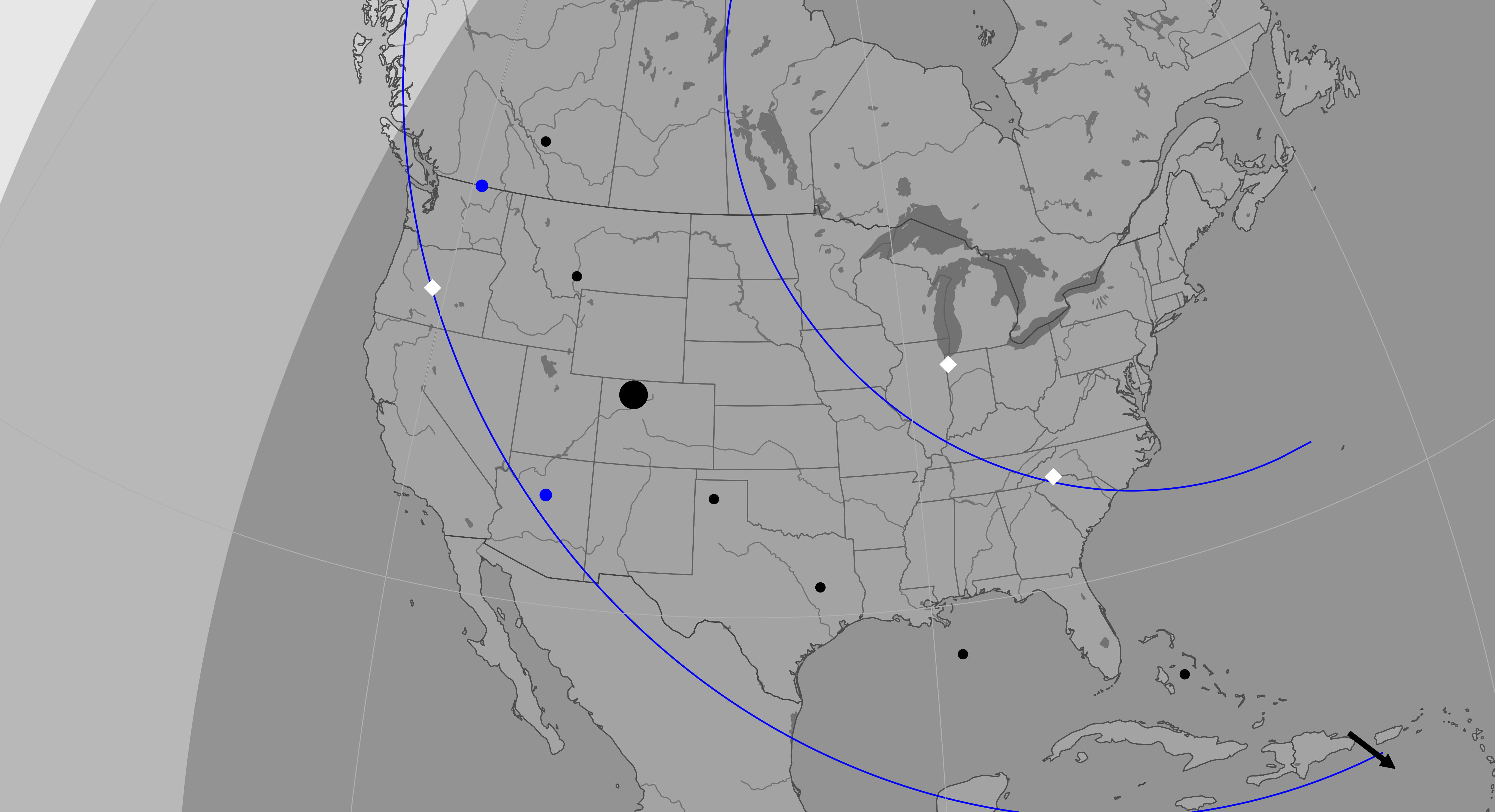}\quad
    \includegraphics[width=\linewidth]{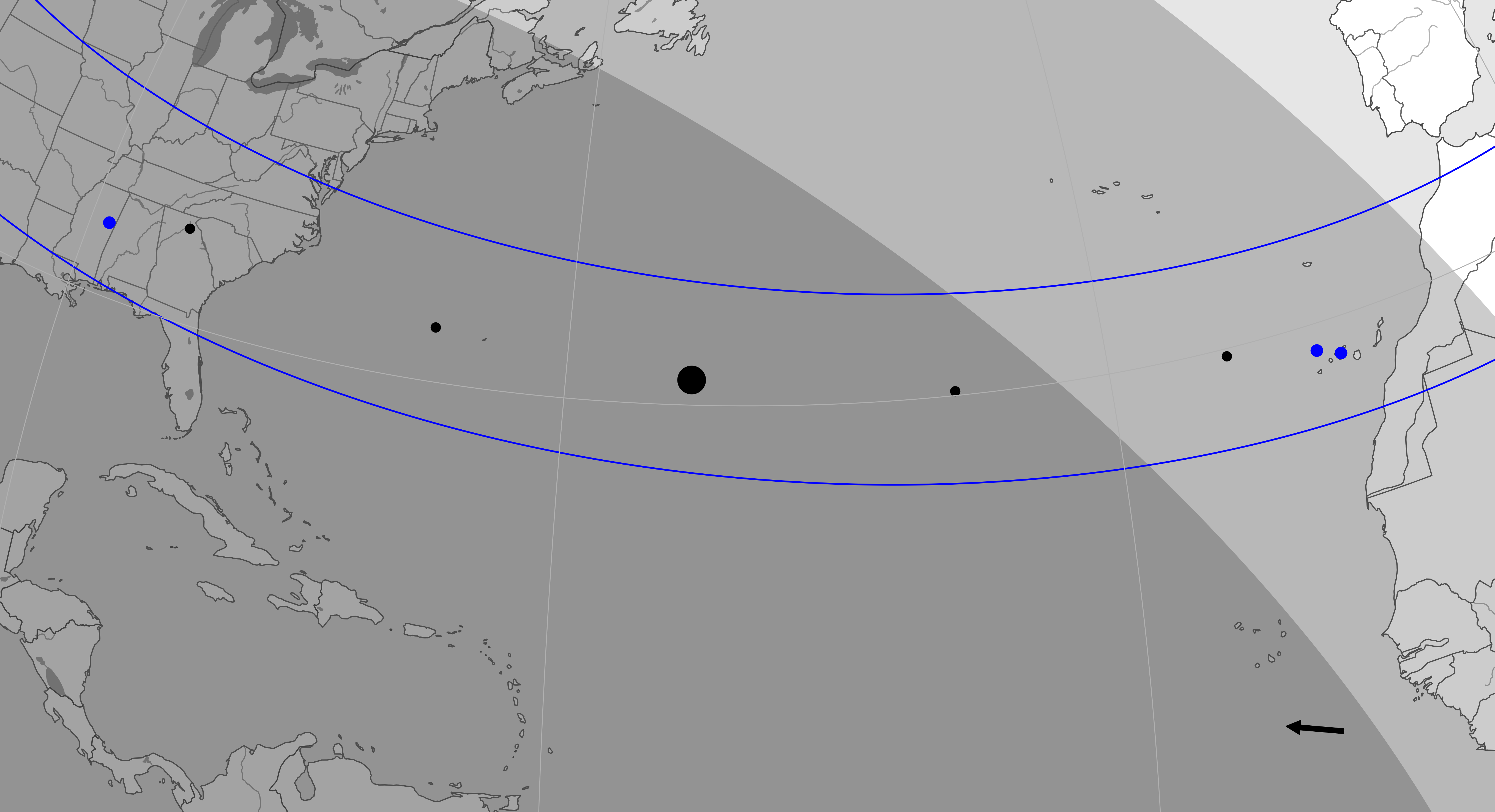}
    \caption{Post-occultation map of the stellar occultations observed on October 14, 2021, from North America (upper panel) and on June 10, 2022, from North America and Europe (lower panel). Solid blue lines and black dots indicate the observed shadow path every minute, and the largest black bullet represents the closest approach instant. The arrow shows the shadow direction over Earth's surface. Blue dots show the positive detections, and white diamonds show the stations with bad weather conditions or technical problems.}
    \label{fig:postmap_2021_2022}
\end{figure}

\onecolumn
\section{Observational circumstances}
\label{appendix1}

The following tables summarize the observational circumstances of each station of the nine stellar occultations presented in this work. For better visualization, the tables were divided into two groups i) the 8 August 2020 event and ii) the other eight stellar occultations. The positive, negative, and overcast locations involved in the 8 August 2020 campaign are listed in Tables \ref{tab:aug_pos_sites}, \ref{tab:aug_neg_sites}, and \ref{tab:aug_overcast_sites}, respectively. Positive and negative observations of the other eight events are present in Tables  \ref{table:other_pos_sites} and \ref{table:other_neg_sites}, respectively.

{\tiny
\begin{longtable}{@{}lccccl@{}}
\caption{Observational circumstances of all observatories that detected the stellar occultation by MS4 on 8 August 2020. The * symbol indicates that this data was taken in drift scan mode.}
\label{tab:aug_pos_sites}\\
\toprule
\textbf{\begin{tabular}[c]{@{}l@{}}Observatory,\\ nearest city, \\ country\end{tabular}} &
  \textbf{\begin{tabular}[c]{@{}c@{}}Latitude ($^\circ$),\\ longitude ($^\circ$),\\ altitude (m)\end{tabular}} &
  \textbf{\begin{tabular}[c]{@{}c@{}}Telescope,\\ aperture (m),\\ filter\end{tabular}} &
  \textbf{\begin{tabular}[c]{@{}c@{}}Time source,\\ instrument\end{tabular}} &
  \textbf{\begin{tabular}[c]{@{}c@{}}Exposure (s),\\ cycle (s),\\ correction (s)\end{tabular}} &
  \textbf{Observers} \\ \midrule
\endfirsthead
\multicolumn{6}{c}%
{{\bfseries Table \thetable\ continued}} \\ 
\hline
\endhead
\begin{tabular}[c]{@{}l@{}}Domaine de La Blaque,\\ Varages,\\ France\end{tabular} &
  \begin{tabular}[c]{@{}c@{}}+43.61239\\ +05.96363\\ 468.0\end{tabular} &
  \begin{tabular}[c]{@{}c@{}}Ritchey-Chrétien\\ 0.5\\ Clear\end{tabular} &
  \begin{tabular}[c]{@{}c@{}}GPS\\ WATEC 910HX\end{tabular} &
  \begin{tabular}[c]{@{}c@{}}0.32\\ 0.32\\ -\end{tabular} &
  \begin{tabular}[c]{@{}l@{}}Jean Lecacheux,\\ Jean-Luc Plouvier\end{tabular} \\ \midrule
\begin{tabular}[c]{@{}l@{}}TAROT Calern,\\ Caussols, \\ France\end{tabular} &
  \begin{tabular}[c]{@{}c@{}}+43.752001\\ +06.923613\\ 1268.0\end{tabular} &
  \begin{tabular}[c]{@{}c@{}}TAROT North\\ 0.25\\ Clear\end{tabular} &
  \begin{tabular}[c]{@{}c@{}}GPS\\ Andor DZ936N-BEX2-DD\end{tabular} &
  \begin{tabular}[c]{@{}c@{}}90.0*\\ 90.0*\\ -\end{tabular} &
  \begin{tabular}[c]{@{}l@{}}Eric Frappa, \\ Alain Klotz\end{tabular} \\ \midrule
\begin{tabular}[c]{@{}l@{}}Méo Station,\\ Grasse,\\ France\end{tabular} &
  \begin{tabular}[c]{@{}c@{}}+43.7546\\ +06.9216\\ 1323.1\end{tabular} &
  \begin{tabular}[c]{@{}c@{}}Ritchey-Chrétien\\ 1.54\\ Clear\end{tabular} &
  \begin{tabular}[c]{@{}c@{}}NTP\\ ZWO ASI1600MM\end{tabular} &
  \begin{tabular}[c]{@{}c@{}}1.0\\ 1.0\\ -\end{tabular} &
  \begin{tabular}[c]{@{}l@{}}Dominique Albanese,\\ Hervé Mariey\end{tabular} \\ \midrule
\begin{tabular}[c]{@{}l@{}}Caussols,\\ Caussols,\\ France\end{tabular} &
  \begin{tabular}[c]{@{}c@{}}+43.753055556\\ +6.921666667\\ 1268.0\end{tabular} &
  \begin{tabular}[c]{@{}c@{}}CFF T400\\ 0.40\\ Clear\end{tabular} &
  \begin{tabular}[c]{@{}c@{}}NTP\\ ZWO 294 MC\\ -\end{tabular} &
  \begin{tabular}[c]{@{}c@{}}3.0\\ 4.0\\ +0.4\end{tabular} &
  \begin{tabular}[c]{@{}l@{}}Raymond Behem,\\ Jean Pierre Prost\end{tabular} \\
\bottomrule
\end{longtable}
\begin{longtable}{@{}lccccl@{}}
\caption{Observational circumstances of all the stations that acquired data of the 8 August 2020 event but did not detect the occultation.  The symbol * indicates that this information is from the observatory website 
.}
\label{tab:aug_neg_sites}\\
\hline
\textbf{\begin{tabular}[c]{@{}l@{}}Observatory,\\ nearest city, \\ country\end{tabular}} &
  \textbf{\begin{tabular}[c]{@{}c@{}}Latitude ($^\circ$),\\ longitude ($^\circ$),\\ altitude (m)\end{tabular}} &
  \textbf{\begin{tabular}[c]{@{}c@{}}Telescope,\\ aperture (m),\\ filter\end{tabular}} &
  \textbf{\begin{tabular}[c]{@{}c@{}}Time source,\\ instrument\\ \end{tabular}} &
  \textbf{\begin{tabular}[c]{@{}c@{}}Exposure (s),\\ cycle (s)\end{tabular}} &
  \textbf{Observers} \\ \hline
\endfirsthead
\multicolumn{6}{c}%
{{\bfseries Table \thetable\ continued}} \\
\endhead
\begin{tabular}[c]{@{}l@{}}Sternwarte Comthurey,\\ Neustrelitz,\\ Germany\end{tabular} &
  \begin{tabular}[c]{@{}c@{}}+53.26608\\ +13.1901666\\ 74.0\end{tabular} &
  \begin{tabular}[c]{@{}c@{}}-\\ 0.18\\ Clear\end{tabular} &
  \begin{tabular}[c]{@{}c@{}}GPS\\ QHY174\\ \end{tabular} &
  \begin{tabular}[c]{@{}c@{}}2.964 \\ 2.964\end{tabular} &
  Konrad Guhl \\ \hline
\begin{tabular}[c]{@{}l@{}}Breitenweg,\\ Herkenrath,\\ Germany\end{tabular} &
  \begin{tabular}[c]{@{}c@{}}+50.993310\\ +07.183794\\ 200.0\end{tabular} &
  \begin{tabular}[c]{@{}c@{}}RC\\ 0.304\\ Clear\end{tabular} &
  \begin{tabular}[c]{@{}c@{}}GPS\\ Watec 120N+\end{tabular} &
  \begin{tabular}[c]{@{}c@{}}2.56\\ 2.56\end{tabular} &
  Bernd Klemt \\ \hline
\begin{tabular}[c]{@{}l@{}}Biesenthal,\\ Biesenthal,\\ Germany\end{tabular} &
  \begin{tabular}[c]{@{}c@{}}+52.759278\\ +13.663\\ 263.0\end{tabular} &
  \begin{tabular}[c]{@{}c@{}}Newtonian\\ 0.3\\ Clear\end{tabular} &
  \begin{tabular}[c]{@{}c@{}}GPS\\ QHY174\end{tabular} &
  \begin{tabular}[c]{@{}c@{}}0.6\\ 0.6\end{tabular} &
  Nikolai Wüensche \\ \hline
\begin{tabular}[c]{@{}l@{}}-\\ Berlin,\\ Germany\end{tabular} &
  \begin{tabular}[c]{@{}c@{}}+52.516111\\ +13.427778\\ 40.0\end{tabular} &
  \begin{tabular}[c]{@{}c@{}}Newtonian\\ 0.254\\ Clear\end{tabular} &
  \begin{tabular}[c]{@{}c@{}}GPS\\ DVTI\end{tabular} &
  \begin{tabular}[c]{@{}c@{}}3.0\\ 3.0\end{tabular} &
  Christian Weber \\ \hline
\begin{tabular}[c]{@{}l@{}}Eppstein-Bremthal,\\ Wiesbaden,\\ Germany\end{tabular} &
  \begin{tabular}[c]{@{}c@{}}+50.13816667\\ +08.364\\ 256.0\end{tabular} &
  \begin{tabular}[c]{@{}c@{}}Schmidt-Cassegrain\\ 0.254\\ Clear\end{tabular} &
  \begin{tabular}[c]{@{}c@{}}GPS\\ QHY-174M\end{tabular} &
  \begin{tabular}[c]{@{}c@{}}3.0\\ 3.0\end{tabular} &
  Oliver Klös \\
\hline
\end{longtable}
}
\begin{table}[!h]
\centering
\caption{Observational circumstances of all the sites that attempted to observe the 8 August 2020 event, but experienced bad weather or technical issues and did not acquire any data.}
\label{tab:aug_overcast_sites}
\resizebox{\textwidth}{!}{%
%
}
\end{table}

\begin{table}[]
\caption{Observational circumstances of all stations that detected 2002 MS4 in a stellar occultation on the other eight events.}
\label{table:other_pos_sites}
\resizebox{\textwidth}{!}{%
%
}
\end{table}

\begin{table}[!h]
\centering
\caption{Observational circumstances of all the stations that did not detect 2002 MS4 or experienced bad weather during the other eight stellar occultations. $^\dagger$ This instrument is described in \cite{Coppejans2013}.}
\label{table:other_neg_sites}
\resizebox{0.9\textwidth}{!}{%
%
 &
  Bill Hanna \\ \hline
\end{tabular}
}
\end{table}

\onecolumn
\newpage
\section{Light curves}
\label{appendix2}
 
Here, we provide the plots of the 80 positive occultation light curves acquired during the nine events observed between 2019 and 2022. They are normalized to the unity, and the time is given in seconds, counting from 00:00:00 (UTC) of the event date. Figures \ref{fig:b1}, \ref{fig:b2}, \ref{fig:b3}, \ref{fig:b4}, and \ref{fig:b5} present the plots from the 8 August 2020 stellar occultation, listed from the northernmost to the southernmost stations (on each column). Figures \ref{fig:b6} and \ref{fig:b7} show the light curves from the other eight events. The black dots present the observational data and the red line is the fitted model. 

\begin{figure}[!h]
    \centering
    \includegraphics[width=0.48\linewidth]{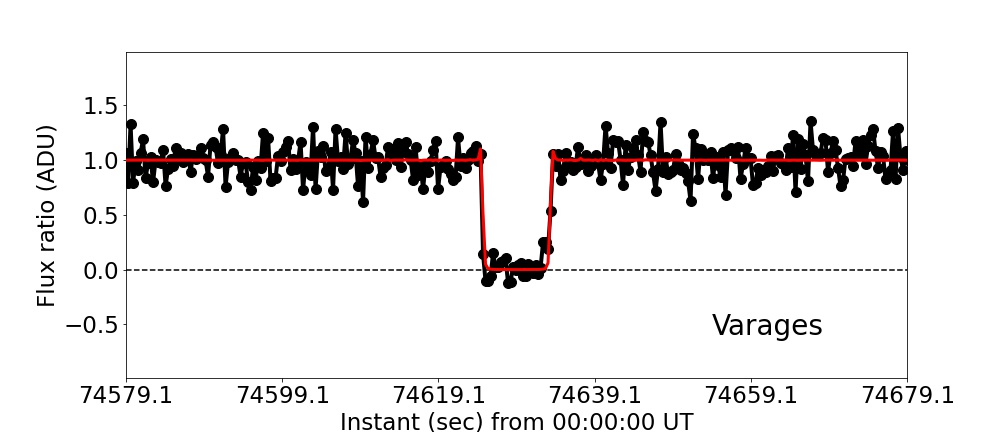}\quad   
    \includegraphics[width=0.48\linewidth]{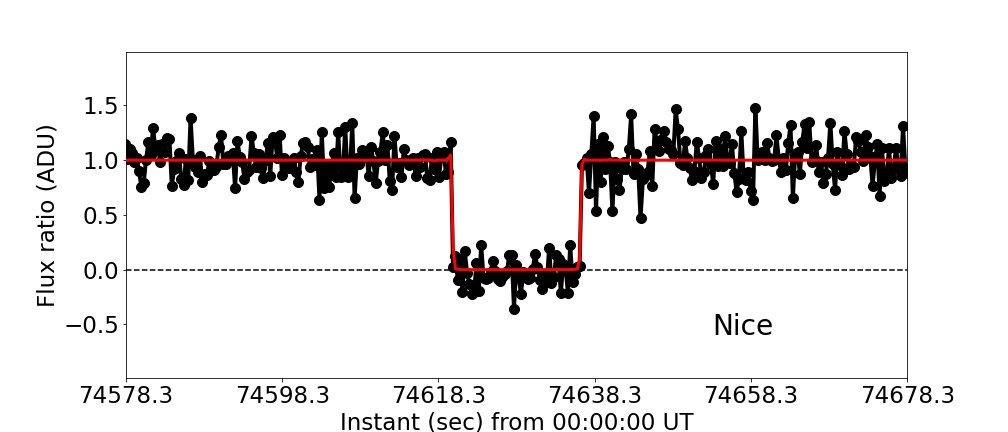}\quad  
    \includegraphics[width=0.48\linewidth]{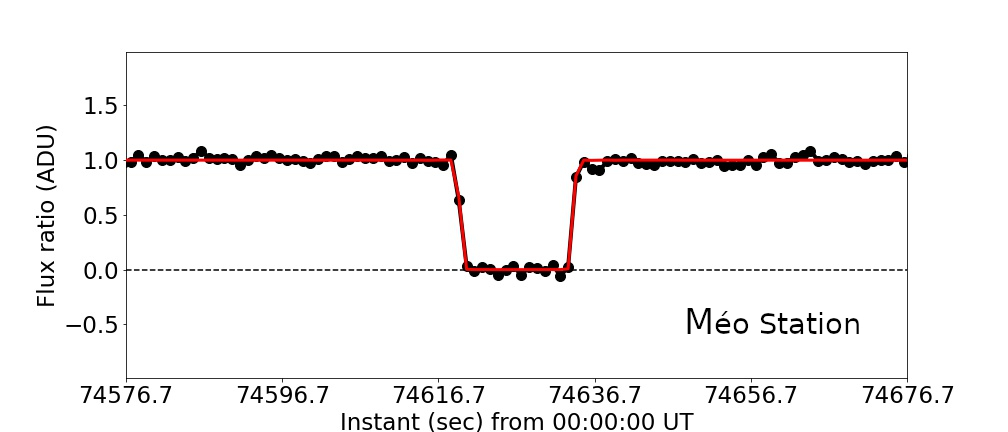}\quad    
    \includegraphics[width=0.48\linewidth]{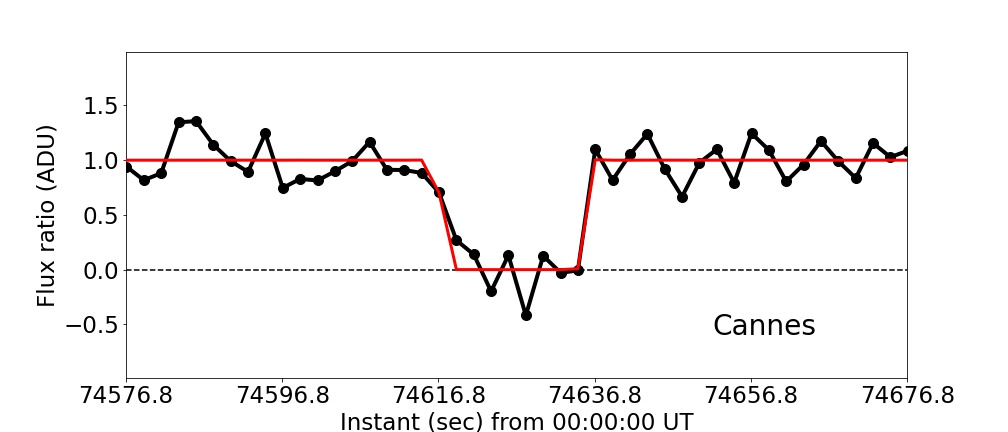}\quad  
    \includegraphics[width=0.48\linewidth]{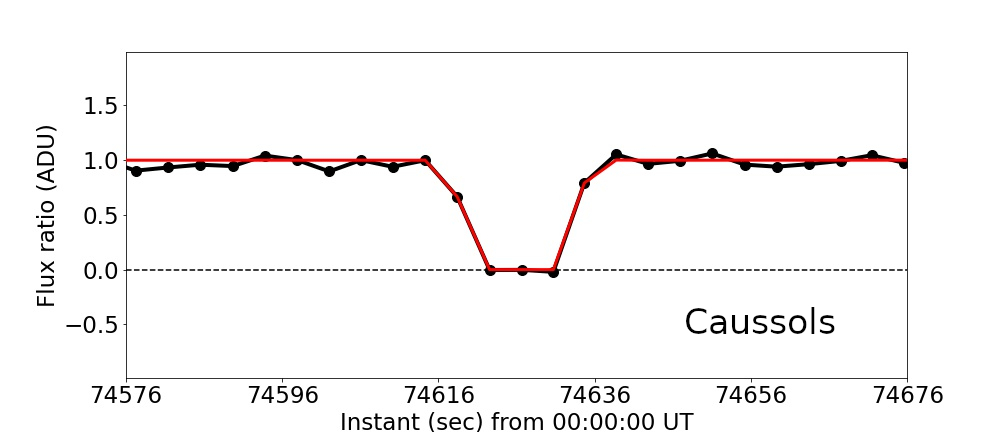}\quad  
    \includegraphics[width=0.48\linewidth]{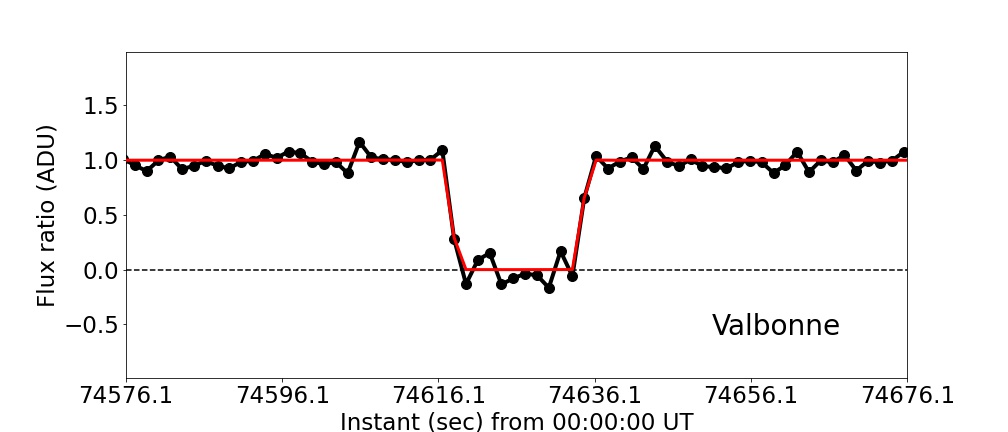}\quad  
    \includegraphics[width=0.48\linewidth]{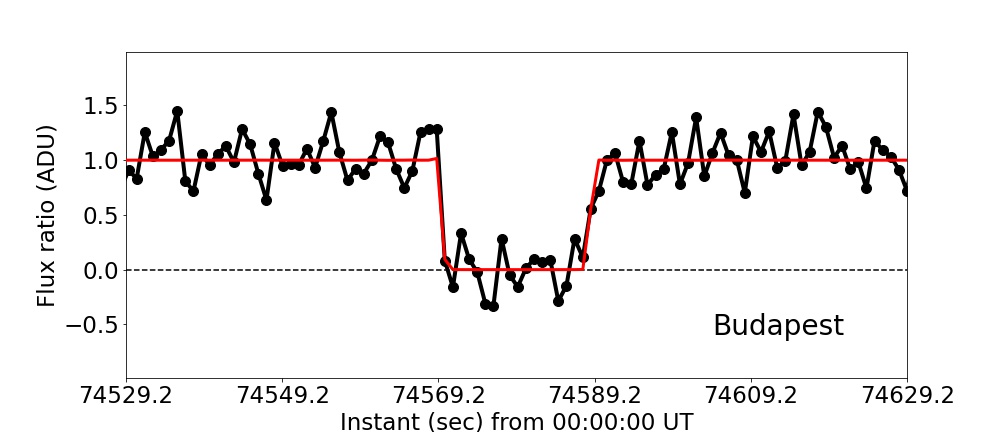}\quad  
    \includegraphics[width=0.48\linewidth]{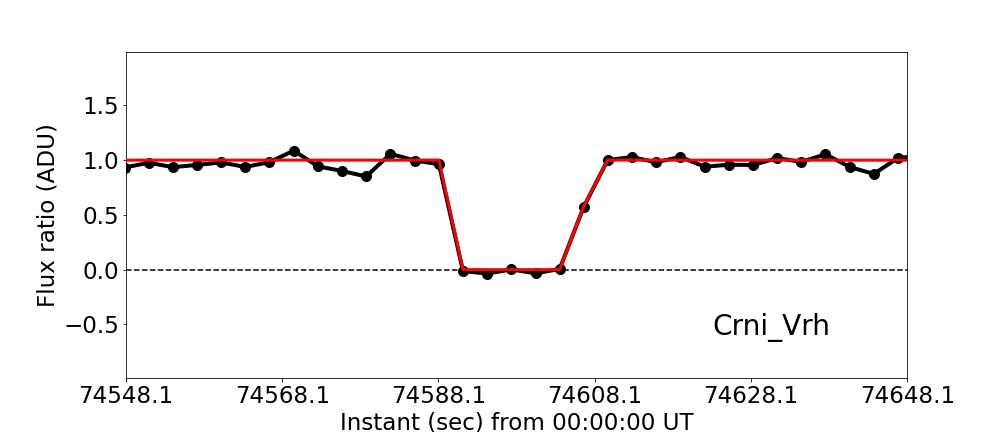}\quad  
    \includegraphics[width=0.48\linewidth]{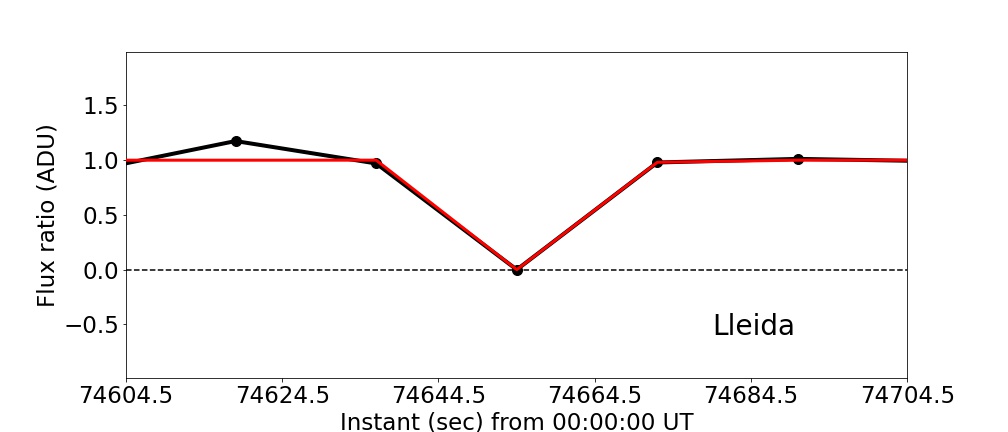}\quad    
    \includegraphics[width=0.48\linewidth]{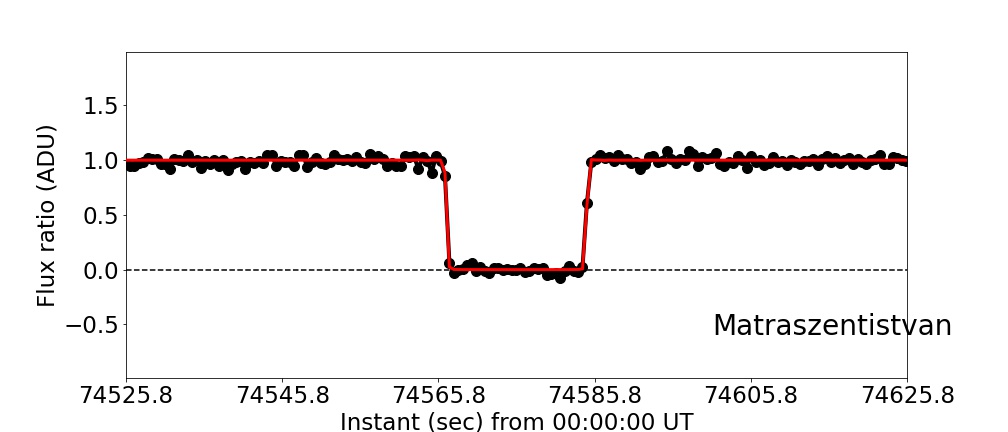}   
    \caption{Sixty-one normalized light curves, centered in the occultation instant, obtained on the 8 August 2020 campaign. The station that acquired the light curve is mentioned in each plot. The black and red points present the observed data and the fitted model, respectively.}
    \label{fig:b1}
\end{figure}

\begin{figure}[!h]
    \centering
    \includegraphics[width=0.48\linewidth]{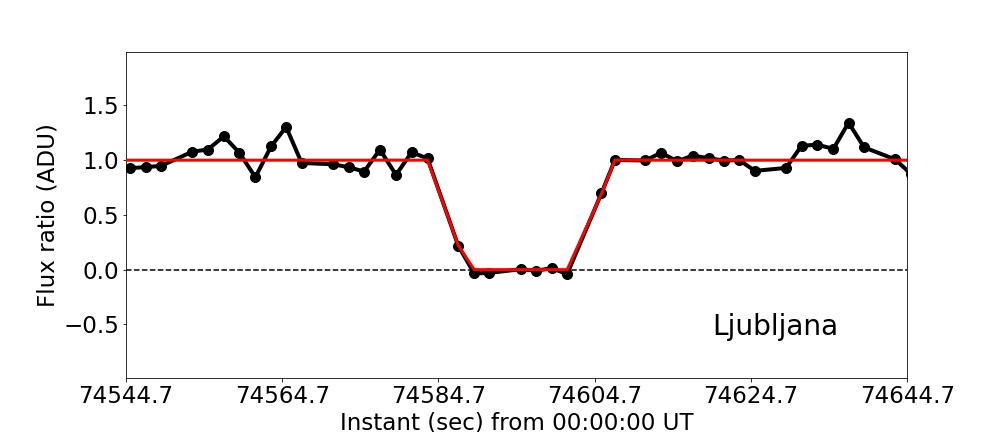}\quad   
    \includegraphics[width=0.48\linewidth]{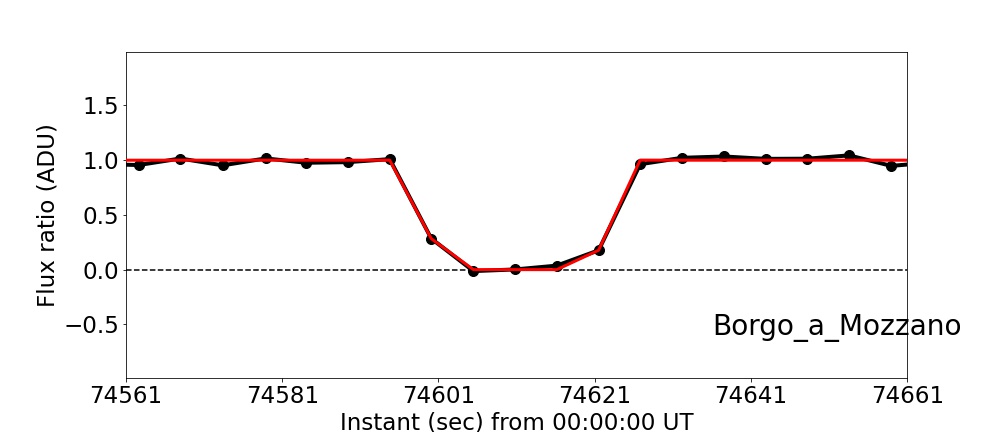}\quad  
    \includegraphics[width=0.48\linewidth]{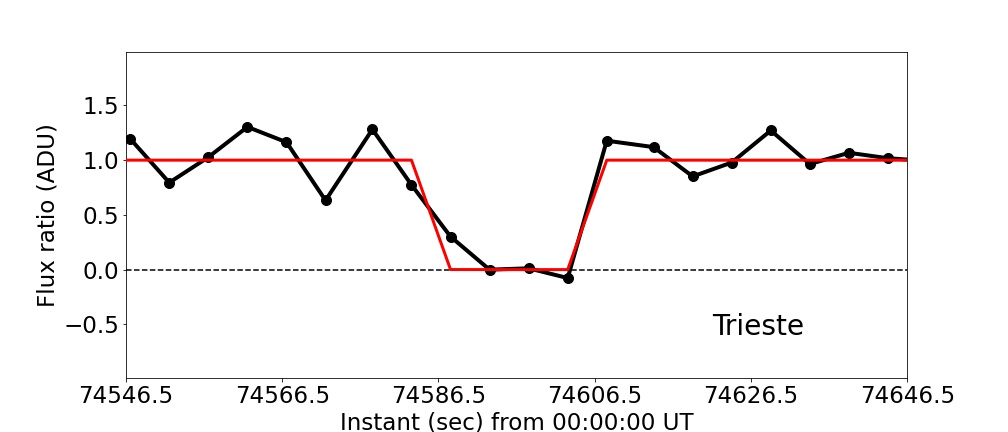}\quad    
    \includegraphics[width=0.48\linewidth]{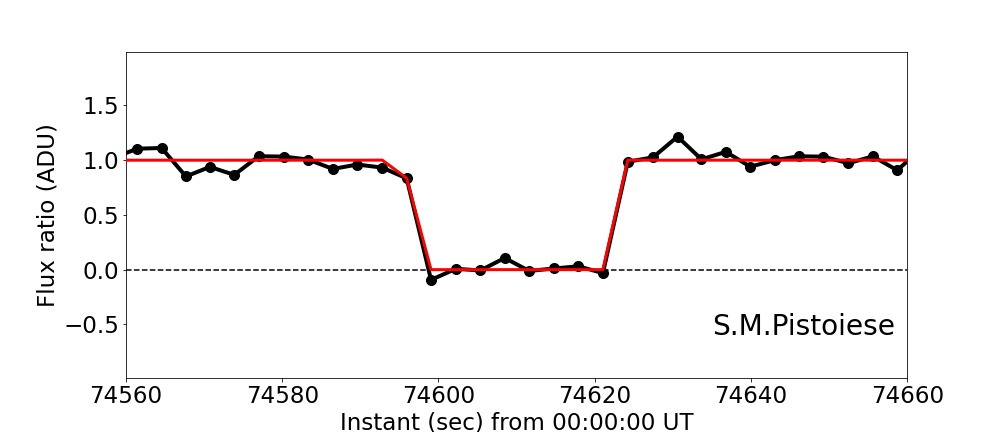}\quad  
    \includegraphics[width=0.48\linewidth]{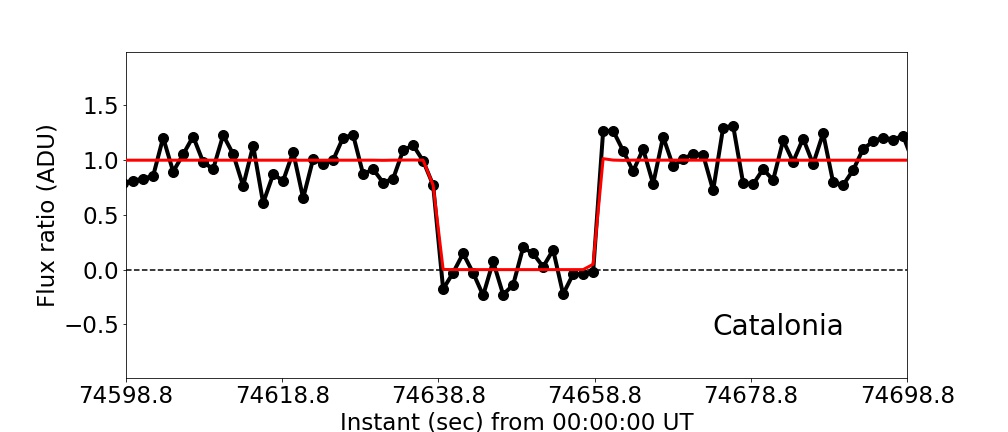}\quad  
    \includegraphics[width=0.48\linewidth]{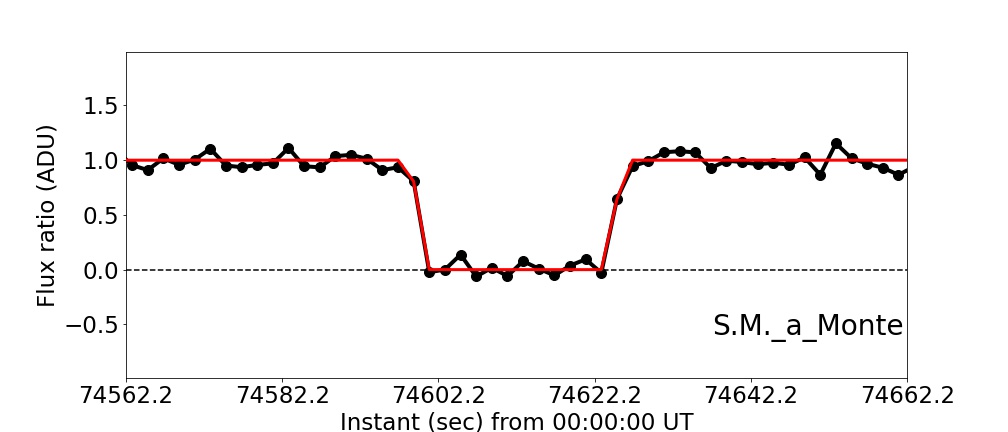}\quad  
    \includegraphics[width=0.48\linewidth]{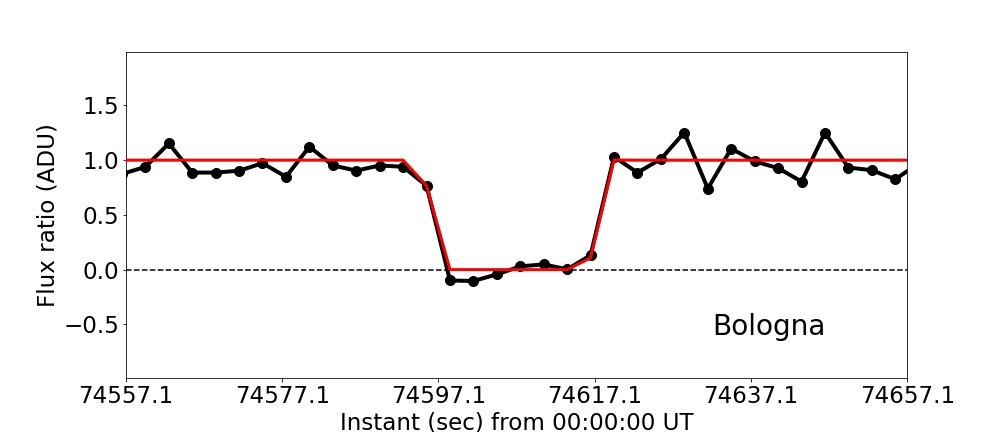}\quad  
    \includegraphics[width=0.48\linewidth]{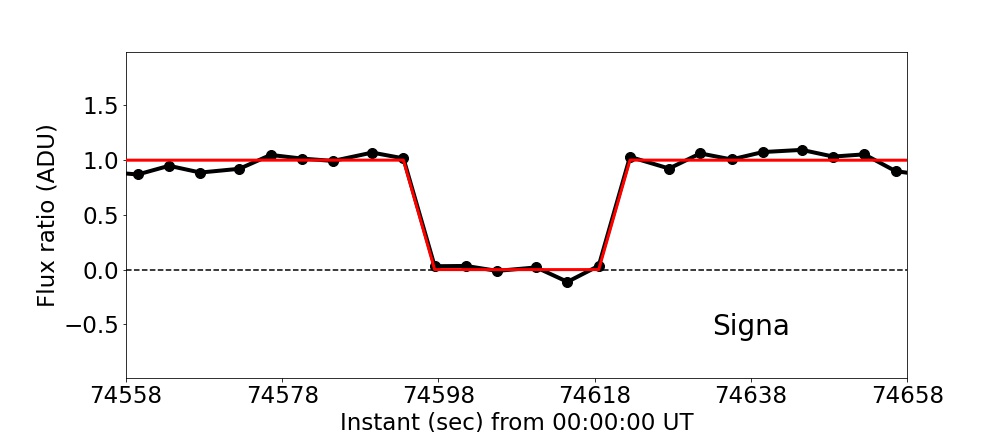}\quad  
    \includegraphics[width=0.48\linewidth]{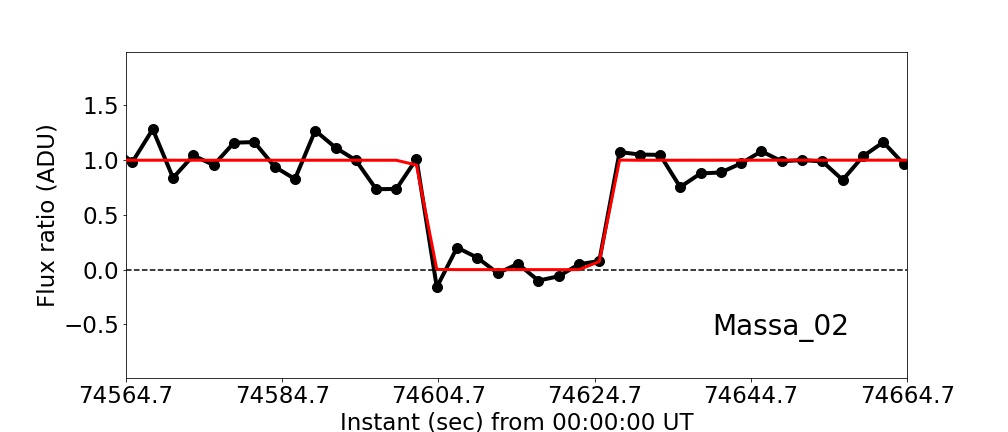}\quad    
    \includegraphics[width=0.48\linewidth]{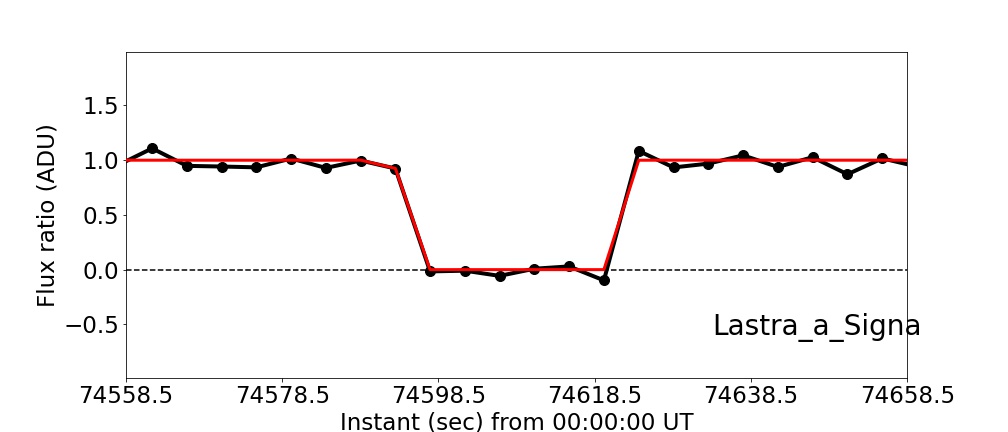}\quad    
    \includegraphics[width=0.48\linewidth]{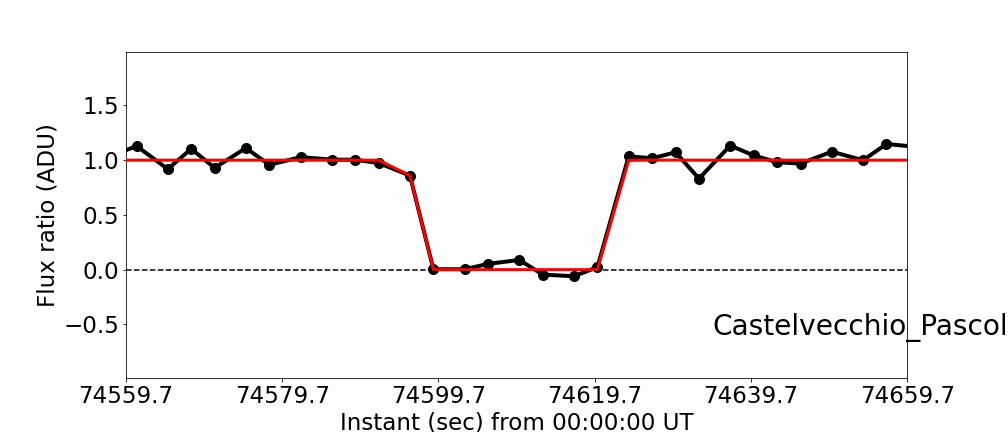}\quad    
    \includegraphics[width=0.48\linewidth]{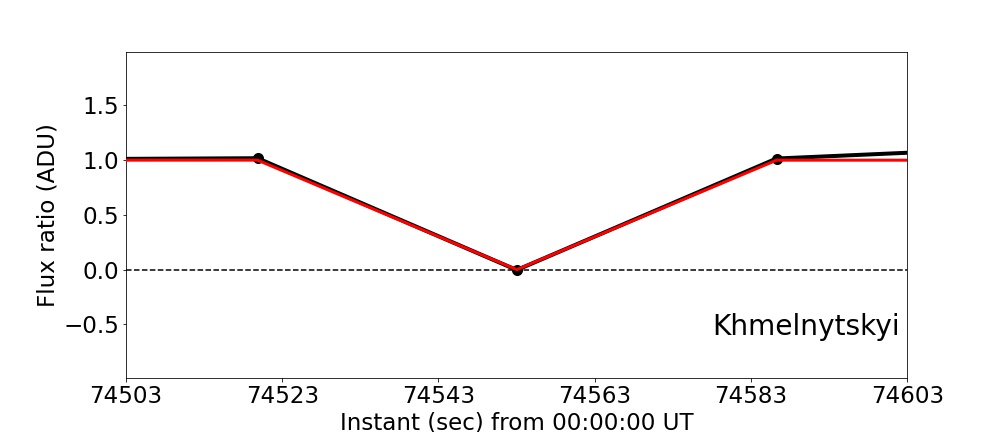}   
    \caption{Continued.}
    \label{fig:b2}
\end{figure}

\begin{figure}[!h]
    \centering
    \includegraphics[width=0.48\linewidth]{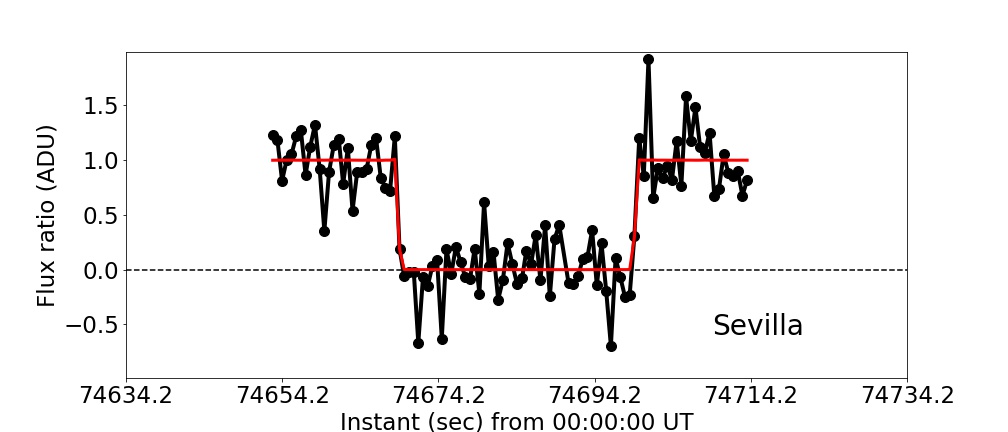}\quad   
    \includegraphics[width=0.48\linewidth]{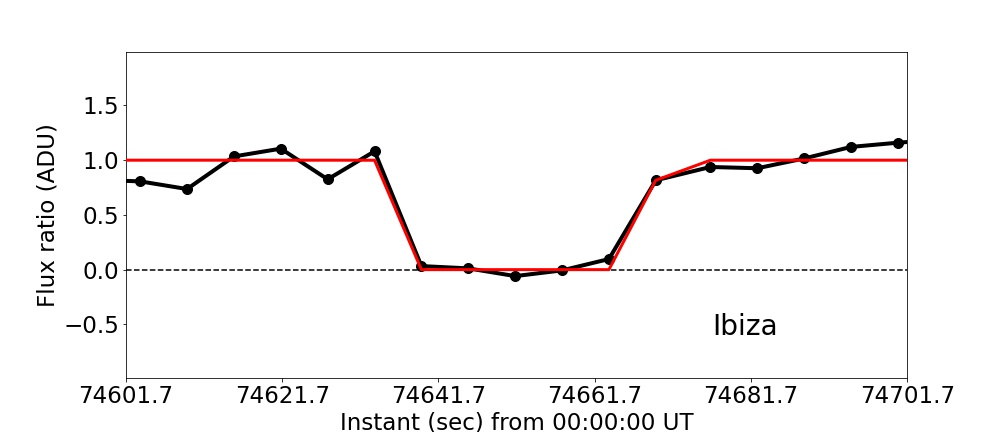}\quad  
    \includegraphics[width=0.48\linewidth]{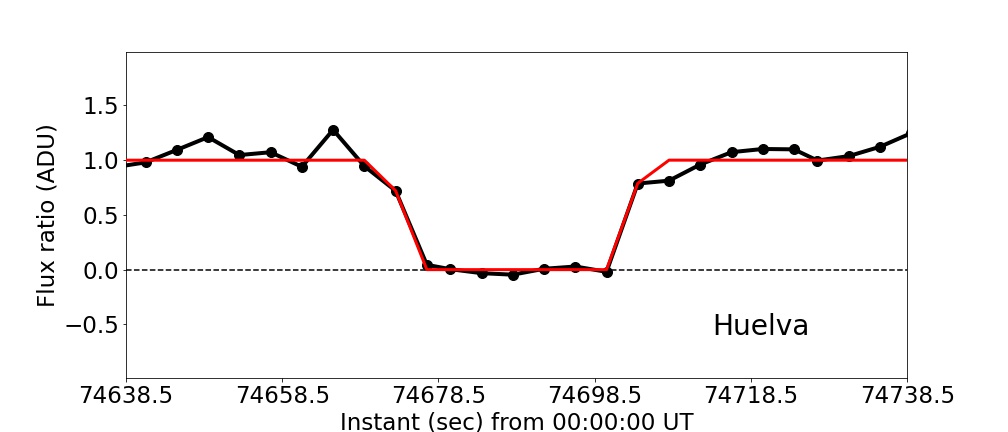}\quad    
    \includegraphics[width=0.48\linewidth]{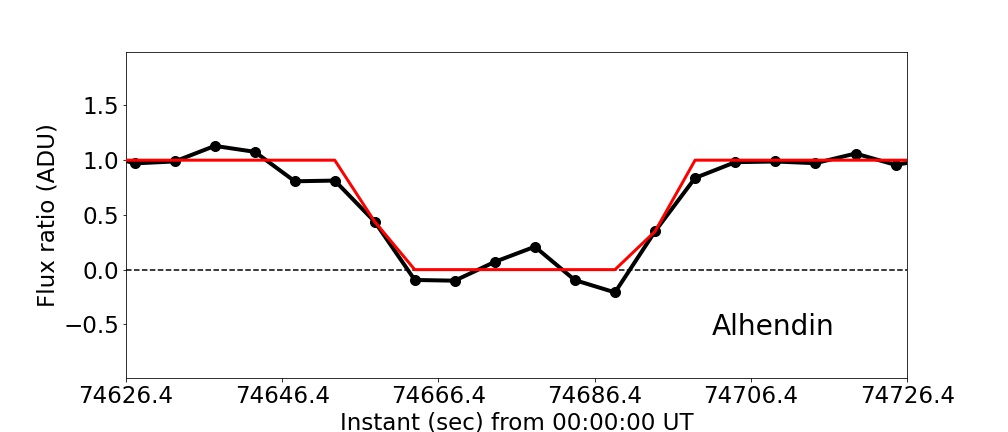}\quad  
    \includegraphics[width=0.48\linewidth]{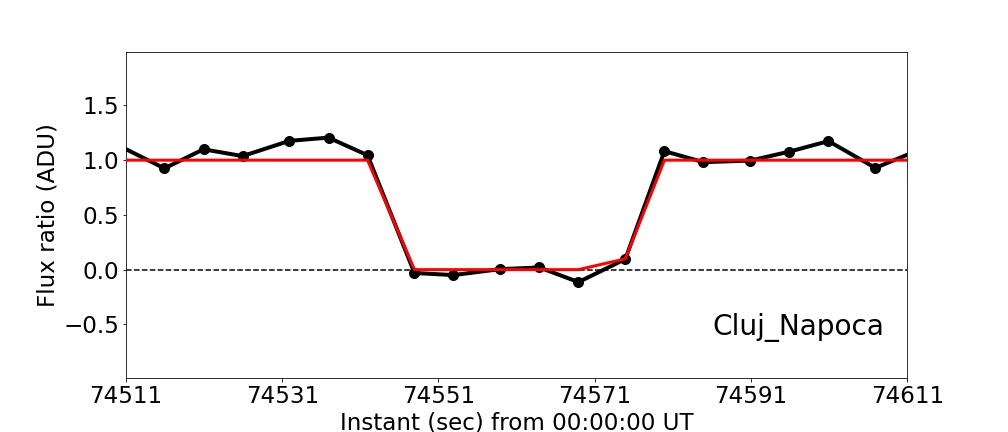}\quad  
    \includegraphics[width=0.48\linewidth]{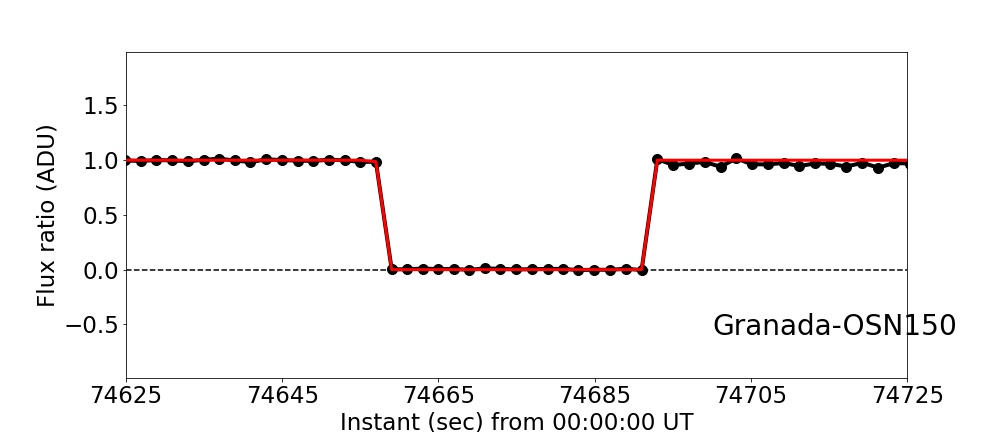}\quad  
    \includegraphics[width=0.48\linewidth]{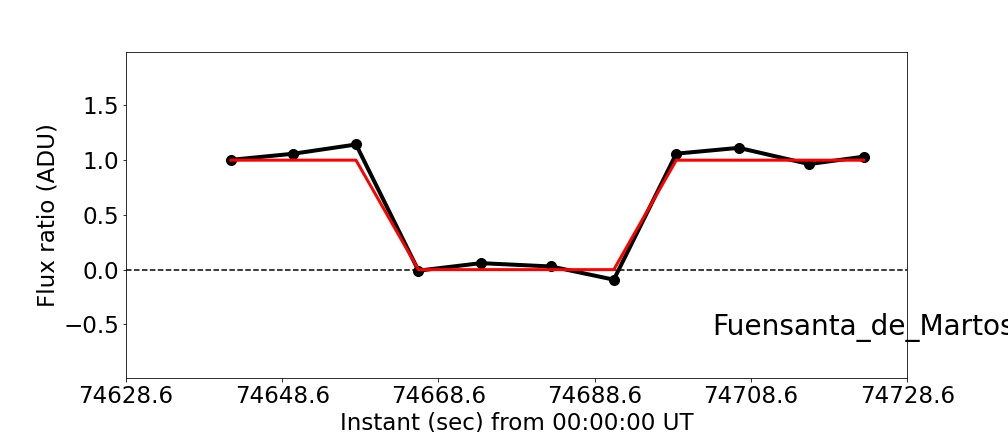}\quad  
    \includegraphics[width=0.48\linewidth]{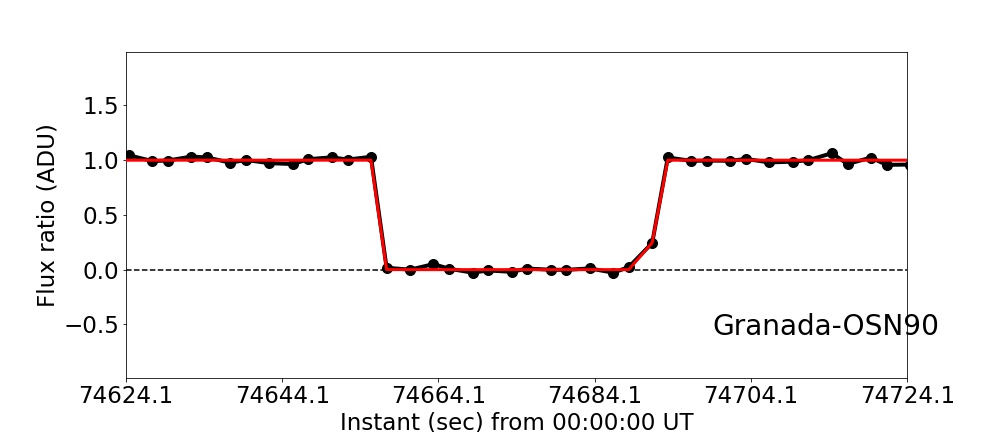}\quad  
    \includegraphics[width=0.48\linewidth]{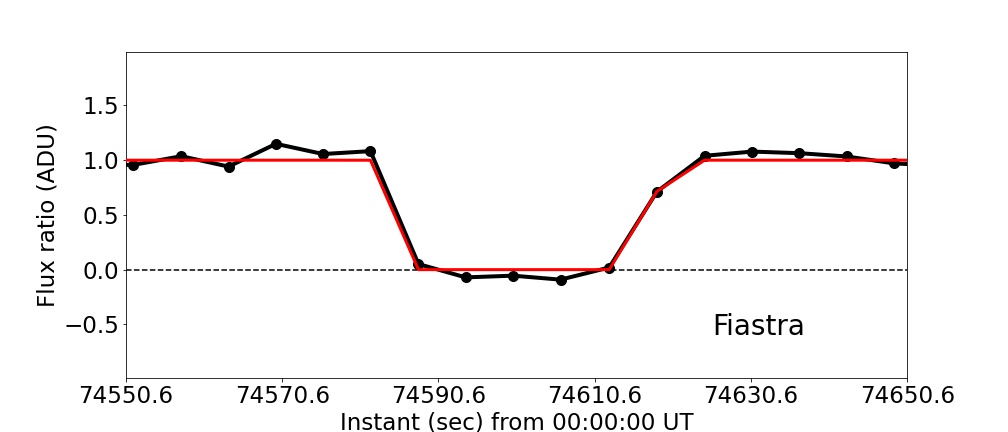}\quad    
    \includegraphics[width=0.48\linewidth]{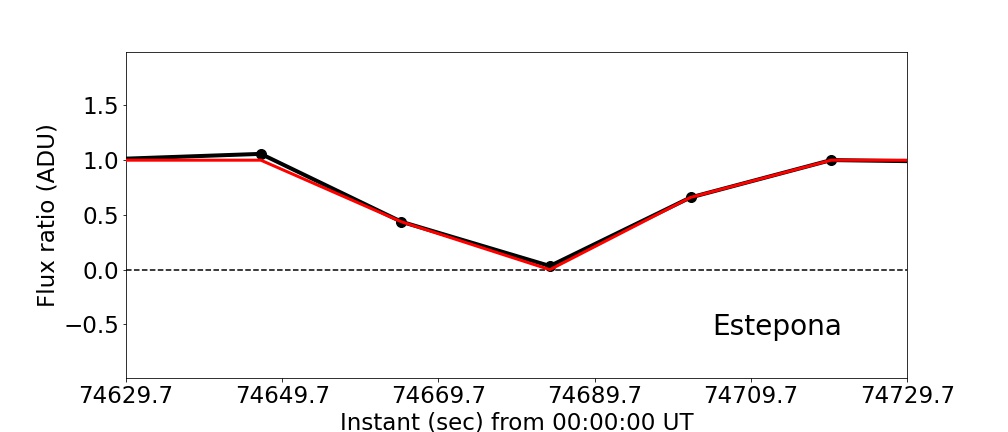}\quad    
    \includegraphics[width=0.48\linewidth]{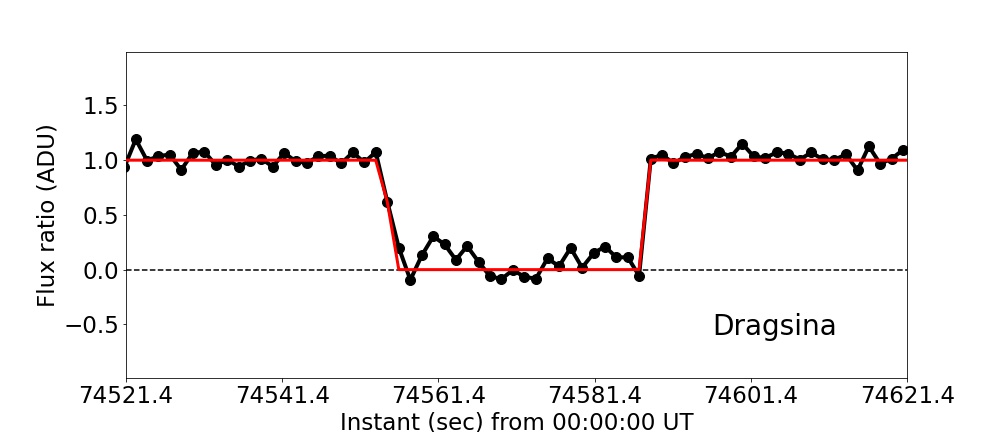}\quad    
    \includegraphics[width=0.48\linewidth]{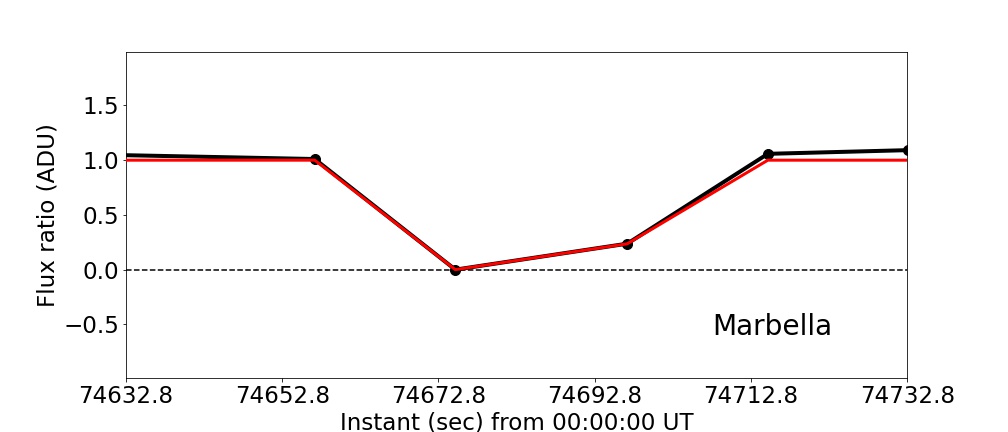}   
    \caption{Continued.}
    \label{fig:b3}
\end{figure}

\begin{figure}[!h]
    \centering
    \includegraphics[width=0.48\linewidth]{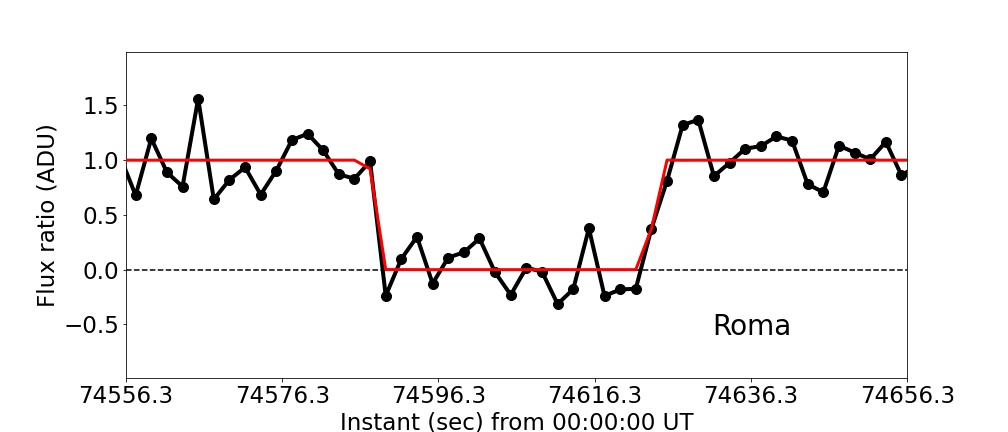}\quad   
    \includegraphics[width=0.48\linewidth]{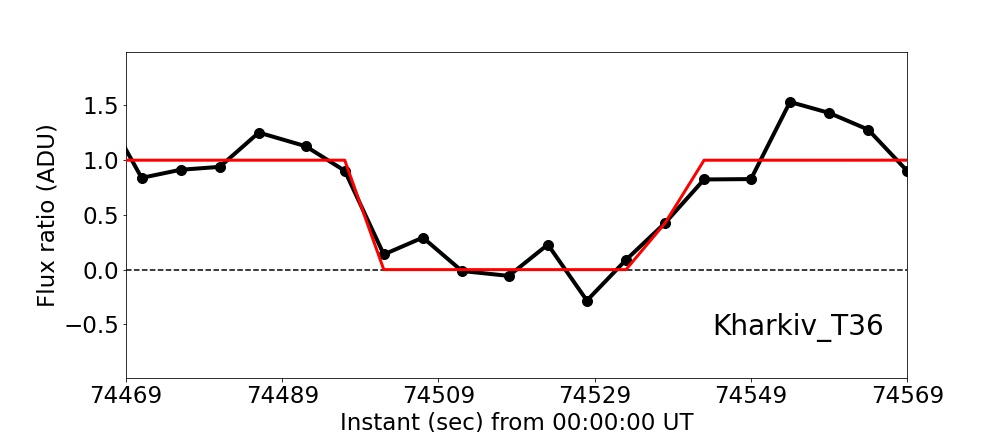}\quad  
    \includegraphics[width=0.48\linewidth]{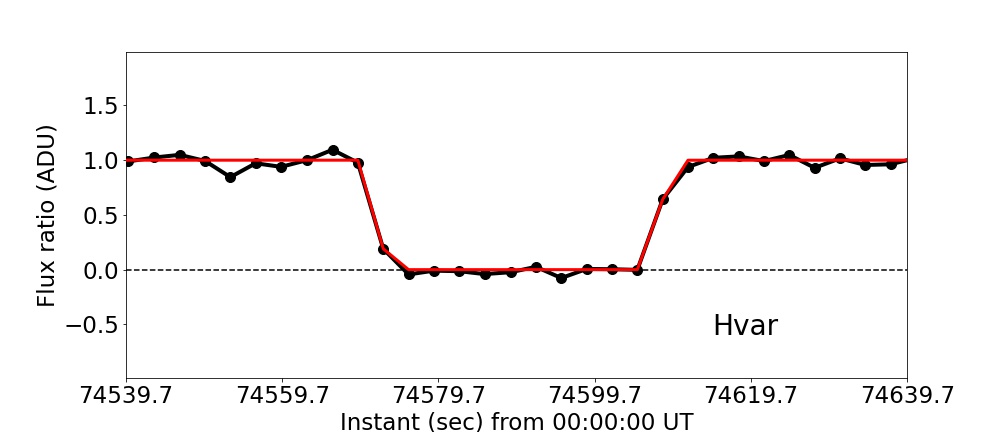}\quad    
    \includegraphics[width=0.48\linewidth]{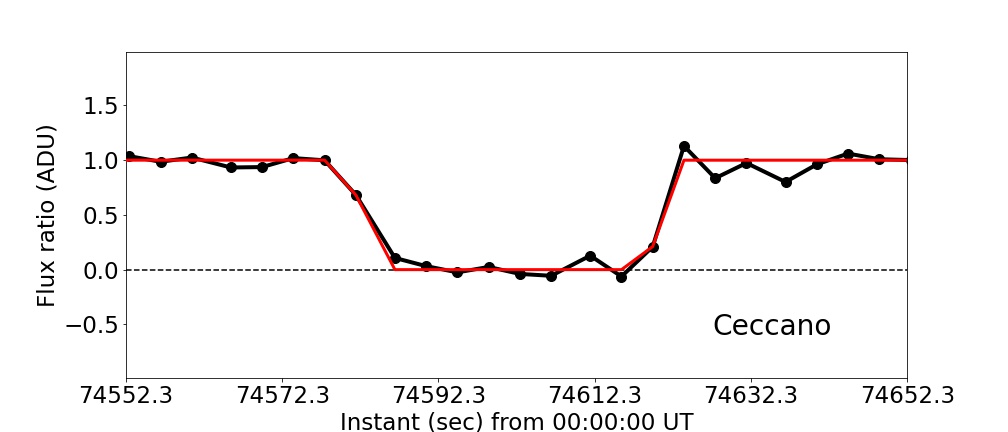}\quad  
    \includegraphics[width=0.48\linewidth]{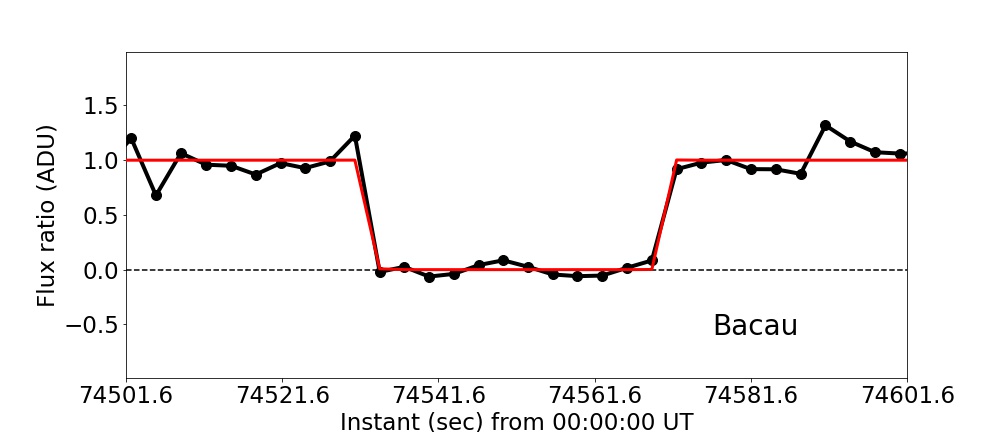}\quad  
    \includegraphics[width=0.48\linewidth]{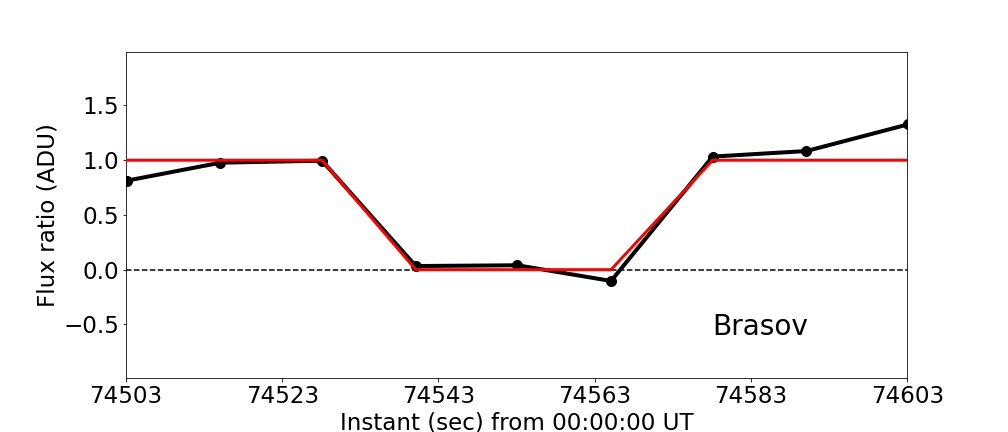}\quad  
    \includegraphics[width=0.48\linewidth]{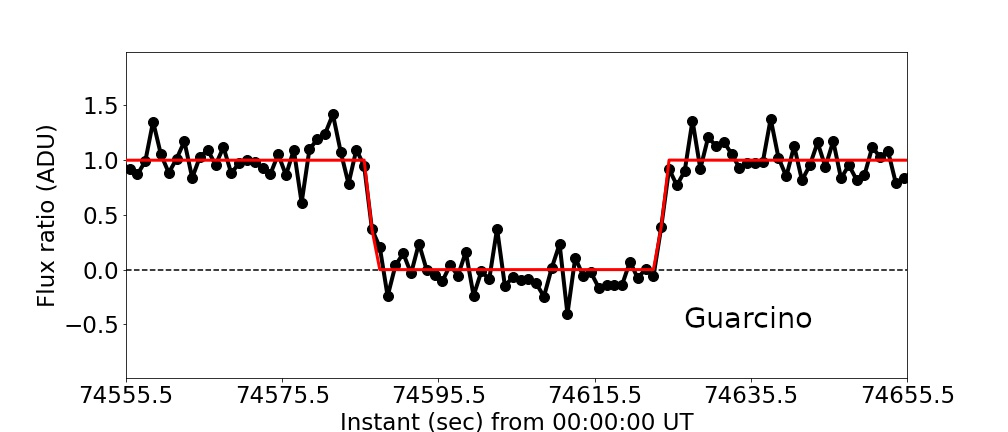}\quad  
    \includegraphics[width=0.48\linewidth]{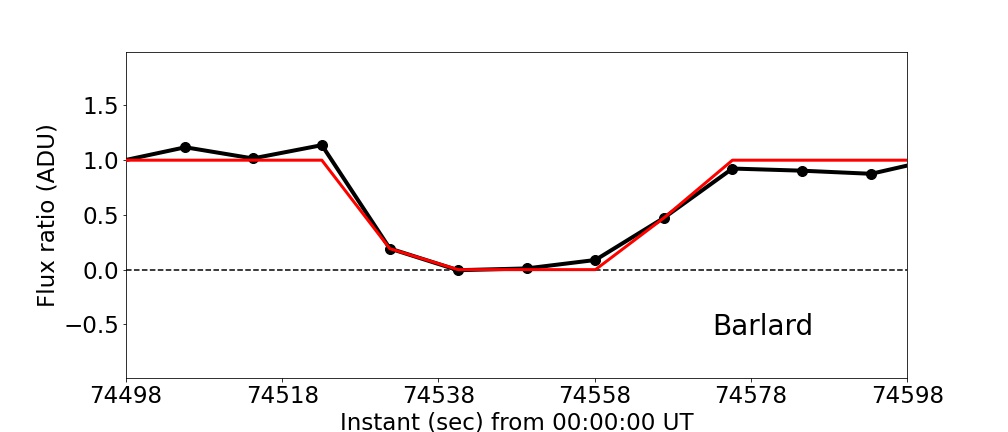}\quad  
    \includegraphics[width=0.48\linewidth]{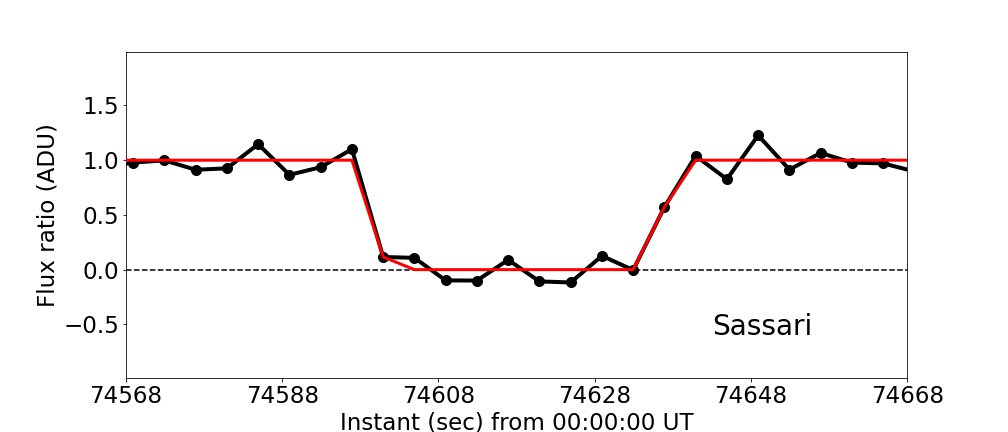}\quad    
    \includegraphics[width=0.48\linewidth]{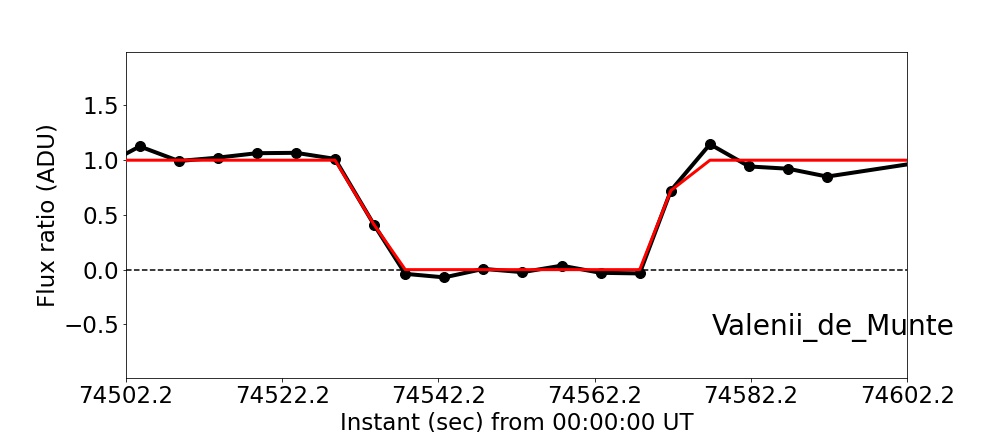}\quad    
    \includegraphics[width=0.48\linewidth]{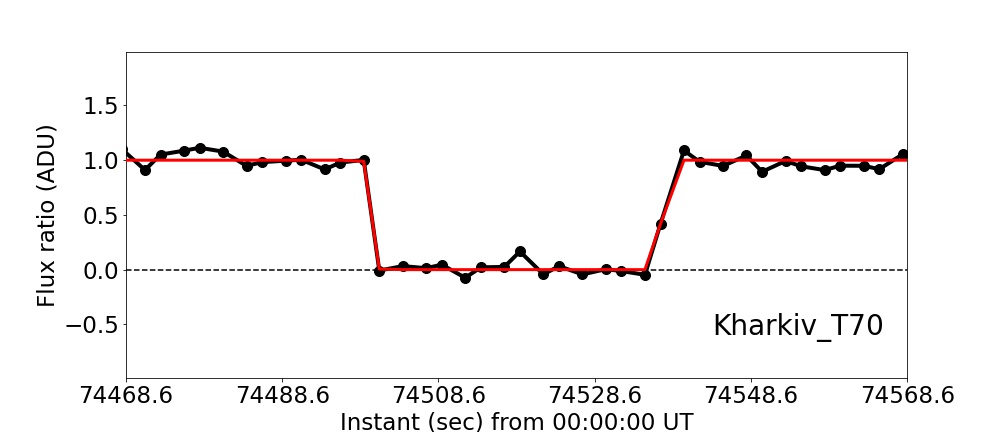}\quad    
    \includegraphics[width=0.48\linewidth]{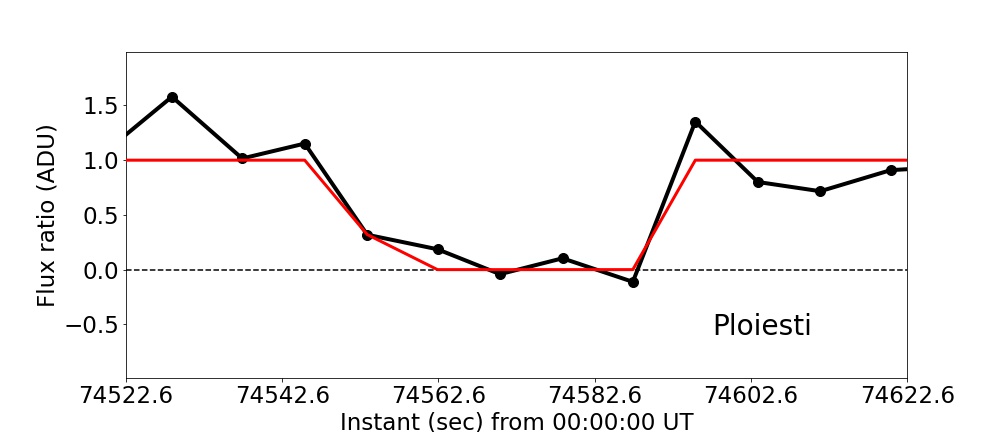}   
    \caption{Continued.}
    \label{fig:b4}
\end{figure}

\begin{figure}[!h]
    \centering
    \includegraphics[width=0.48\linewidth]{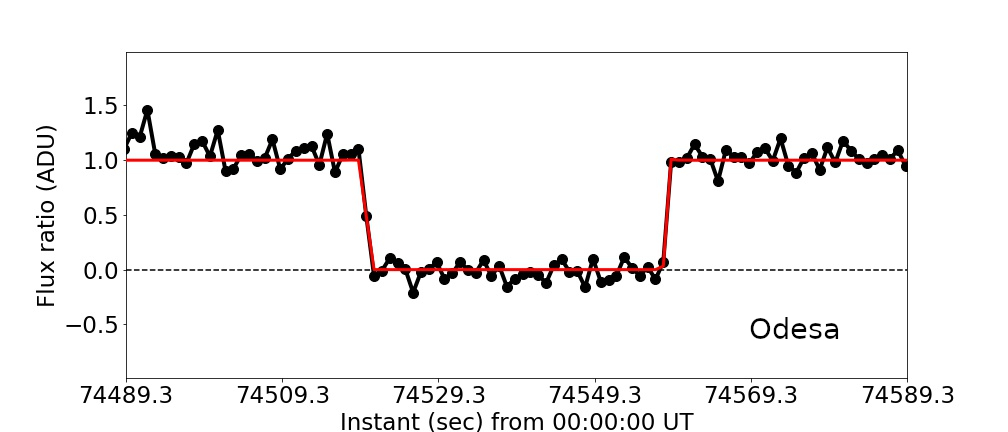}\quad   
    \includegraphics[width=0.48\linewidth]{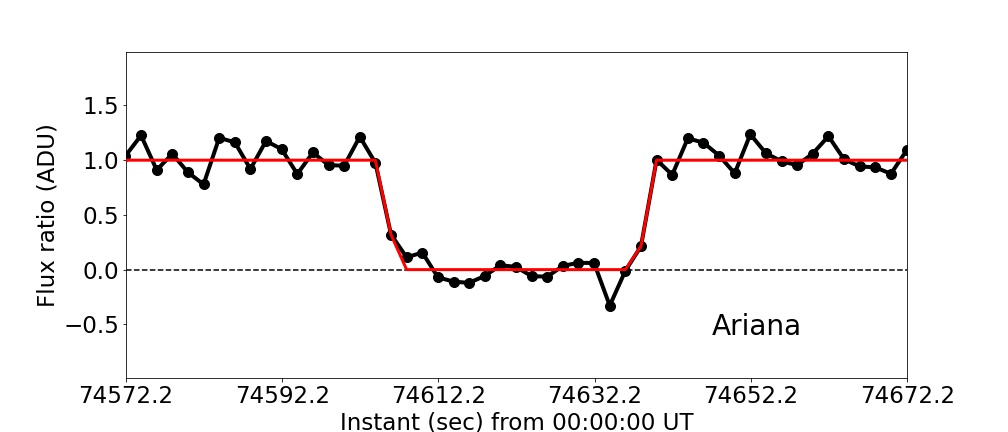}\quad  
    \includegraphics[width=0.48\linewidth]{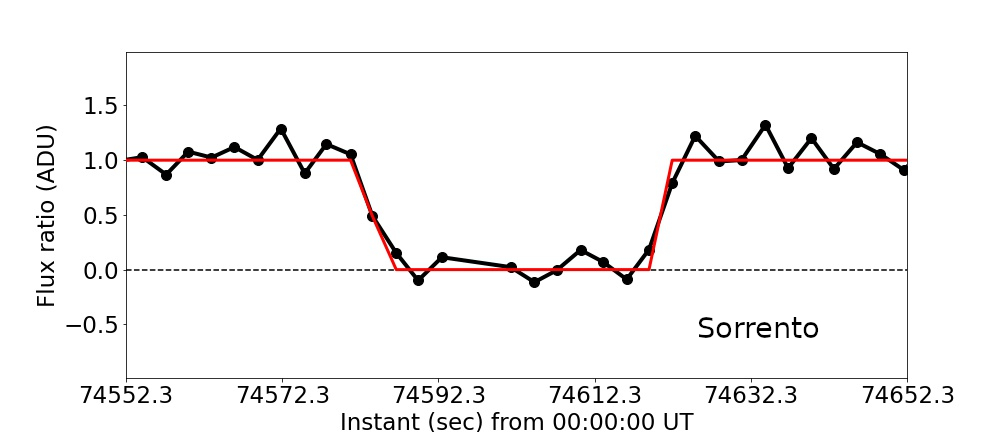}\quad    
    \includegraphics[width=0.48\linewidth]{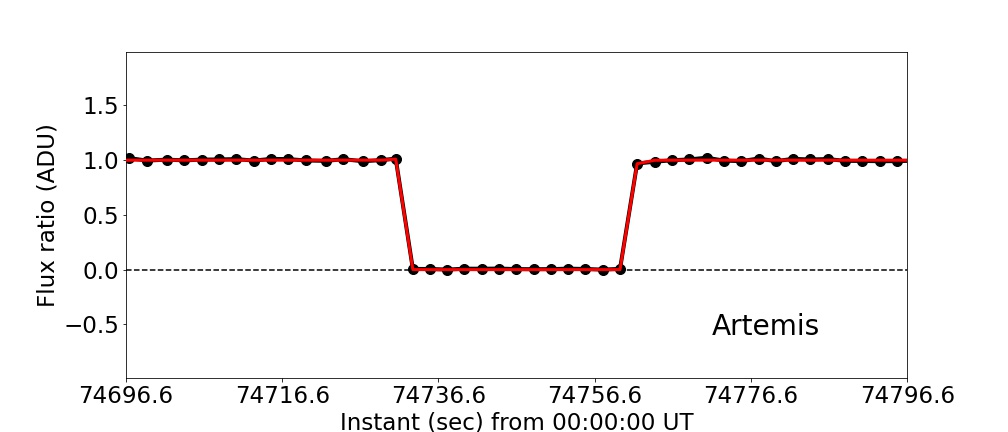}\quad  
    \includegraphics[width=0.48\linewidth]{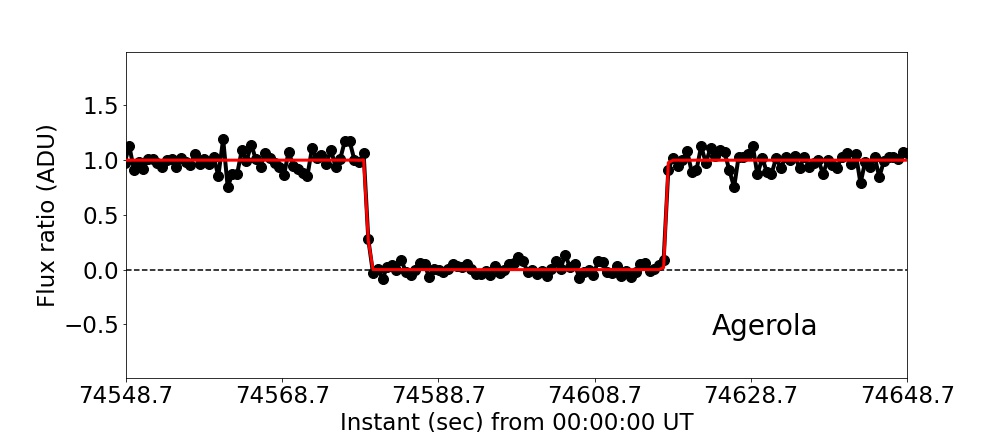}\quad  
    \includegraphics[width=0.48\linewidth]{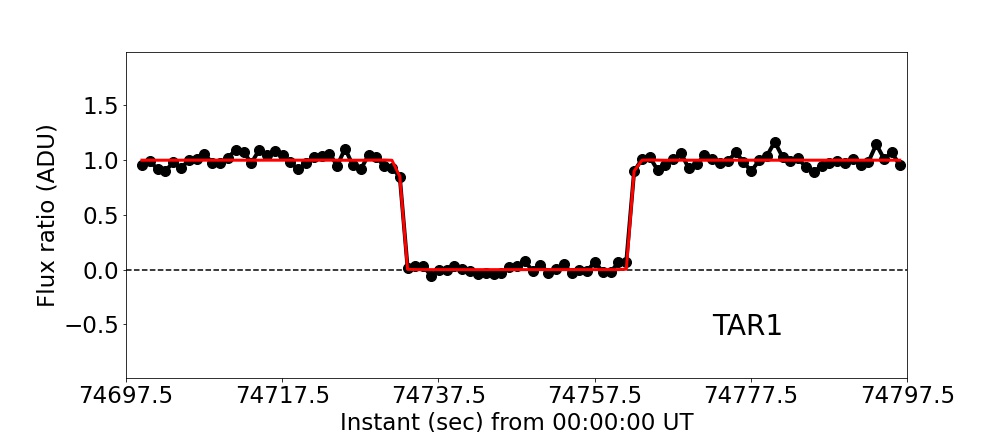}\quad  
    \includegraphics[width=0.48\linewidth]{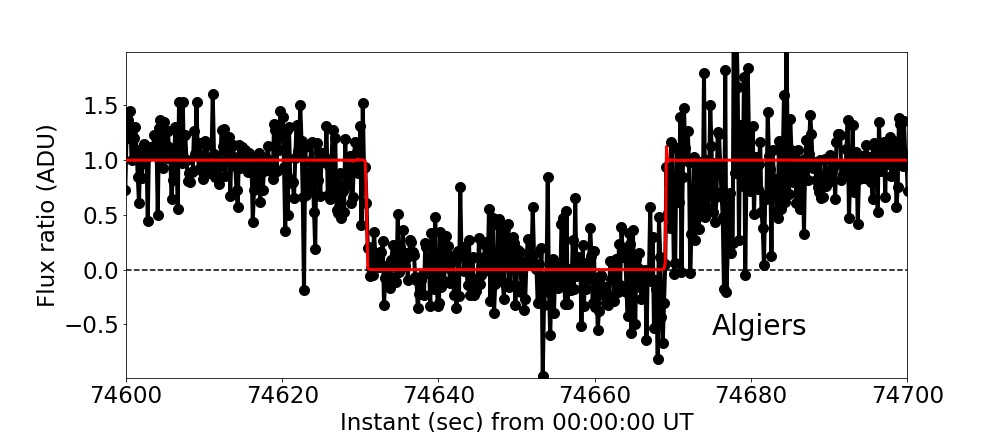}\quad  
    \includegraphics[width=0.48\linewidth]{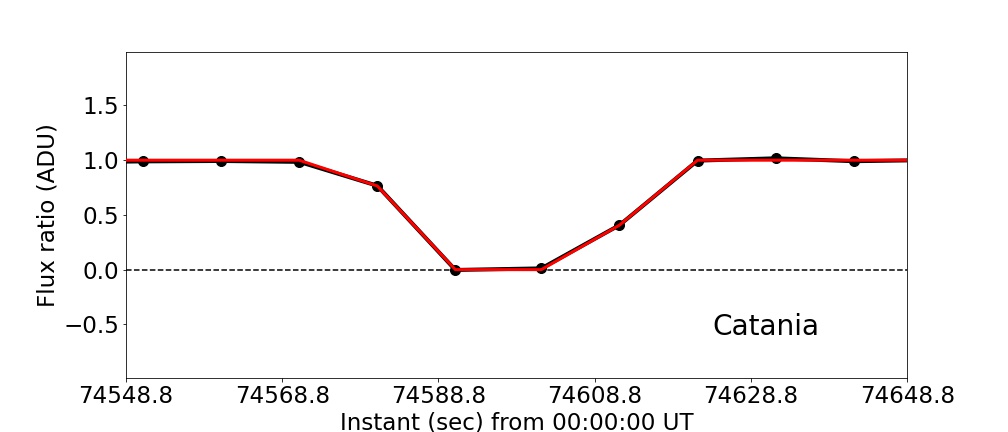}\quad  
    \includegraphics[width=0.48\linewidth]{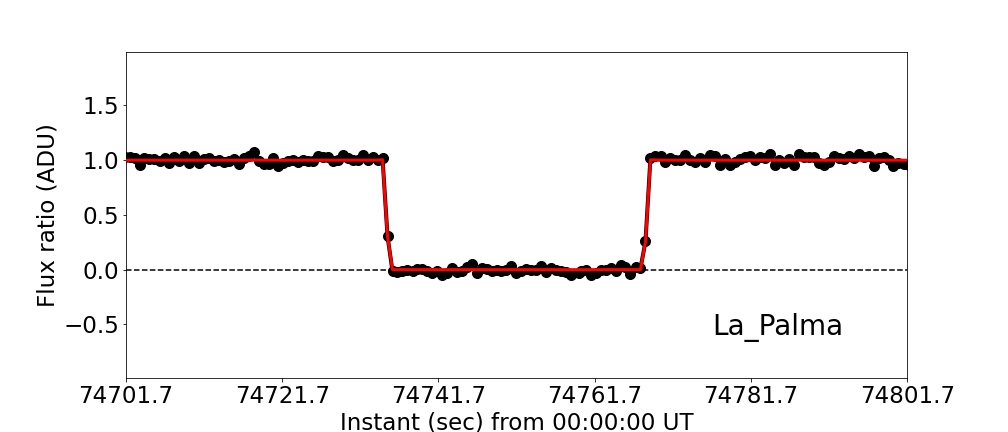}\quad    
    \includegraphics[width=0.48\linewidth]{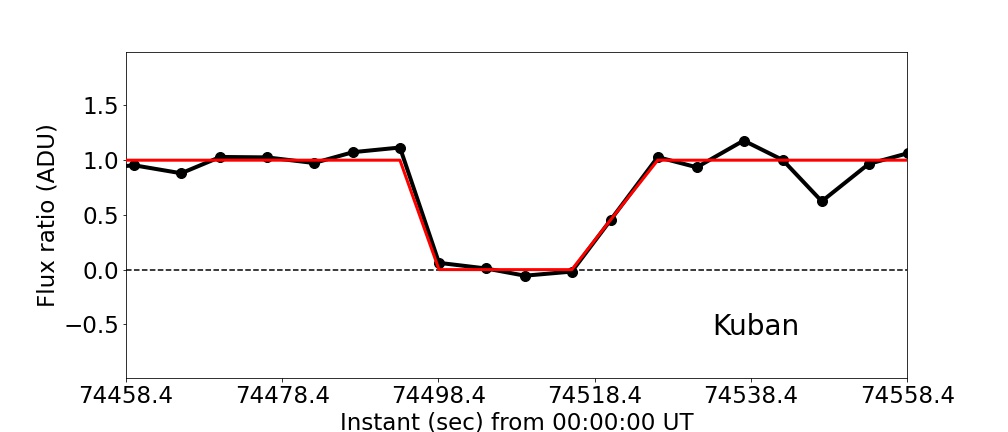}\quad    
    \includegraphics[width=0.48\linewidth]{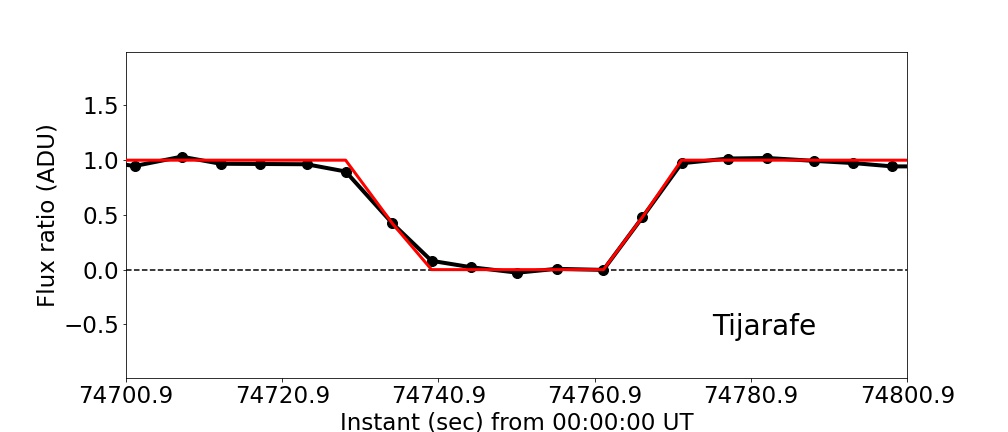}\quad    
    \includegraphics[width=0.48\linewidth]{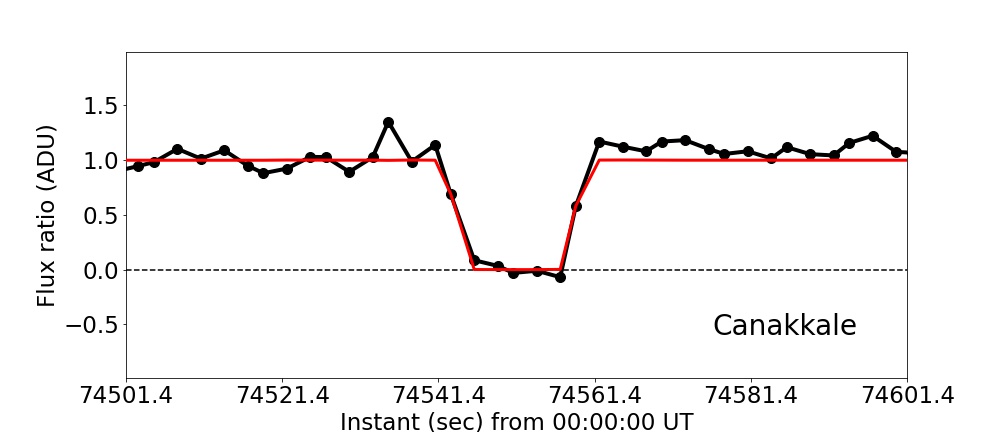}   
    \caption{Continued.}
    \label{fig:b5}
\end{figure}

\onecolumn

\begin{figure}[!h]
    \centering
    \includegraphics[width=0.48\linewidth]{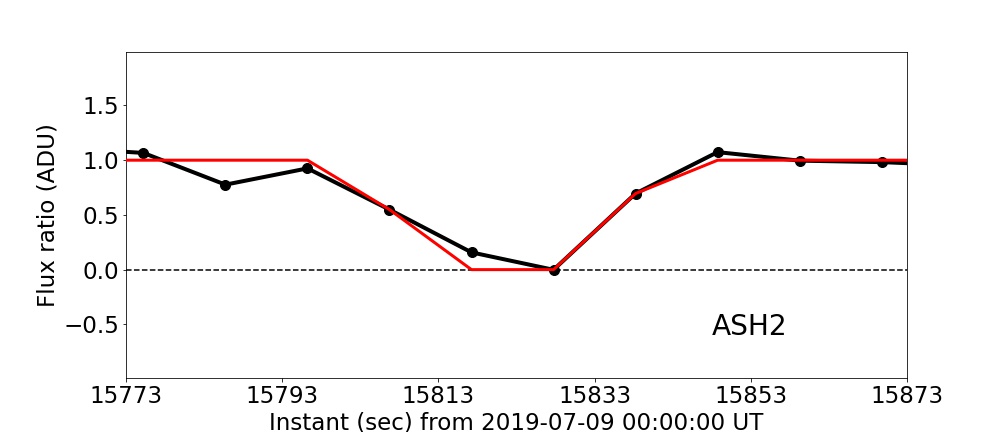}\quad   
    \includegraphics[width=0.48\linewidth]{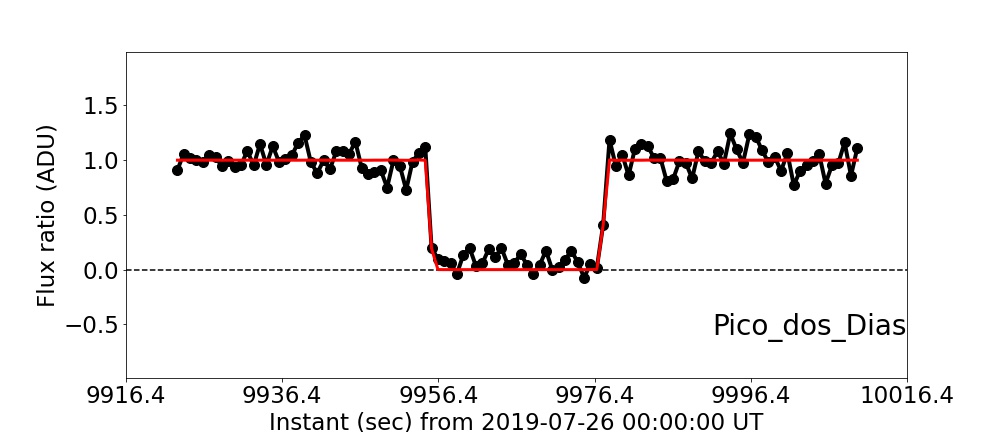}\quad  
    \includegraphics[width=0.48\linewidth]{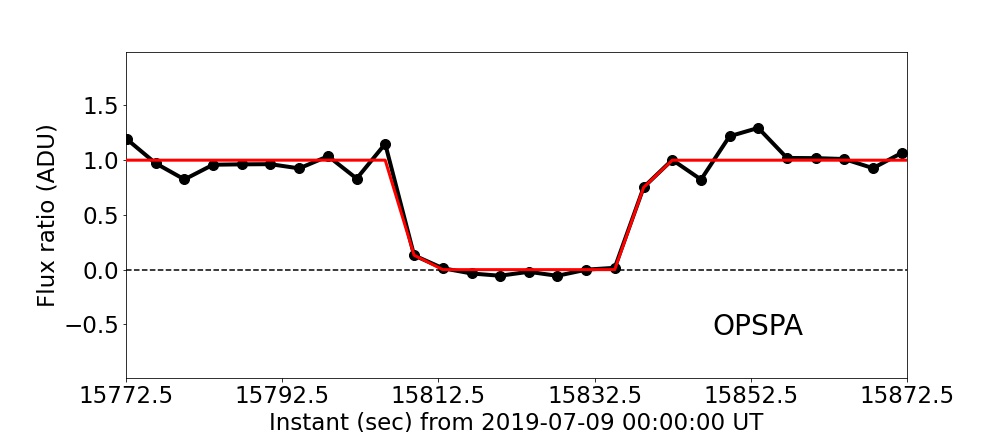}\quad    
    \includegraphics[width=0.48\linewidth]{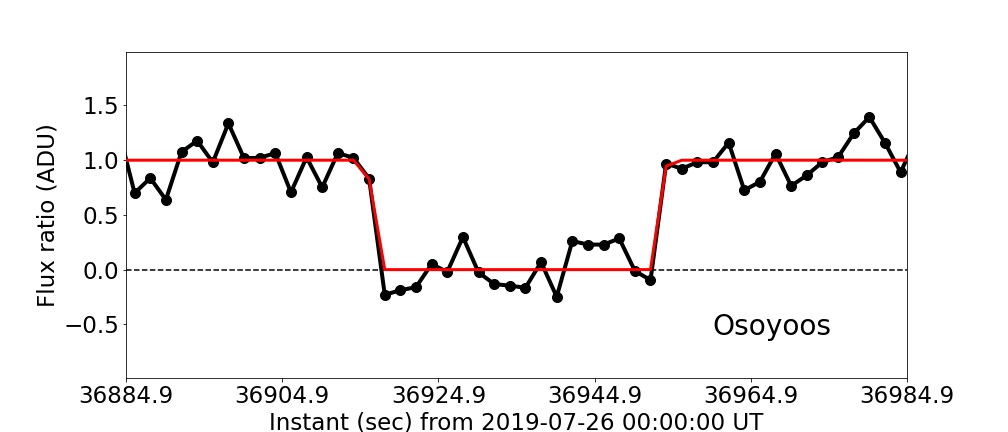}\quad  
    \includegraphics[width=0.48\linewidth]{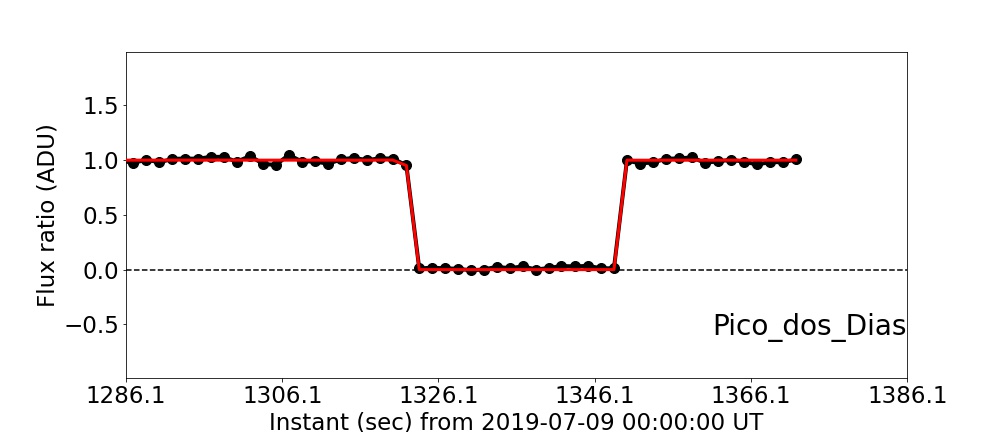}\quad  
    \includegraphics[width=0.48\linewidth]{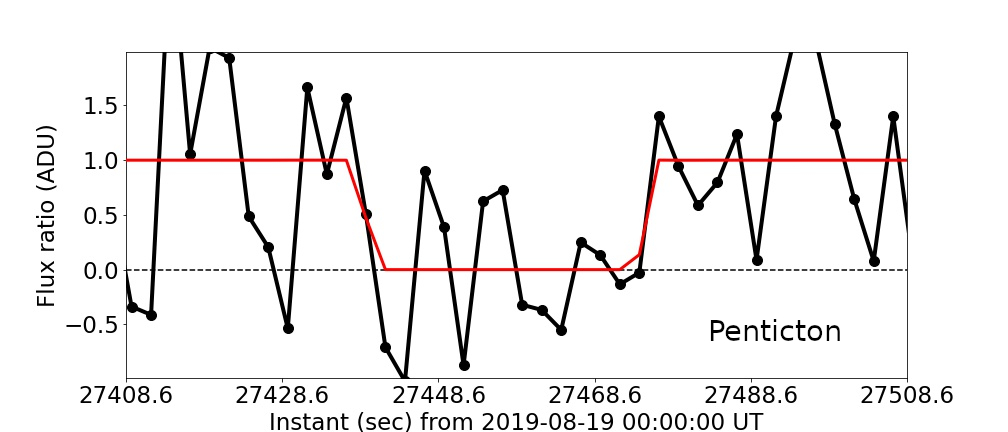}\quad  
    \includegraphics[width=0.48\linewidth]{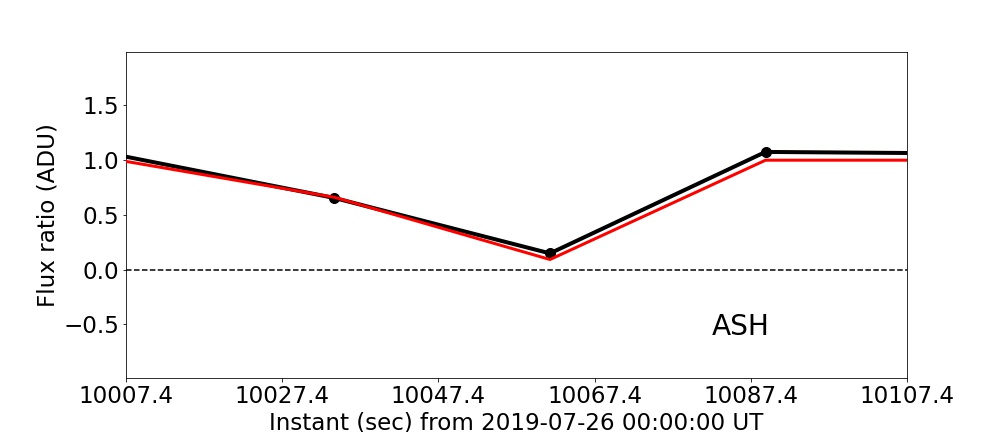}\quad  
    \includegraphics[width=0.48\linewidth]{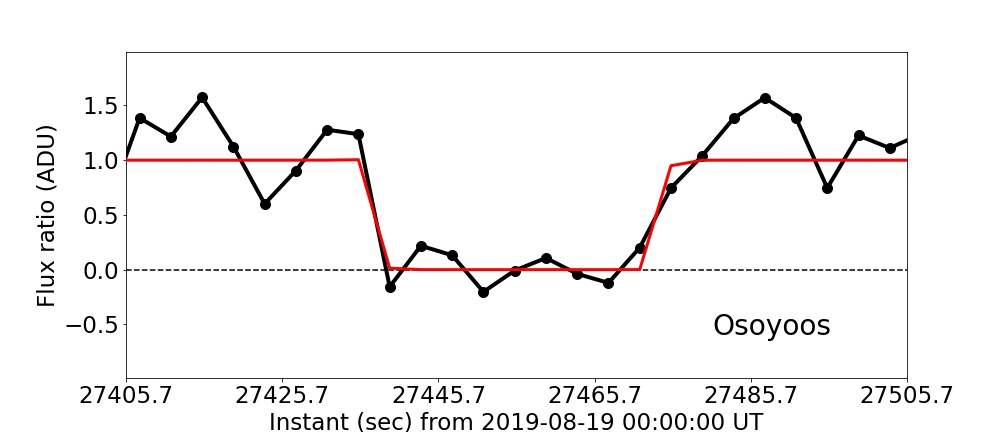}\quad  
    \includegraphics[width=0.48\linewidth]{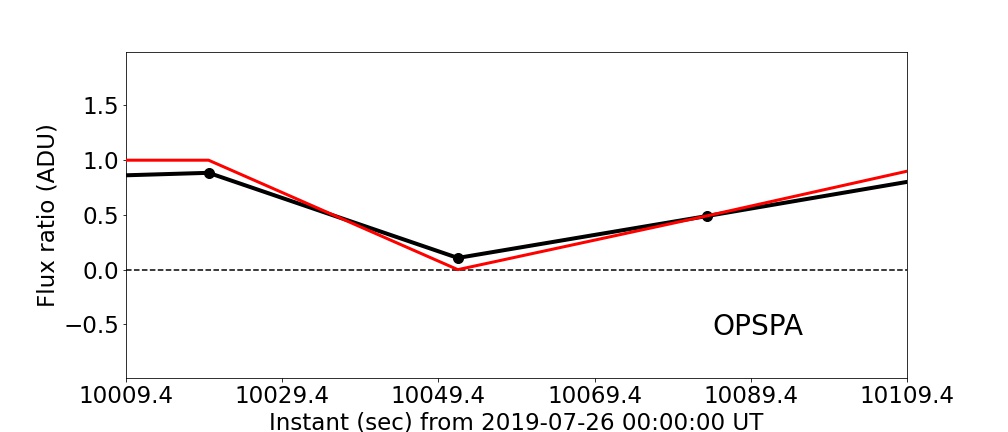}\quad    
    \includegraphics[width=0.48\linewidth]{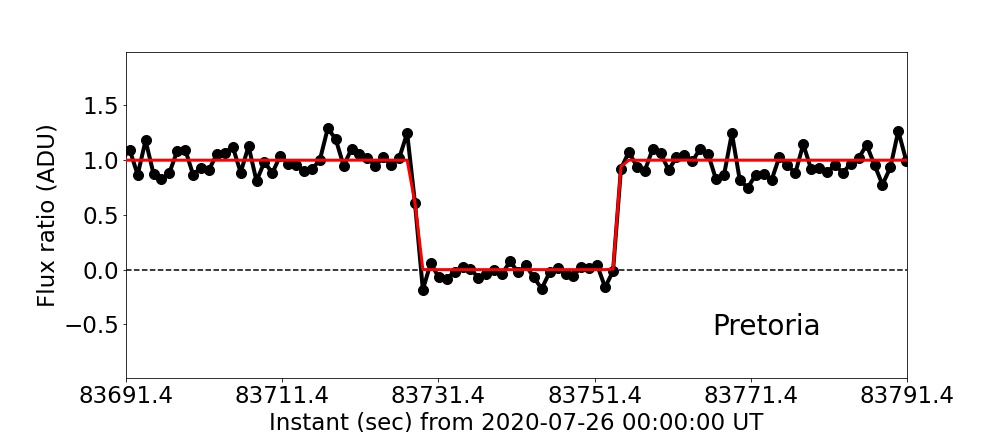}\quad    
    \includegraphics[width=0.48\linewidth]{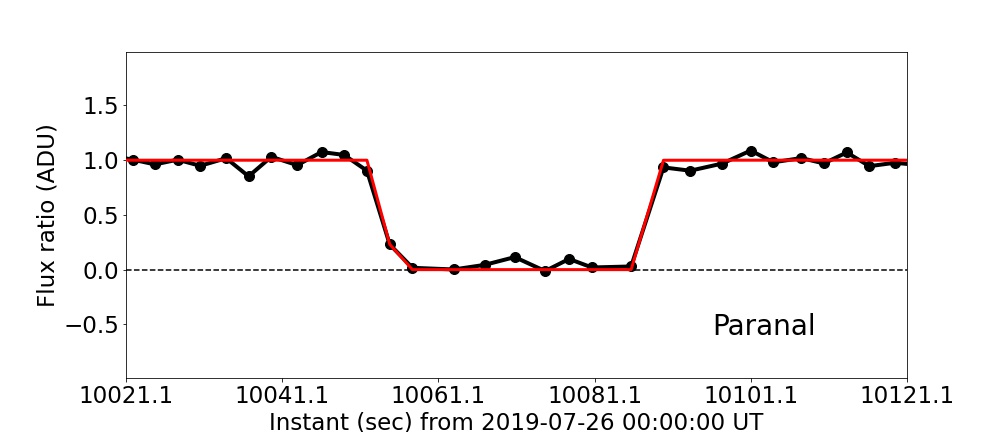}\quad    
    \includegraphics[width=0.48\linewidth]{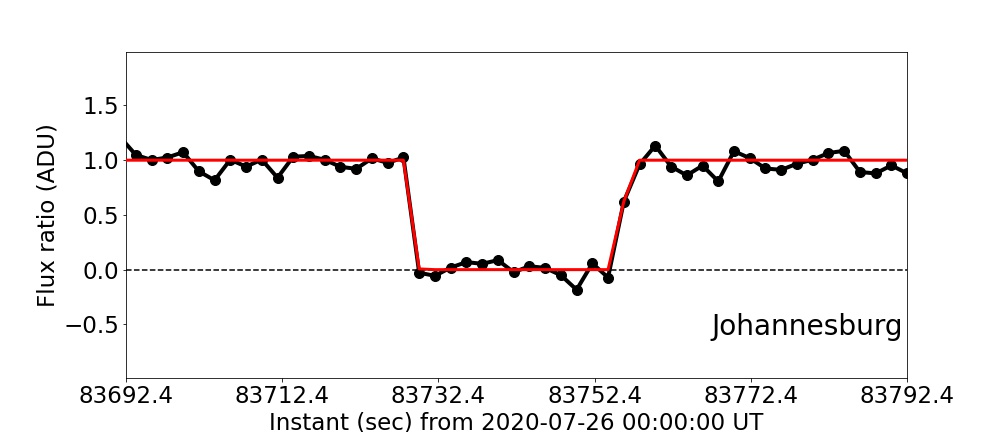}   
    \caption{Observed (black points) and calculated (red line) light curves for each site that observed a stellar occultation by 2002 MS$_4$, except the 8 August 2020 multichord event. See table \ref{table:other_pos_sites} for observational details.}
    \label{fig:b6}
\end{figure}

\begin{figure}[!h]
    \centering
    \includegraphics[width=0.48\linewidth]{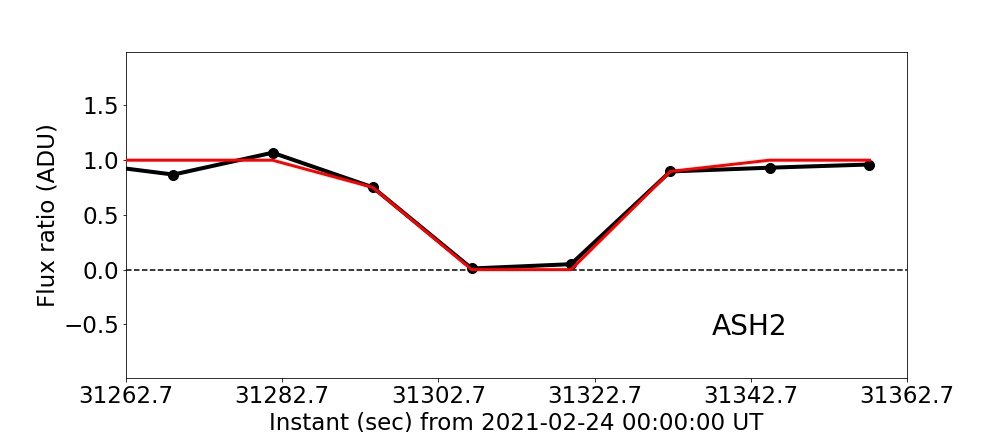}\quad   
    \includegraphics[width=0.48\linewidth]{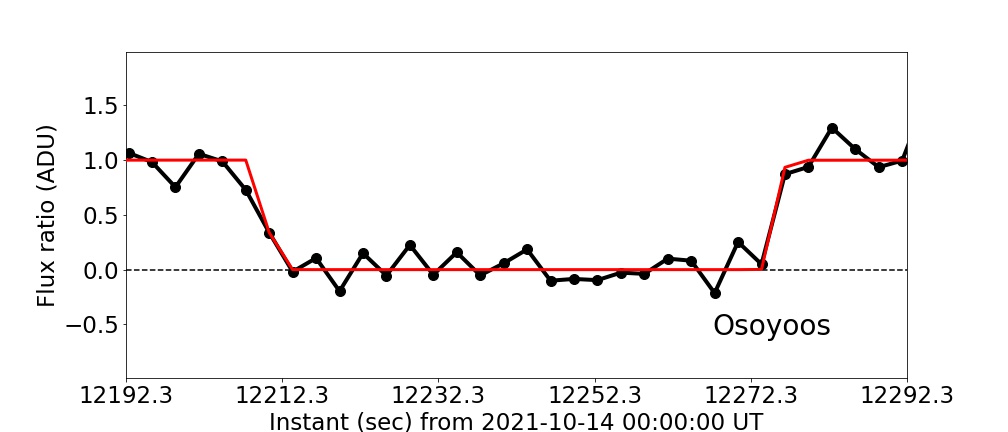}\quad  
    \includegraphics[width=0.48\linewidth]{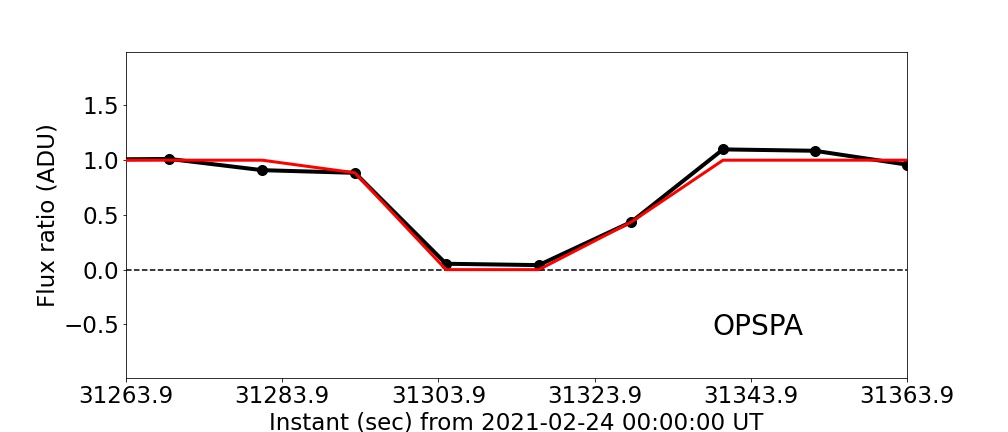}\quad   
    \includegraphics[width=0.48\linewidth]{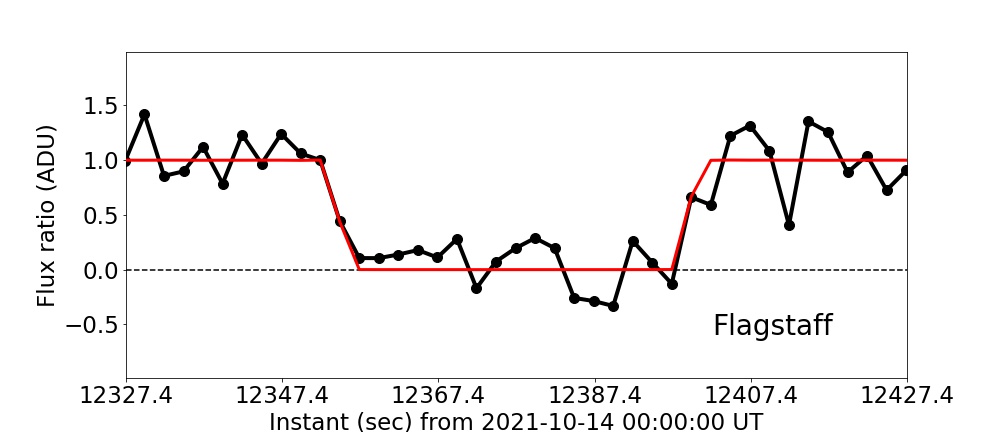}\quad   
    \includegraphics[width=0.48\linewidth]{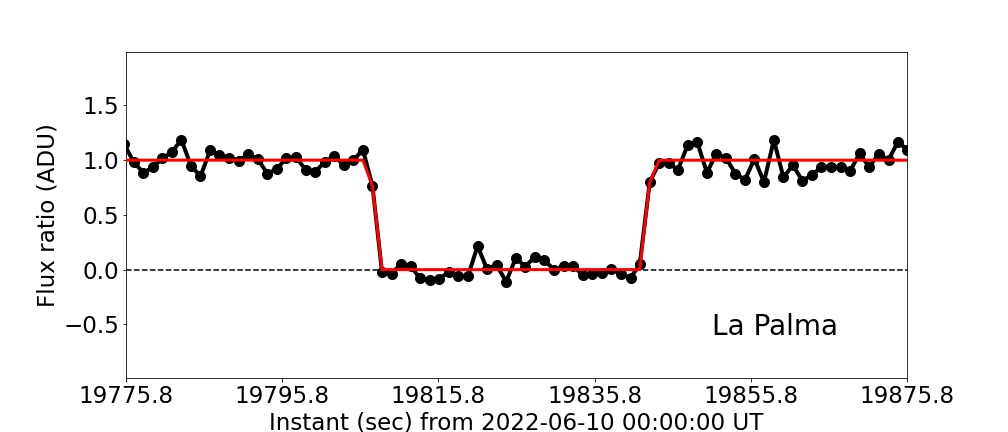}\quad  
    \includegraphics[width=0.48\linewidth]{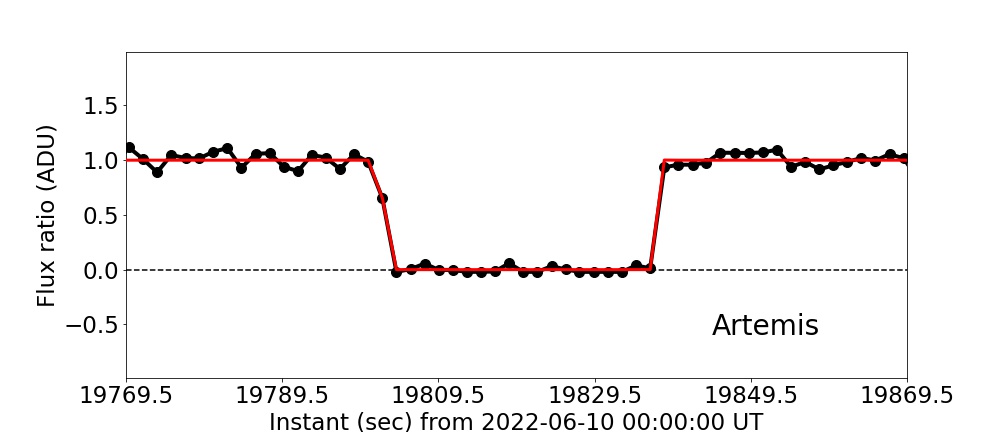}\quad   
    \includegraphics[width=0.48\linewidth]{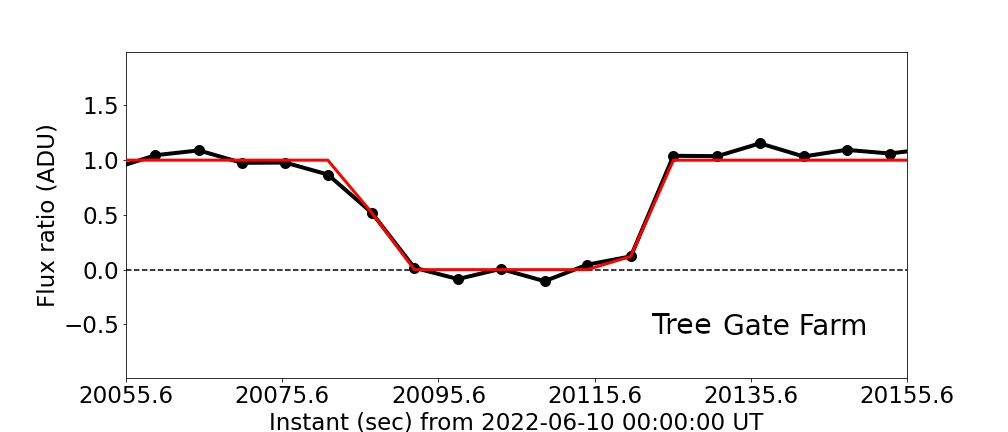}  
    
    \caption{Continued.}
    \label{fig:b7}
\end{figure}

\onecolumn
\newpage
\section{The unknown satellite hypothesis}
\label{appendix3}

\begin{figure}[!htb]
    \centering\includegraphics[width=\linewidth]{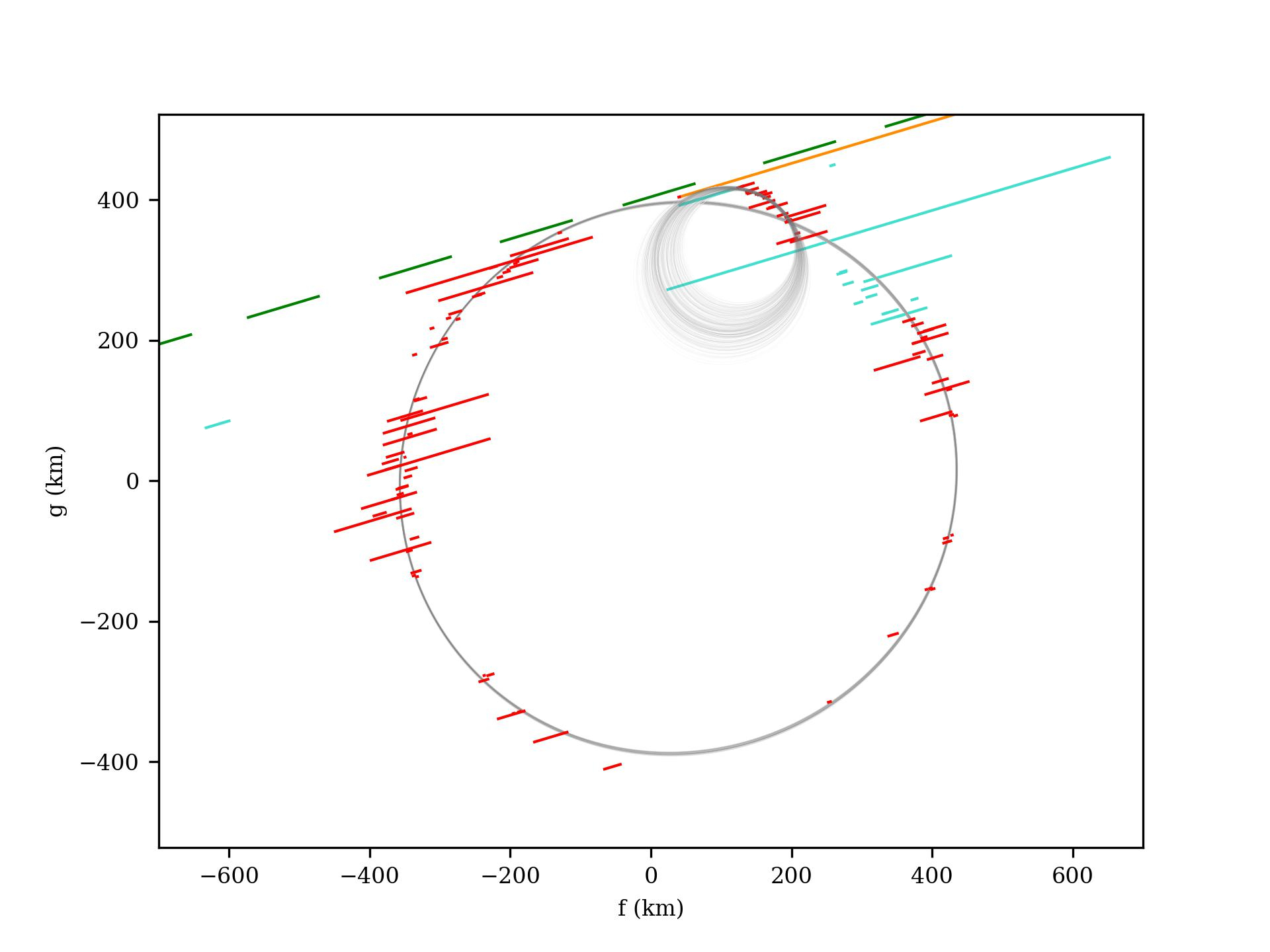}
    \caption{The gray regions show the 1$\sigma$ fitted limb for the main body and the putative satellite. Green segments show each exposure acquired from Montsec station. This data set was the closest negative chord at North. The orange segment shows the negative part of the Varages light curve. Red segments are the observed immersion and emersion ($1\sigma$) instants used for the limb fitting. The turquoise segments are the immersion and emersion ($1\sigma$) instants not used in the fits. }
    \label{fig:satellite_hip}
\end{figure}

Here we present a short discussion about a hypothesis raised by the referee of a mutual event by an unknown satellite causing the `bulge' on the observed limb of 2002 MS4 during the 8 August 2020 occultation. To explore such a possibility, I used all points between 5º and 25º to fit a circular limb of the putative satellite and the points between 50º and 340º to fit the main body limb. The satellite solutions were filtered by the negative part of the Varages light curve (orange line in Fig. \ref{fig:satellite_hip}). As a result, we derived a diameter of $\approx$ 788 km for the main body and $\approx$ 213 for the putative satellite. Considering both projected areas at the sky plane, the area's equivalent diameter would be about 808 km. This result does not explain the diameter obtained from thermal measurements. Unless the putative satellite has a significant oblateness or if it irradiates more in the thermal than the 2002 MS$_4$ surface. 

\end{appendix}

\end{document}